\RequirePackage[colorlinks,allcolors=blue]{hyperref} 

\documentclass[]{agujournal2019}
\usepackage{siunitx}
\usepackage[export]{adjustbox}
\usepackage{url} 
\usepackage{lineno}
\usepackage[finalnew]{trackchanges} 
\usepackage{soul}

\draftfalse

\journalname{JGR: Planets}

\begin{document}

%
%

\title{Structure\remove{and Energetics} of Jupiter's High-Latitude Storms:  Folded Filamentary Regions Revealed by Juno}

%
%




\authors{L.N. Fletcher\affil{1}, Z. Zhang\affil{2}, S. Brown\affil{3}, F.A. Oyafuso\affil{3}, J.H. Rogers\affil{4}, M.H. Wong\affil{5, 6}, A. Mura\affil{7}, G. Eichst\"{a}dt\affil{8}, G.S. Orton\affil{3}, S. Brueshaber\affil{9}, R. Sankar\affil{5}, C. Li\affil{10}, S.M. Levin\affil{3}, F. Biagiotti\affil{7}, T. Guillot\affil{11}, A. P. Ingersoll\affil{2}, D. Grassi\affil{7}, C.J. Hansen\affil{12}, S. Bolton\affil{13}, J.H. Waite\affil{14}}

\affiliation{1}{School of Physics and Astronomy, University of Leicester, University Road, Leicester, LE1 7RH, UK}
\affiliation{2}{California Institute of Technology, Pasadena, CA, USA}
\affiliation{3}{Jet Propulsion Laboratory, California Institute of Technology, 4800 Oak Grove Drive, Pasadena, CA 91109, USA}
\affiliation{4}{British Astronomical Association, London, UK.}
\affiliation{5}{Space Sciences Laboratory, University of California, Berkeley CA, USA}
\affiliation{6}{SETI Institute, Mountain View, CA, 94043-5139, USA}
\affiliation{7}{Istituto di Astrofisica e Planetologia Spaziali, Istituto Nazionale di Astrofisica, Roma, Italy}
\affiliation{8}{Independent scholar, Stuttgart, Germany}
\affiliation{9}{Michigan Technical University, Houghton, MI, USA.}
\affiliation{10}{University of Michigan, Ann Arbor, MI, USA.}
\affiliation{11}{Universiti\'{e} C\^{o}te d'Azur, OCA, Lagrange CNRS, 06304 Nice, France}
\affiliation{12}{Planetary Science Institute, Tucson, AZ, USA}
\affiliation{13}{Southwest Research Institute, San Antonio, Texas, TX, USA.}
\affiliation{14}{Department of Physics and Astronomy, The University of Alabama, Tuscaloosa, AL, USA}






\correspondingauthor{Leigh N. Fletcher}{leigh.fletcher@le.ac.uk}




\begin{keypoints}
\item Jupiter’s Folded Filamentary Regions (FFRs) dominate the appearance and lightning activity of cyclonic belts.
\item FFRs change from being microwave-bright at shallow pressures, to microwave-dark below the water cloud, and are detected to at least 100 bar.
\item FFRs promote moist convection and are the main source of lightning, particularly in belts near $52-57^\circ$N and poleward of $66^\circ$N.
\end{keypoints}

%
%

%
%


\begin{abstract}

Sprawling, turbulent cloud formations dominate the meteorology of Jupiter's mid-to-high latitudes, known as Folded Filamentary Regions (FFRs).  A multi-wavelength characterisation by Juno reveals the spatial distribution, vertical structure, and energetics of the FFRs.  The cloud tops display multiple lobes of stratiform aerosols, separated by darker, cloud-free lanes, and embedded with smaller eddies and high-altitude cumulus clouds. These cyclonic FFRs are microwave-bright in shallow-sounding wavelengths ($p<5$ bars) and microwave-dark in deep-sounding wavelengths ($p>10$ bars), with the transition potentially associated with the water condensation layer (6-7 bars).  Associating microwave contrasts with temperature anomalies, this implies despinning of cyclonic eddies above/below their mid-planes.  Despite deep roots (being detectable in wavelengths sounding $\sim100$ bars), they are ``pancake vortices'' with horizontal extents at least an order of magnitude larger than their depth.  In the northern hemisphere, FFRs are most common in cyclonic belts poleward of $40^\circ$N (all latitudes are planetocentric), particularly a North Polar Filamentary Belt (NPFB) near $66-70^\circ$N that defines the transition from organised belts/zones to the chaotic polar domain.  This distribution explains the high lightning rates from $45-80^\circ$N, peaking in a belt poleward of $52.3^\circ$N, which may trace the availability of water for moist convection.  Many observed lightning flashes can be associated to specific FFRs containing bright storms, but some FFRs display no activity, suggesting quiescent periods during storm evolution.  Analogies to Earth's oceanic eddies suggest that cyclones deform isentropic surfaces at their midplanes, raising deep water-rich layers upwards to promote moist convection, release latent heat, and inject clouds into the upper troposphere.

\end{abstract}

\section*{Plain Language Summary}

Swirls of chaotic, billowing white clouds dominate JunoCam images of Jupiter’s mid-to-high latitudes, casting shadows against the darker clouds below.  Their appearance is reminiscent of filamentary eddies in Earth’s oceans, revealed by contrasts in temperature, salinity, and phytoplankton blooms.  We use Juno’s suite of instruments spanning reflected sunlight (visible) and thermal emission (infrared and microwave) to determine common characteristics and 3D structure of these `Folded Filamentary Regions’ (FFRs), finding that they have deep roots (visible down to $\sim100$ bars), and display a change from being microwave-bright (i.e., warm and/or \add{ammonia-poor}) to microwave-dark (cool and/or ammonia-rich) as we probe deep below the water cloud.  FFRs dominate the weather systems at high latitudes:  they are mostly found in Jupiter’s belts (particularly those suspected to have the largest availability of water), and are often associated with lightning, explaining the high rates of flashes and sferics detected at high latitudes.  FFRs form a `North Polar Filamentary Belt’ that marks the transition into the polar domain.  These cyclonic eddies may be the key driver for Jupiter’s moist convective eruptions, and thus lightning, latent heat release, and internal heat flux at high latitudes.

%
%



\section{Introduction}
\label{intro}

A glimpse at the chaotic, sprawling cloud patterns at Jupiter's high latitudes \cite{17orton, 22rogers} might evoke comparisons with eddies and swirls in Earth's oceans, made visible by phytoplankton blooms, temperature, and salinity contrasts \cite<e.g.,>[]{23taylor}.  Unlike the organised system of belts and zones, the complexity and variety of Jupiter's high-latitude storms makes their characterisation daunting, and yet they appear to dominate the meteorology of Jupiter's mid-to-high latitudes, potentially providing the key route for vertical energy transport and heat release by moist convection.  \remove{These chaotic clouds hint at pent-up convective available potential energy waiting to be released, waiting for convection to be triggered.} 

In this survey we attempt to make sense of Jupiter's high-latitude meteorology, revealing the \add{spatial} distribution of `folded filamentary regions' (FFRs), their vertical structure, and their connection to lightning, leveraging Juno's multi-wavelength capabilities from the visible, infrared, and microwave to probe these features in three dimensions.  Juno investigations of Jovian meteorology have thus far focussed on global studies at large scales \cite<e.g., exploring belt/zone variability in environmental conditions and lightning activity,>[]{17li, 18brown, 20grassi, 21fletcher, 25sankar}, local studies of small scales \cite<e.g., individual vortices or storm eruptions,>[]{20wong, 21bolton, 22hueso, 25biagiotti, 25brueshaber}, or studies restricted to a single altitude \cite<e.g., at the cloud-tops,>[]{22rogers}.  The Juno record from 2016 to 2024 now allows us to explore how small-scale meteorological structures - the FFRs - shape the generalised meteorology of the northern hemisphere.  

\subsection{FFR Studies before Juno}
Pioneer imaging revealed filamentary high-latitude clouds, and the transition into the polar domains near $\sim65^\circ$ (all latitudes are planetocentric) in both hemispheres \cite{76gehrels}, but it was the Voyager flybys that enabled preliminary surveys of their complexity. \citeA{79mitchell} presented an overview of Voyager-1 imaging, describing the region poleward of $36^\circ$N as having ``large areas of puffy unorganised clouds and other regions of folded ribbon-like clouds.''  They categorised these cyclonic regions as `Type I' and contrasted them with `Type II' cyclones that lacked internal detail (e.g., sectors forming between mid-latitude anticyclonic ovals).  \citeA{79ingersoll} described the `folded filament' pattern as ``rapidly varying, disorganised, turbulent," with a short lifetime, and identified their occurrence in cyclonic shear zones (i.e., in belts). \citeA{79smith} noted the eruption of bright clouds within a compact oval-like cyclone near $38^\circ$S (the South South Temperate Belt, SSTB), with the Voyager imaging sequence showing the cyclone taking on a filamentary structure over a matter of days (their Fig. 8).  

There were attempts to connect these filamentary clouds to the occurrence of Jovian lightning, firstly by identifying the bands associated with lightning activity, and then attempting to connect lightning to individual cloud formations.  Jovian lightning appeared to be most optically active near $49-50^\circ$N \cite{04ingersoll}, from Voyager \cite{91magalhaes, 92borucki}, New Horizons \cite{07baines}, and Galileo \cite{99little}.  Voyager identified bands of lightning near $49^\circ$ and $60^\circ$N, but saw ``no unusual cloud features'' associated with these detections \cite{92borucki}.  New Horizons observed high-latitude lightning poleward of $60^\circ$N, and near $52^\circ$N \cite{07baines}, noting the relationship to cyclonic shear regions.   Indeed, all these nightside spacecraft image sets showed lightning flashes to be located in cyclonic regions -- or on the westward jets alongside them -- in regions that are dominated by FFRs.  Focussing on individual storms, \citeA{99little} tracked a thunderstorm at $49^\circ$N in Galileo images from the dayside to the nightside, noting that the lightning flashes occurred close to an ``isolated bright cloud,'' which we now identify as a bright storm within an FFR (their Fig. 6).  Similarly, \citeA{04dyudina} correlated Cassini/ISS images of clouds (dayside) and lightning flashes (nightside), finding that flashes coincided with a very bright storm in an FFR at $34^\circ$N and similar, larger eruptions in belts at $24^\circ$N and $13^\circ$S.

\subsection{Meteorology from Juno}
Juno's combination of a highly-inclined orbit, with close flyby passes (perijoves, PJ), and multi-wavelength remote sensing has enabled a step-change in our characterisation of Jupiter's high latitude meteorology.  \citeA{17orton} noted that FFRs were ``amorphously shaped regions'' and the brightest of the northern features seen by JunoCam during PJ1 (August 2016), ranging from ``compact'' to ``chaotic and sprawling'' with sizes as large as $\sim7000$ km.  They noted a particularly extended FFR region between $54-59^\circ$N (referred to as the N5 domain, see below).   At southern mid-latitudes, Juno and ground-based observers monitored the development of convective plumes within cyclones in the South Temperate Belt \cite{20inurrigarro, 22hueso}, reminiscent of those seen in the SSTB by Voyager.  

\citeA{22rogers} used JunoCam data to explore zonal jets and FFRs at Jupiter's southern high latitudes, identifying ``an irregular (but stable) belt of chaotic cyclonic regions'' poleward of the sinuous S6 jet ($64^\circ$S), and more FFRs scattered in the $70-80^\circ$S domain and in the S4 domain (poleward of the $48.6^\circ$S jet). Clouds within FFRs were bright and reflective in JunoCam's methane band (889 nm), suggesting aerosols reaching higher altitudes than the surroundings, which has been recently confirmed by JIRAM \cite<Jovian Infrared Auroral Mapper,>[]{17adriani} spectroscopy of an FFR near $40^\circ$N \cite{25biagiotti}. \citeA{22rogers} showed that although the interval between perijoves (53 days at the start of the mission) was too long to track individual FFRs, the use of amateur images (where FFRs are seen as ``irregular, pale patches'') with 2-to-4-day intervals indicated that FFRs were drifting westwards over time.  JunoCam maps of the southern FFRs, taken 77-88 minutes apart, revealed strong cyclonic circulation around the periphery of the FFRs, with local variations and subtle internal motions.  Similar synergy between JunoCam and ground-based images has enabled long-term tracking of cyclonic features in the S2 domain \cite{24rogers_epsc} and high northern domains \cite{22rogers_epsc}.

Juno has also proven adept at studying Jovian lightning activity, using flashes observed in the UV \cite{20giles} and optical \cite{20becker}, `whistlers' and Jupiter dispersed pulses (JDPs) detected by the Waves instrument in the radio \cite{18imai, 18kolmasova, 19imai}, and lightning sferics detected in the microwave \cite{18brown}.  \citeA{18brown} identified peaks in lightning activity in cyclonic belts (near 45$^\circ$N, 56$^\circ$N, 68$^\circ$N and 80$^\circ$N), but without identifying the specific features responsible for the discharges.  Clusters of lightning flashes observed during PJ6 \cite{18brown} were linked to deep (potentially water) clouds observed by Hubble in cyclones from $45-50^\circ$N \cite{20wong}, and FFRs and other areas of active moist convection were found to be marked by a convective cloud signature containing juxtaposed high-altitude clouds, clearings, and deep clouds \cite{20imai, 23wong}. Transient Luminous Events (TLEs) observed by Juno in the ultraviolet \cite{20giles} were linked to FFRs observed in the visible by JunoCam and Hubble.  And at lower latitudes, Juno has identified lightning flashes in much larger convective plumes in notable outbreaks at $8^\circ$N \cite{25brueshaber}.  The survey presented here aims to test whether FFRs are the dominant source of lightning at high latitudes to explain the distribution observed by \citeA{18brown}.

\subsection{Overview of FFR Survey}
In this work, we present a multi-wavelength characterisation of Jupiter's northern FFRs, revealing their vertical structure for the first time.  Section \ref{junocam-jiram} uses visible imaging from JunoCam and 5-$\mu$m-imaging from JIRAM to characterise the morphology of FFRs observed at the cloud tops.  Section \ref{mwr} searches for microwave signatures of FFRs, comparing data from JunoCam and the Microwave Radiometer (MWR).  We determine the spatial distribution of FFR activity in the northern hemisphere, and discover that FFRs are always microwave-bright in shallow-sounding observations (i.e., above the water cloud), but microwave-dark at wavelengths sensing deeper than the water cloud.  We show that FFRs are commonly located in a polar band that we call the ``north polar filamentary belt'' (NPFB, poleward of $66.0^\circ$N), marking the transition from the organised domains of mid-latitudes to the polar domain.  Section \ref{lightning} relates MWR lightning detections to visible and microwave maps of FFR activity, revealing that the majority of Jupiter's northern hemisphere lightning activity is associated with these FFRs, and that the FFRs' preferential location in cyclonic belts explains the prevalence of lightning detections in belts.  Section \ref{discussion} then discusses the implications of the microwave inversion, cloud morphology, and lightning activity for the vertical structure of FFRs, and their importance for energy transport in Jupiter's high latitudes.  The paper ends with a comparison to Saturn's convective storms, and to surface and subsurface eddies that are ubiquitous in Earth's oceans.


\subsection{Belt/Zone Nomenclature}
In this paper, we adopt labels N1 through N7 to describe the northern domains in Jupiter's banded structure \cite<e.g.,>[]{17rogers_baa}.  Each domain sits between two prograde jets, and has a weaker retrograde jet near its centre, separating a cyclonic belt (equatorward side) from an anticyclonic zone (poleward side).  The northern temperate belts have been identified as microwave-bright in the upper troposphere \cite{20oyafuso, 21fletcher}, warm in the mid-infrared sounding 0.1-0.6 bars \cite{16fletcher_texes, 24bardet}, and sometimes darker in colour compared to the neighbouring zones.  The subdivision into domains is natural, as eastward jets provide strong potential vorticity (PV) boundaries, such that PV within a domain is largely homogenised by mixing \cite{06read_jup, 11marcus}, leading to a `PV staircase.'  Regions surrounding the westward jets in the centre of the domain are known to fulfil the necessary (but not sufficient) criteria for barotropic \cite{79ingersoll} and baroclinic \cite{06read_jup} instabilities, possibly resulting in the eddies and storms discussed in this work.  We use Cassini/ISS cloud tracking measurements of prograde jets \cite{03porco} to define the southern edges of these domains as follows (all latitudes are planetocentric):  N1 ($21.3^\circ$N, containing the North Temperate Belt and Zone), N2 ($31.6^\circ$N, containing the North North Temperate Belt and Zone), N3 ($39.0^\circ$N), N4 ($43.3^\circ$N), N5 ($52.3^\circ$N), N6 ($61.0^\circ$N), and N7 ($66.0^\circ$N, equatorward of the broad north polar region, NPR).  See \ref{appendix1} for the use of similar nomenclature for the southern domains.




\section{JunoCam and JIRAM}
\label{junocam-jiram}


\subsection{JunoCam Morphology}
\label{junocam}
\begin{figure}[ht]
\makebox[\textwidth]{%
\noindent\includegraphics[width=1.3\textwidth]{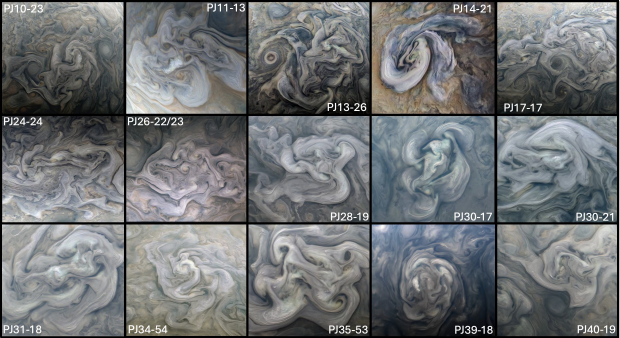}
}%
\caption{JunoCam observations of FFRs.  All images are credited to NASA/JPL-Caltech/SwRI/MSSS, and were processed by Kevin M. Gill.  Each panel is labelled by the perijove number and image number. Illumination, spatial scale, and lightning conditions change from panel to panel, and these should not be taken as natural colour.   Bright white storms are notable in PJ11-13, PJ30-17, PJ39-18, PJ35-53, PJ31-18, PJ34-54, and PJ30-21, while PJ26-22/23 and PJ24-24 are examples of more quiescent FFRs. 
}
\label{junocam_ffrs}
\end{figure}

\begin{figure}[!h]
\makebox[\textwidth]{%
\noindent\includegraphics[width=1.3\textwidth]{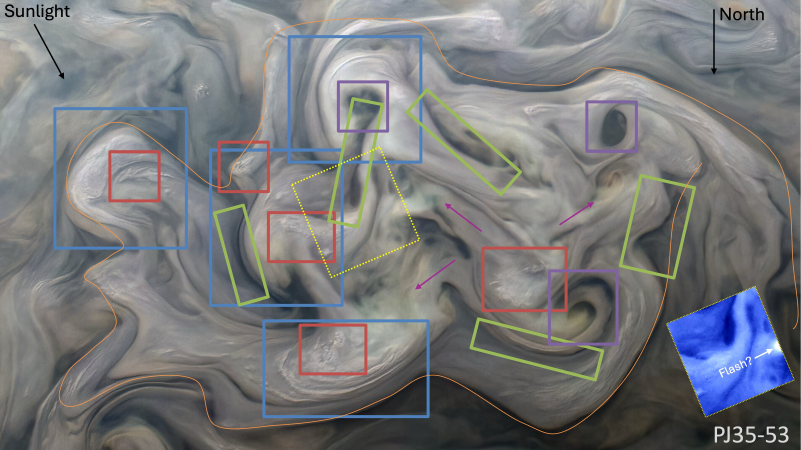}
}%
\caption{View of a single FFR in PJ35, located near $50^\circ$N (the N4 domain), $290^\circ$W on 2021-Jul-21 (Credit: NASA/JPL-Caltech/SwRI/MSSS/Kevin M. Gill).  The north pole is downwards, the direction of sunlight approximately SSE as indicated.  Key features are identified as stratiform lobes (blue boxes), dark lanes (green boxes), discrete eddies/whorls (purple boxes), cumulus clusters (red boxes), and green hazes (magenta arrows), all contained within the broad area of the FFR (orange outline).  Note that cyclonic motion is anticlockwise, and anticyclonic motion is clockwise.  This image spans approximately $5\times10^\circ$ in latitude and longitude.  The blue inset image shows a single frame with a suspected lightning flash at the edge, and represents the yellow dotted box in the image centre.}
\label{junocam_morphology}
\end{figure}

Visible-light imaging from JunoCam has provided exceptional views of Jovian meteorological features throughout the mission  \cite<e.g.,>[]{17orton_juno, 22rogers, 24sankar, 25brueshaber}.  The JunoCam instrument and its operation are described in \citeA{17hansen}, and the image processing pathway is described in \citeA{22rogers}. JunoCam is a visible-light camera that uses `push-frame' imaging, relying on spacecraft spin to build spatial coverage without any moving camera mechanisms. JunoCam's CCD has four adjacent colour filters associated with four $1648\times128$ pixel regions on the detector.  The horizontal field of view is $58^\circ$ ($\sim633$ $\mu$rad/pixel), with a $5^\circ$ region spanned by the four filter strips.  Each image is acquired in broadband red, green, and blue filters, or in a narrow-band filter centred on the methane absorption band at 889 nm.  The images are all processed and projected into cylindrical and polar maps by G. \add{Eichst\"{a}dt}; the individual maps are then merged into global or regional composite maps.  For the present paper, some images have also been processed and map-projected by other citizen scientists as credited.  Although a quantitative assessment of FFRs in JunoCam data will be the subject of a future study, we provide a qualitative description of common FFRs morphologies with reference to the montage of images in Figs. \ref{junocam_ffrs} and \ref{junocam_morphology}. 

Fig. \ref{junocam_ffrs} reveals the exceptional variety displayed by this class of features.  Some (e.g., PJ39-18) appear compact and quasi-circular in shape, somewhat similar to the organised appearance of the circumpolar cyclones.  Others have powerful swirls of bright cloud material \cite<PJ14-21, nicknamed the `Nautilus',>{24guillot_egu}.  But the majority of the FFRs are sprawling chaotic masses of overlapping and swirling cloud patterns, with smaller dark eddies superimposed onto larger cloud filaments, and criss-crossed by darker lanes. The extent of the turbulence varies from feature to feature, and the short duration of JunoCam imaging means that short-term motions (expected to be cyclonic based on the morphology) and long-term evolution remain poorly constrained.  

Fig. \ref{junocam_morphology} annotates just one example (PJ35 image JNCE\_2021202\_35C00053, targeting $56.1^\circ$N, $288^\circ$W, taken at 2021-07-21T08:05 from an altitude of 8,862 km) to describe the morphology of an FFR in the N4 domain.  The FFR is mostly encompassed by the thin orange line which, given the `tail' to the right-hand side of the image, implies anti-clockwise motion (i.e., cyclonic rotation in the northern hemisphere).  The feature measures approximately $4^\circ$ latitude by $9^\circ$ longitude ($\sim4800\times6900$ km).  Blue boxes indicate white aerosols in \textit{stratiform clouds}, appearing to spread radially outwards from the centre of the FFR like hammer-head shaped lobes.  These hammer-head clouds suggest expansion and possible eddying both cyclonically (to the left) and anticyclonically (to the right).  Dark shadows can be seen NNW of the FFR (sunlight is coming in from the SSE), suggesting elevated aerosols in the stratiform lobes casting shadows on the deeper background clouds. The filamentary structure emerges from multiple lobes apparently interacting with one another.  Green boxes indicate \textit{dark lanes} that separate the white lobes, and purple boxes indicate discrete \textit{eddies/whorls} in the cloud layer - whether these are darker aerosols or an absence of clouds will be discussed below.  These small eddies range from $0.5-1.0^\circ$ in size ($\sim$335-770 km width).  Red boxes reveal \textit{clusters of cumulus clouds} embedded within the stratiform lobes.  Such embedded cumulus are seen across Jupiter at the $\sim10$-km scale of the highest resolution JunoCam images \cite<`pop-up' clouds,>[]{19hansen_agu}, and are suggestive (but not conclusive) of moist convection.  Furthermore, the organisation of cumulus into curvilinear features are suggestive (but not conclusive) of squall lines.  In many (but not all) FFRs, there is at least one very dense, very bright white patch of these cumuliform – or perhaps cumulonimbus – clouds, hereinafter referred to as `bright white storms,' and we will show evidence that these are likely thunderstorms.  Adjacent to these storms, magenta arrows point towards a \textit{greenish tinge} observed in several FFRs (e.g., PJ30-17, PJ30-21, PJ14-21, PJ13-26).  This is spatially indistinct, but might indicate spectrally-unusual gases or aerosols brought into the cloud layer by convective upwelling within the FFR.  The green haze appears in images processed by multiple citizen scientists, although the visibility depends on the degree of saturation in the colour composition.  

The raw JunoCam data features a bright spot on the edge of a single colour frame that is tentatively identified as a lightning flash (first spotted by citizen scientist Brian Swift).  The spot is considerably brighter than the surrounding clouds, and visually extended (as might be expected for a lightning flash filtered through overlying cloud and haze layers).  This spike has been removed from most citizen-science compositions, but is suggestive of a single lightning flash within the centre of the FFR ($50^\circ$N, $288^\circ$W), coincident with the greenish hazes in the image centre.  It is rare to see visible lightning at these latitudes due to the strong solar illumination, and we believe this to be the first visible lightning flash reported in this domain in JunoCam images.


\begin{figure}[!h]
\makebox[\textwidth]{%
\noindent\includegraphics[width=1.3\textwidth]{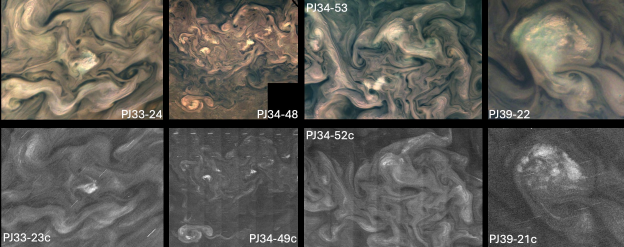}
}%

\caption{Four examples (PJ33-39) where JunoCam RGB images (top row) were immediately preceded by CH$_4$-band images of the same feature (bottom row), revealing the vertical structure of the features within the FFRs.  Visibly bright cumulus-like clusters (bright white storms) are higher than the surrounding stratiform clouds; visibly dark lanes are deeper than the stratiform lobes.  All images processed by Gerald Eichst\"{a}dt, and credited to NASA/JPL-Caltech/SwRI/MSSS. }
\label{junocam_ch4}
\end{figure}

The RGB images give the impression of a 3D view, with overlapping filaments.  However, assessment of the vertical structure of these FFRs requires extension to the infrared.  JunoCam's 890-nm channel sounds a strong CH$_4$ absorption, sensing aerosols in the upper troposphere, but unfortunately the FFR in Fig. \ref{junocam_morphology} was not observed in this filter.  In Fig. \ref{junocam_ch4} we display a gallery of four examples (PJ34-39) where JunoCam RGB images were immediately preceded by CH$_4$-band images, revealing (i) maximum brightness in the cumulus clusters suggesting they reach higher than the surrounding stratiform lobes; (ii) moderate brightness of the stratiform lobes compared to (iii) low brightness in the dark lanes and whorls, suggesting the lowest cloud layers.  These are cropped images that have not been mapped onto a latitude/longitude grid (the observing angle for the FFRs changed from one image to the next, so images have been approximately co-aligned).  The CH$_4$-band images exhibit banding artefacts related to the overlaps of adjacent JunoCam frames.  Finally, the greenish haze can be observed in PJ33-24, PJ34-53 and PJ39-22, diffuse but surrounding some of the highest features observed in the CH$_4$-band images.

\subsection{Comparing JunoCam and JIRAM}
\label{jiram}

JunoCam senses scattered sunlight from aerosols in Jupiter's upper troposphere and lower stratosphere, but to get a sense of the clouds in the deeper troposphere we require 5-$\mu$m `M-band' imaging from Juno's JIRAM instrument \cite{17adriani}. JIRAM acquires rectangular images as Juno spins at 2 rpm, using a compensating mirror at the telescope entrance to `despin' the rapid rotation, and a beam splitter to split the light into the imager and spectrometer channels. The M-band images have a wavelength of $4.78\pm0.48$ $\mu$m, and the $128\times432$ pixel detector results in a field of view of $1.75^\circ \times 5.94^\circ$ (i.e., $\sim240$ $\mu$rad/pixel).  We assemble many hundreds of individual JIRAM frames into latitude-longitude maps and polar projections for each perijove, using SPICE \cite<via SpiceyPy,>[]{20annex} to assign the geometry to each pixel \cite{18fletcher_waves}.  The spatial resolution of the images varies dramatically during the perijove, being closest over northern latitudes (where coverage is sparse) and most distant over the southern pole.  

\begin{figure}[!h]
\makebox[\textwidth]{%
	\includegraphics[width=1.3\textwidth]{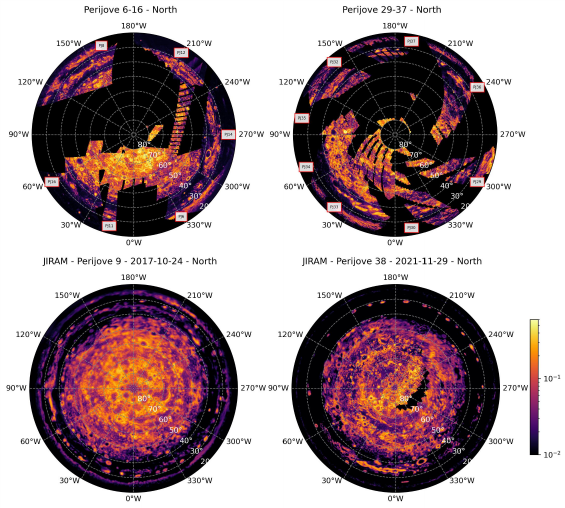}
}%
\caption{North polar projections (equidistant azimuthal) of JIRAM observations at 5 $\mu$m to showcase examples of spatial coverage $20-90^\circ$N.  Top row provides a combination of high-resolution coverage from multiple perijoves with limited spatial coverage (overlapping observations poleward of $80^\circ$N have been deliberately omitted).  Bottom row shows examples of individual perijoves with poorer spatial resolution but near global coverage of the northern hemisphere.  Brightness units are spectral radiance (Wm$^{-2}$sr$^{-1}\mu$m$^{-1}$), as shown by the logarithmic colour scale.  These data have not been corrected for limb darkening effects.}
\label{jiram_coverage}
\end{figure}

The challenge of investigating FFRs using JIRAM data is exemplified by Fig. \ref{jiram_coverage}, which demonstrates the coverage of the northern hemisphere under a range of conditions.  The top row shows montages of observations from multiple perijoves, PJ6-16 on top left, and PJ29-37 top right, as examples where the close proximity to Jupiter restricts the spatial coverage of northern mid-latitudes.  The bottom row shows examples of improved spatial coverage for individual perijoves (PJ9 and PJ38) at the expense of lower spatial resolution (i.e., mapping sequences taken as Juno was approaching the north pole).  Given the limited coverage and the gaps between perijoves, it is impossible to track one FFR from one perijove to the next - this is also confirmed for the southern hemisphere in \ref{appendix1}, where the spatial coverage is more extensive.   Nevertheless, some broad trends are apparent in Fig. \ref{jiram_coverage}:  the cloud cover at latitudes equatorward of $40^\circ$N (i.e., the tropical and N1-N2 domains) obscures 5-$\mu$m emission, making it appear dark and only broken by elongated `barge-like' structures and bright rings around anticyclones in the N2 domain ($30-40^\circ$N).  Conversely, thinner clouds permit brighter 5-$\mu$m emission for the N3-N7 domains.  This diffuse 5-$\mu$m glow poleward of $40^\circ$N \cite<due to reduced opacity in clouds and hazes at $p<0.5$ bars,>[]{23wong} serves as the background for the smaller-scale meteorological features - dark patches (associated with FFRs, as we explore below), bright-ringed anticyclonic ovals \cite{10depater_jup}, and the octagon of circumpolar cyclones \cite{18adriani_jiram} at the north pole.

\begin{figure}[!t] 
\makebox[\textwidth]{%
    \includegraphics[width=1.5\textwidth]{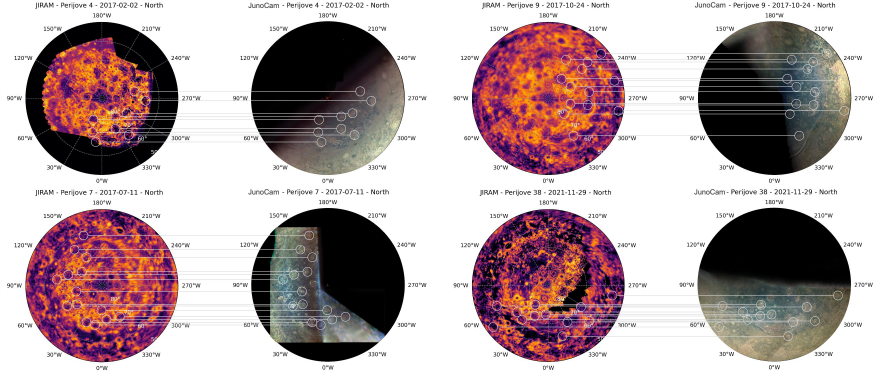}
}%
\caption{Comparison of polar equidistant azimuthal projections (50-90$^\circ$N) from JIRAM 5-$\mu$m imaging and JunoCam RGB imaging, selecting four perijoves (4, 7, 9 and 38) with the best overlap in spatial coverage.  Two grey circles, connected by a horizontal line, are used to guide the eye for matching FFR features (only a subset of features are identified in this way).}
\label{jiram-junocam-polar}
\end{figure}

Given that JunoCam requires illuminated conditions, whereas JIRAM does not, there were only limited opportunities to identify FFRs that were observed by both instruments. Moreover, when they were, the JIRAM images were usually obtained on the inbound trajectory before high-resolution JunoCam imaging was enabled.  Fig. \ref{jiram-junocam-polar} displays equidistant azimuthal north polar projections of JunoCam and JIRAM data side by side, focussing on the northern hemisphere poleward of $50^\circ$N.  We select only those perijoves with the most extensive northern coverage - PJ4, 7, 9 and 38 \cite<maps of all PJs are available in our supporting data,>[]{25fletcher_data}.  Composite JunoCam maps are assembled following \citeA{22rogers}, with qualitative modifications to brightness gradients and contrast to improve overlaps between images. Circles are used to compare FFR features in both images, revealing all FFR features to be dark silhouettes against the diffuse 5-$\mu$m glow that characterises the northern hemisphere.  

However, there are many more 5-$\mu$m-dark features in the JIRAM maps than can be explained as visibly bright FFRs, and the JunoCam images are only suitable for identifying reflective FFRs over $\sim100^\circ$ of longitude, meaning that neither JunoCam nor JIRAM are suitable for tracking global FFR statistics in a single perijove.  Initial attempts at using JIRAM alone to determine the distribution of FFRs with latitude (not shown) hinted at large number of FFRs occurring in a broad band at the edge of the polar domain (the `north polar filamentary belt' or NPFB poleward of the N7 jet at $66^\circ$N), compared to an absence of FFRs in a `bland zone' within the N6 domain ($63-66^\circ$N), but these statistics were hampered by the challenge of uniquely identifying FFRs in JIRAM data.

\begin{figure}[!h]
\makebox[\textwidth]{%
    \includegraphics[width=1.3\textwidth]{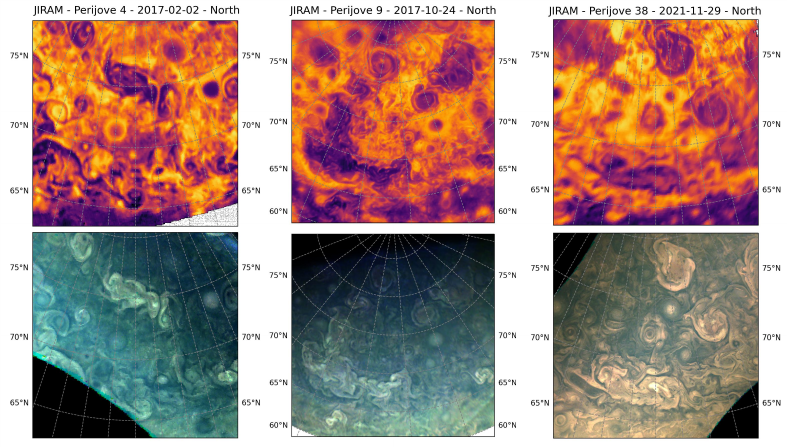}
}%
\caption{Orthographic projections of JIRAM and JunoCam images providing close-ups of FFRs on PJ4, PJ9 and PJ38, giving examples of how the visible-light morphology is related to 5 $\mu$m emission.  Note that although JIRAM has the higher spatial resolution ($\sim240$ $\mu$rad/pixel), the JunoCam observations ($\sim633$ $\mu$rad/pixel) were taken when Juno was closer to Jupiter.}
\label{jiram-junocam-ortho}
\end{figure}

As a final comparison, we zoom in on a subset of features observed by both JIRAM and JunoCam in Fig. \ref{jiram-junocam-ortho}, demonstrating an excellent correspondence between features observed by these two instruments.  This comparison confirms the visibly bright FFRs as dark at 5 $\mu$m, with thick clouds obscuring the bright emission emerging from the deeper atmosphere \cite<confirmed by higher optical depths measured within an FFR by JIRAM spectroscopy,>[]{25biagiotti}.  Sprawling multi-lobed FFRs, particularly those seen near $65-70^\circ$N during PJ9, show that the visibly dark lanes are actually gaps in the clouds, permitting 5 $\mu$m flux to escape.  The same is true of the smaller whorls/eddies embedded within the stratiform clouds, such as those seen in the FFR between $75-80^\circ$N during PJ38.  Anticyclonic white ovals that are unrelated to the FFRs also appear dark and cloudy at 5 $\mu$m, but display less internal structure:  they are broad and featureless ovals of cloud, lacking that filamentary structure that is common in the cyclonic FFRs.  Nevertheless, the complexity of the JIRAM images in Fig. \ref{jiram-junocam-ortho} confirms that identification of FFRs at 5 $\mu$m will be challenging.  We therefore turn to Juno's microwave radiometer to search for a reliable FFR detector.

\section{Juno Microwave Radiometer (MWR)}
\label{mwr}
Juno's Microwave Radiometer \cite{17janssen} measures the angular dependence of Jupiter's microwave brightness as Juno spins (2 rpm), using six antennae/channels with wavelengths between 1.4 and 50 cm sensitive to thermal emission, with opacity  primarily from extremely pressure-broadened NH$_3$ and H$_2$O features.  Channels 2-6 (1.4-24 cm) are all co-boresighted, channel 1 (50 cm) points in a different direction.  Channel 6 (1.4 cm) has the narrowest beam ($11^\circ$ beam half-power width), the remaining shallow sounding channels 3-5 (3.5-11.5 cm) have a beam of $12^\circ$, whereas the deepest sounding channels 1-2 (24-50 cm) have a larger width of $20^\circ$ \cite{17janssen, 20zhang}.  This means that discrete atmospheric features like FFRs are more likely to be observed at shorter wavelengths.  Previous studies exploring Jupiter's discrete features have focussed on hotspots, rifts, and storms at tropical latitudes \cite{20fletcher, 25brueshaber, 25moeckel}; cyclones and anticyclones at mid-latitudes \cite{21bolton}; and the circumpolar cyclones \cite{24hu_agu}.  This study focusses on using MWR to detect and characterise the vertical structure of FFRs.

\subsection{Nadir-equivalent brightness temperature}  

Microwave-bright features in Jupiter's northern temperate latitudes were initially identified in the raw antenna temperatures $T_A$ \cite{22fletcher_epsc}, but as these are convolved with the MWR antenna pattern, the brightness contrasts were difficult to interpret.  Antenna temperatures depend both on the angular dependence of the thermal emission and the details of how the MWR antenna pattern intersects with Jupiter's surface. 

The deconvolution of the antenna pattern to generate `pencil-beam' brightness temperatures ($T_B$) from $T_A$ follows the recipes explained in \citeA{20zhang}, incorporating all perijoves up to PJ61 to iteratively estimate the zonal (i.e., north-south) brightness contrasts.  In the first iteration, the $T_A$ are used to derive a first estimate of the zonal-mean atmospheric $T_B$ model as follows:  (i) we screen $T_A$ for all perijoves in each latitude bin to remove spurious contaminations from synchrotron radiation and microwave opacity from high-density electron regions formed by auroral electron precipitation \cite{20hodges, 25bhattacharya}, and then fit a quadratic polynomial \add{in emission angle} to the data in each latitude bin following the formulation of \citeA{20oyafuso}.  Secondly (ii), we screen out any outliers that lie more than a standard deviation away from the median value over all perijoves for that latitude and emission angle.  Thirdly (iii), these filtered data are fitted with a new quadratic, which we use as our zonally- and perijove-averaged $T_B$ model.  Fourthly (iv), the $T_B$ model is convolved with the MWR beam to compute the residual between modelled and measured $T_A$ for each observation.  This residual is then added to the zonally-averaged $T_B$ to become a new estimate of $T_B$, replacing the zonal-mean $T_A$ in step (i) for subsequent iterations.  We repeat steps (i) through (iv) over multiple iterations until the zonal-mean $T_B$ no longer changes \cite<i.e., the solution converges,>[]{20zhang}.

\begin{figure}[t]
\centering
\includegraphics[width=\textwidth]{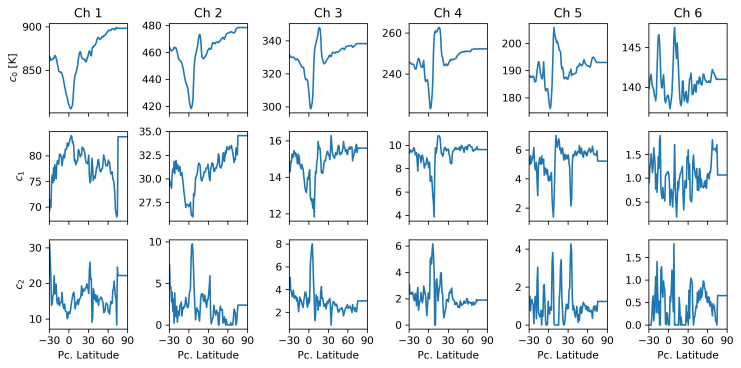}
\caption{Coefficients used to determine the nadir $T_B$ (coefficient $c_0$) and limb darkening ($c_1$, $c_2$) for the perijove-averaged atmosphere, using all data from PJ1-61, showing results between $30^\circ$S and $90^\circ$N, with brightness gradients poleward of $60^\circ$N being presented for the first time.  These are updated compared to PJ1-12 calculations in \citeA{20oyafuso}, but latitudinally-uniform values are used polewards of $75^\circ$N.}
\label{MWR_Ccoeffs}
\end{figure}

\begin{figure}[t]
\centering
\includegraphics[width=\textwidth]{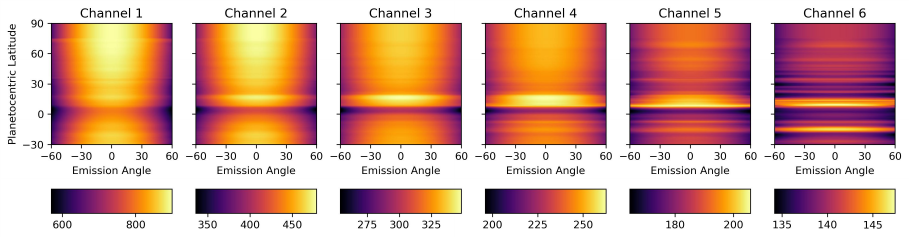}
\caption{Brightness temperatures [K] from $30^\circ$S to $90^\circ$N shown for all six MWR channels, estimated iteratively using all perijoves from PJ1 to PJ61.  These are shown as a function of emission angle to demonstrate limb darkening, and are symmetric about $0^\circ$.  Removal of this background belt/zone structure enables mapping of FFRs.}
\label{MWR_nadirTB}
\end{figure}

The coefficients of the final $T_B$ model over all perijoves are shown in Fig. \ref{MWR_Ccoeffs}, where $c_0$ corresponds to the nadir-equivalent brightness temperature, and $c_1$ and $c_2$ capture the extent of limb darkening \cite{20oyafuso}.  Although it would be preferable to treat each perijove independently, individual PJs are not always optimised for MWR - for example, Juno's `gravity orbits' (during which Juno's high-gain antenna was held to a fixed Earth-pointing orientation) sometimes yield such a small range of emission angles for MWR that limb-darkening coefficients would be impossible to estimate.  Each latitude has its own combination of coefficients in Fig. \ref{MWR_Ccoeffs}, except poleward of $75^\circ$N, where we adopt a uniform value for the coefficients.  Here the presence of the polar cyclone (PC), and the smaller number of datapoints per latitude bin, prevent us from developing a meaningful latitudinally-dependent $T_B$ fit. The perijove-averaged $T_B$ as a function of latitude and emission angle is shown for each channel in Fig. \ref{MWR_nadirTB}, which was used to define the correction factor needed to transform off-nadir $T_B$ to the nadir equivalent $T_B$ (i.e., emission angle of $0^\circ$).  Global $T_B$ maps combining all the MWR observations can be seen in \ref{appendix2}, with synchrotron/auroral contaminations primarily affecting channel 1, but also evident in channels 2 and 3 (i.e., longward of 10 cm wavelength).  In the next Section, we subtract the perijove-averaged nadir-equivalent $T_B$ from every point, resulting in maps that show $\Delta T_B$ from the zonal average.



\subsection{MWR as an FFR Detector}

MWR observes Jupiter in narrow longitudinal swaths at each perijove, beginning with broader coverage over the north pole before becoming narrower at low latitudes (near perijove).  The PJ latitude moved from $3.8^\circ$N in August 2016 (PJ1) to $33.1^\circ$N in February 2022 (PJ40) and $51.3^\circ$N by May 2024 (PJ61), enabling improving views of northern latitudes.  However, the orbital geometry between PJ38 and PJ51 produced more limited longitudinal coverage, so we initially focus only on PJs between May 2017 (PJ6) and October 2021 (PJ37) in this part of the study, approximately spanning Juno's prime mission (July 2016 to July 2021).

\begin{figure}[!htbp]
\makebox[\textwidth]{%
	\includegraphics[width=1.5\textwidth]{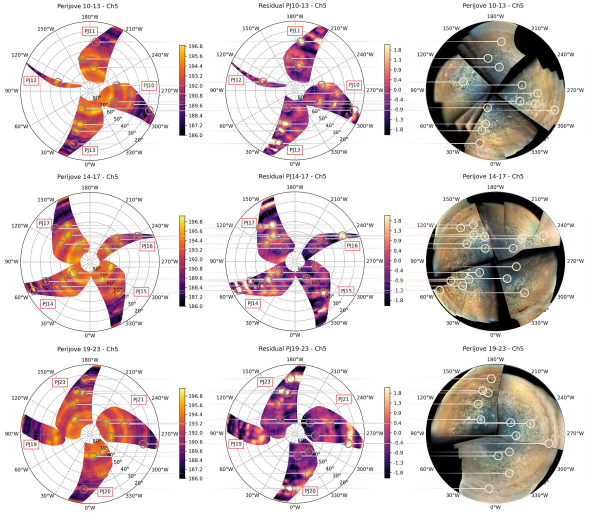}
}%
\caption{MWR Channel 5 (3.0 cm, sounding 1.6 bars) compared to JunoCam observations over the northern latitudes at the same perijoves.  Perijoves are shown in approximate `quartets' to span a broad range of longitudes over a $\sim6$-month period:  PJ10-13 in the top row, PJ14-17 in the middle row; PJ19-23 in the bottom row.  The viewing geometries are somewhat different for PJ12 (a ``tilt'' orbit where the spin axis was adjusted to compensate for Jupiter's rotation) and PJ19 (where Juno's spin orientation was ``cross-track'').  There was no JunoCam imaging of the northern hemisphere at PJ19. The left column shows the nadir-equivalent $T_B$, the central column shows the $\Delta T_B$ (with red indicating microwave-bright features), the right column shows JunoCam polar projections for comparison.  Horizontally-connected circles are used to qualitatively connect FFR features visible in both instruments (note that microwave-bright `brown barges' near $35^\circ$N are not circled) - a more quantitative assessment of FFR locations is included in the main text.  A version without the annotation circles is available in our supporting data \cite{25fletcher_data}.}
\label{ch5-junocam1}
\end{figure}

\begin{figure}[!htbp]
\makebox[\textwidth]{%
\includegraphics[width=1.5\textwidth]{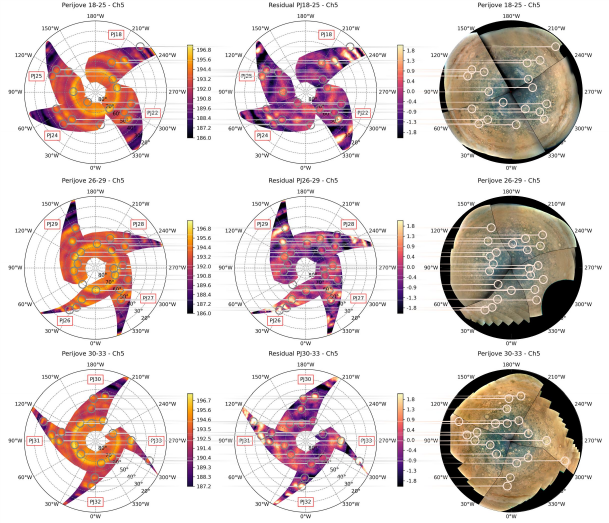}
}%
\caption{As for Fig. \ref{ch5-junocam1}, but for PJ18-25 (top row), PJ26-29 (middle row), and PJ30-33 (bottom row).  A version without the annotation circles is available in our supporting data \cite{25fletcher_data}.}
\label{ch5-junocam2}
\end{figure}

During the prime mission, the orbital geometry was such that PJ longitudes were approximately $90^\circ$ apart on consecutive perijoves, so we display maps of nadir-equivalent $T_B$ (and $\Delta T_B$) in `quartets' of perijoves.  A representative selection is shown in Figs. \ref{ch5-junocam1}-\ref{ch5-junocam2}, using JunoCam visible-light observations of the same perijoves to connect microwave-bright features (as seen in Channel 5, 3.0 cm, sounding contrasts at 1.6 bars) to visibly bright FFRs.  Channel 5 was selected for display as contrasts are smaller in channel 4 (5.75 cm sounding 5 bars), and channel 6 is somewhat noisier (1.4 cm, sounding 0.65 bar).  The depths sounded at nadir by each channel are based on contribution functions calculated in Fig. 6 of \citeA{21fletcher}.  MWR data are only shown for emission angles smaller than $75^\circ$ and a spacecraft range to Jupiter's 1-bar surface smaller than 2 $R_J$.

Figs. \ref{ch5-junocam1}-\ref{ch5-junocam2} show Jupiter's banded structure of microwave-bright belts and microwave-dark zones \cite{17li, 17ingersoll, 21fletcher, 21duer}, culminating in a microwave-bright band in the N7 domain (i.e., the north polar filamentary belt poleward of the N7 prograde jet at $66^\circ$N).  By subtracting the perijove-averaged atmosphere at every latitude to get $\Delta T_B$, we can see that there are significant longitudinal perturbations to the brightness (these are sometimes visible in the $T_B$ maps, too).  Circles connected by horizontal lines provide a qualitative guide to show how microwave-bright features are aligned with FFRs observed with JunoCam.  We stress that this is typical but not universal:  some microwave-bright features do not have clearly contrasted visible-light counterparts, and some visible FFRs do not exhibit strong microwave contrasts.  Furthermore, elongated `brown barges' \add{(which occur primarily in the N2 domain near $35^\circ$N)} are also microwave-bright, such as the barge captured by \citeA{21bolton} during PJ19.  Nevertheless, the correspondence is good, particularly when considered over the 24 independent perijoves shown in Figs. \ref{ch5-junocam1}-\ref{ch5-junocam2}.  One specific example is the `Nautilus'-like filamentary region captured during PJ14 at $41^\circ$N, $65^\circ$W in Fig. \ref{ch5-junocam1} \cite{24guillot_egu}, which is shown in more detail in Fig. \ref{junocam_ffrs}.  We can conclude that microwave-bright features in the N3 through N7 domains are typically associated with visible FFRs, either due to an absence of ammonia opacity, or due to elevated atmospheric temperatures.

\subsection{FFR Microwave Brightness Inversion}

\begin{figure}[!h]
\makebox[\textwidth]{%
\includegraphics[width=1.5\textwidth]{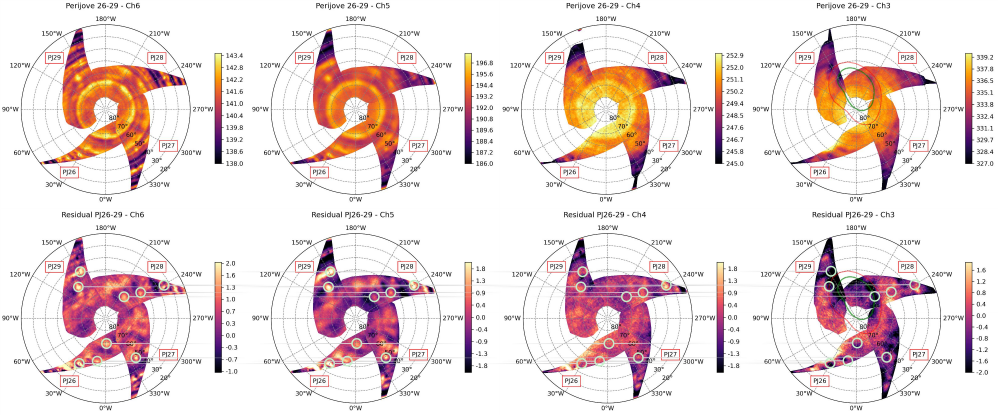}
}%
\caption{Quartets of nadir-equivalent $T_B$ [K] for MWR channels 3-6 (top row) and the residual $\Delta T_B$ once the perijove-averaged brightnesses are subtracted.  Light green circles are used as a qualitative guide to show how features appear in each channel, revealing a change in microwave contrast from shallow-sounding to deep-sounding channels.  Note that cold plasma associated with the northern auroral oval caused decreased microwave brightness at the longest wavelengths - these are apparent as a dark region in channel 3 between the green oval \cite<the location of the main auroral oval derived by>[]{17nichols} and the red oval \cite<the locus of Io footprint locations from>[]{03grodent} - small scale channel-3 features should not be trusted in this location. A version without the annotation circles is available in our supporting data \cite{25fletcher_data}.}
\label{mwr_pj26-29}
\end{figure}

We now expand our use of MWR to consider other channels, to understand the vertical structure of FFRs in the troposphere.  A comparison of channels 3-6 for one quartet (PJ26-29) is shown in Fig. \ref{mwr_pj26-29} - channels 1 and 2 could not be used to study FFRs during the prime mission, as the spacecraft distance to northern mid-latitudes was too great for these broad beams to resolve the compact FFR signatures.  Qualitatively, we can see that microwave-bright features in channel 5 (1.6 bars) have counterparts in the other channels - they remain bright in shallow-sounding channel 6 (0.65 bars), their contrast is reduced but still apparent in channel 4 (5 bars), but there is an interesting transition in channel 3 (11.5 cm, sounding down to 14 bars).  Here, there is a hint that the microwave-bright features at $p<5$ bars have microwave-dark counterparts at $p>10$ bars - we shall better quantify this correlation below.  

\begin{figure}[!h]
\makebox[\textwidth]{%
\includegraphics[width=1.5\textwidth]{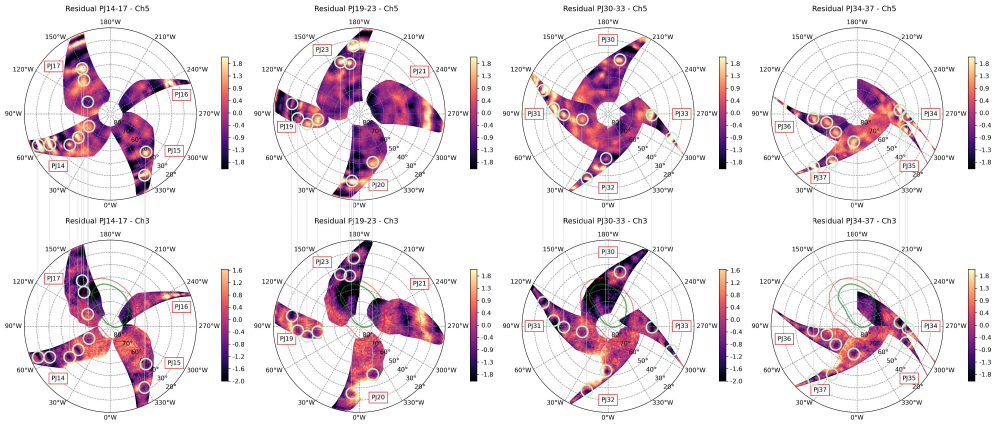}
}%
\caption{Comparing $\Delta T_B$ residuals [K] for MWR channel 5 (3.0 cm, sounding 1.6 bars in the top row) and channel 3 (11.5 cm, sounding 14 bars in the bottom row), revealing the inversion in microwave brightness from the shallow atmosphere to the deep atmosphere.  Circles connected by vertical lines guide the eye for similar features.  The auroral oval \cite<green>[]{17nichols} and Io footprints \cite<red,>[]{03grodent} are shown to indicate auroral-affected regions. A version without the annotation circles is available in our supporting data \cite{25fletcher_data}.}
\label{mwr_c5-c3}
\end{figure}

Further comparisons of channels 3 and 5 are shown in Fig. \ref{mwr_c5-c3}, revealing that the FFR brightness contrasts change from being microwave-bright to microwave-dark for the vast majority of identified features from PJ14 to PJ37.  This inversion in microwave brightness has been previously observed for Jupiter's belts and zones \cite{17ingersoll, 21fletcher}, anticyclones, and cyclones \cite{21bolton}, and now appears to be a feature of FFRs as well.  Indeed, the inversion for the $34.3^\circ$N cyclone studied by \cite{21bolton} during PJ19 can be seen again in the 2nd column of Fig. \ref{mwr_c5-c3}, showing that this is a property of both elongated brown barges and the visibly bright FFRs.  From this initial survey we conclude that the structures associated with FFRs extend down to at least the 14-bar level (channel 3), far below the altitude of the visible clouds (0.5-2 bars), and that the brightness inversion occurs between 5 and 14 bars.  We return to the question of whether the structure can extend deeper in Section \ref{deep}.

Maps of the $\Delta T_B$ residuals in Fig. \ref{mwr_pj26-29}, particularly those in Channel 6, display a regular pattern of striations with temperature contrasts at the 0.1-K level.  These stripes are parallel to Juno's spin axis, and reveal a data processing artefact in the time domain (i.e., on successive spins).  In processing the MWR data, we apply a window that smooths over the gain during multiple consecutive spins, which leads to correlated noise that is typically smaller than the white noise floor.  However, given that white noise is smallest at the shortest wavelengths, this artefact is most apparent in Channel 6, and only when we plot the residual from the zonally-averaged brightness temperatures.  The same processing window is used for all channels, and could be further optimised with a full recalibration of the data to minimise these stripes.  Instead, we retain the stripes at the 0.1-K level as an estimate of the trustworthiness of features in the residual maps.

\subsection{FFR Location Analysis}

With the zonally-averaged structure removed, the $\Delta T_B$ residual maps provide an excellent tool for identifying discrete meteorological features as microwave-dark and microwave-bright spots. At this stage we are agnostic of whether the features are genuinely FFRs, or some other structure (e.g., cyclones, brown barges, etc.), and attempt to identify the location, $T_B$ and $\Delta T_B$ for each feature.  We omit channels 1 and 2 in this analysis, due to their lower spatial resolution and more extreme contamination from synchrotron and auroras.  Perijoves between PJ6 and 40 were considered, although PJ9 and PJ38 were omitted due to the narrow longitudinal swaths provided on those orbits, resulting in 264 unique features being identified.

Maps were converted to an orthographic projection centred on $45^\circ$N to maximise visibility of the bright/dark spots.  Spots were identified manually using the Channel-5 $\Delta T_B$ (sounding 1.6 bars), as this provided the clearest contrasts with respect to the background.  This involved clicking on the approximate centres of each feature and recording their properties in channels 3-6, repeating the process twice to ensure robustness of results.  Although an automated process would be preferable, this manual process proved better for discarding data artefacts.  To mitigate this subjectivity, the standard deviation in a $2^\circ \times 2^\circ$ box around each feature was used to determine whether $T_B$ contrasts were significantly different from zero.  \add{Nevertheless, this manual process, coupled with MWR's incomplete spatial coverage, could introduce bias:  some microwave-bright anomalies may be unrelated to FFRs (e.g., the barges at lower latitudes), and those observed from greater distance may have too small a contrast to be detectable.  We therefore focus on qualitative trends in the discussion that follows.}

\begin{figure}[!h]
\makebox[\textwidth]{%
\includegraphics[width=1.3\textwidth]{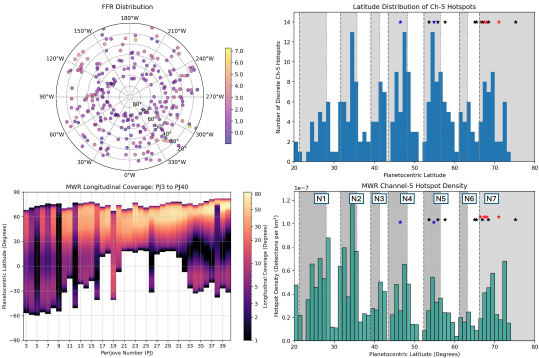}
}%
\caption{\textit{Top Left:} Spatial distribution of discrete features identified in MWR channel 5 across all perijoves from 6 to 40, excluding the polar domain poleward of $75^\circ$N.  The colour bar shows $\Delta T_B$, the brightness temperature residual compared to the zonally- and temporally-averaged atmosphere.  \textit{Top Right: } Latitudinal distribution of microwave-bright features detected in MWR channel 5.  \add{\textit{Bottom left: } The longitudinal width sampled by each perijove as a function of latitude, used to convert the raw detection frequency into a hotspot density in the \textit{bottom right}, accounting for the changing surface area with latitude.  In both histograms,} eastward (westward) jets are indicated by vertical dashed (dotted) lines, and cyclonic belts are shaded, from the N1 domain on the left to the N7 domain (polar region) on the right.  Stars indicate latitudes of lightning flash detection by Juno's Stellar Reference Unit between PJ11-17 \cite{20becker} - red for shallow $p<2$ bar lightning; blue for 2-4 bar lightning; and black where no half-width was measured.}
\label{ffrmap}
\end{figure}

The locations and channel-5 contrasts of each feature are mapped and compiled into a histogram in Fig. \ref{ffrmap}.  The $\Delta T_B$ contrasts become significantly smaller at higher latitudes, reflecting either (a) the reduced MWR spatial resolution at higher latitudes (perijove latitudes varied from $8.5^\circ$N at PJ6 to $33.1^\circ$N at PJ40); and potentially (b) a real decrease in $T_B$ contrast for more northerly features.  These are challenging to disentangle until later in the Juno mission, when FFR contrasts in the north polar domain will be more visible.  \add{We also convert the $\Delta T_B$ detections into a hotspot density to account for the changing spatial coverage as a function of latitude.  The longitudinal span of each perijove is calculated as a function of latitude and converted into an areal coverage (Fig. \ref{hotspot_density}).  The summed hotspot detections are then divided by the total area observed at each latitude (assuming that every hotspot detection is independent) to calculate the density in \mbox{Fig. \ref{ffrmap}.}}

The latitudinal distributions in Fig. \ref{ffrmap} are compared to the locations of Jupiter's eastward and westward jets \cite{03porco}. Grey shaded regions are cyclonic belts poleward of prograde jets, and we find microwave-bright features (cyclones, barges, and FFRs) to be most common in these belts.  There are notably fewer microwave-bright features identified in anticyclonic zones (poleward of retrograde jets).  The relationship is not perfect:  there are suggestions that bright spot activity in the N2 and N4 domains peak towards the poleward side of the belt; and bright spot activity in the N3 domain is actually higher in the N3 zone than the N3 belt.  Poleward of the N7 jet ($66^\circ$N), bright spot detections are distributed broadly with latitude in the polar domain, potentially up to $75^\circ$N - this is the North Polar Filamentary Belt (NPFB), where FFRs appear to dominate the activity at the boundary between the zonal structure of the mid-latitudes and the more chaotic structure of the polar domain.  Features in the polar domain close to the circumpolar cyclones (CPCs) will be the topic of a future study \cite{24hu_agu}.

\subsubsection{Correlations between MWR Channels}

\begin{figure}[!h]
\centering
\includegraphics[width=\textwidth]{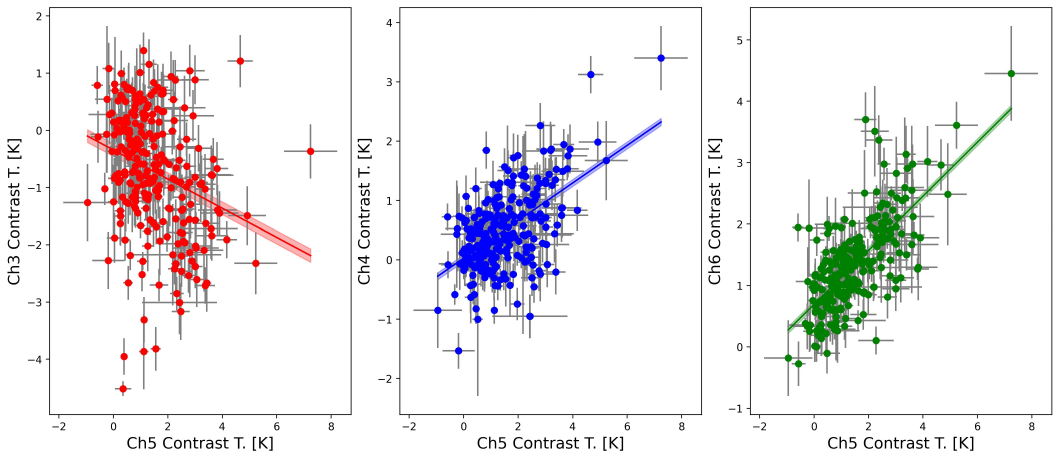}
\caption{Correlations between channel-5 brightness and channels 3, 4 and 6, revealing positive correlations in shallow-sounding channels (4 and 6), and weak negative correlation in the deep-sounding channel 3.  Each FFR has a measurement as a dot with error bars (based on the standard deviation in a $2^\circ \times 2^\circ$ box), and linear regression lines (and their shaded uncertainties) have been added.  Note that extreme outliers in channel 3 (related to synchrotron emission) were removed.  Regression slopes and Pearson correlation coefficients are provided in the main text.}
\label{mwr_correlations}
\end{figure}

We attempted to quantify the correlations between residuals in channels 3-6 that have been previously identified by eye in Fig. \ref{mwr_c5-c3}.  Fig. \ref{mwr_correlations} provides scatter plots comparing the microwave-bright features in channel 5 with the $\Delta T_B$ measured in other channels.  The shallow-sounding channels 4-6 are positively correlated, with linear regression slopes of $0.32\pm0.03$ (channel 4 versus 5) and $0.44\pm0.03$ (channel 6 versus 5), whereas deep-sounding channel 3 is anti-correlated with channel 5 (slope of $-0.25\pm0.06$) as expected from our earlier qualitative comparisons.  The Pearson $r$-values (a measure of the \add{strength of the} correlation between the datasets) are -0.27, 0.55 and 0.70 for channels 3, 4 and 6 respectively (with respect to channel 5), with $p$-values \add{(a measure of the significance of the correlation)} of $10^{-6}$, $10^{-22}$ and $10^{-40}$ respectively, strong evidence against the null hypothesis (no correlation) in each case.  \add{Note that the small p-values are due to the large number of FFR samples, $N\sim260$.}  As a result, microwave-bright features observed in shallow-sounding channels are likely to have microwave-dark features in deep-sounding channel 3.  

\subsection{Depth of FFRs}
\label{deep}

\begin{figure}[!h]
\makebox[\textwidth]{%
\includegraphics[width=1.5\textwidth]{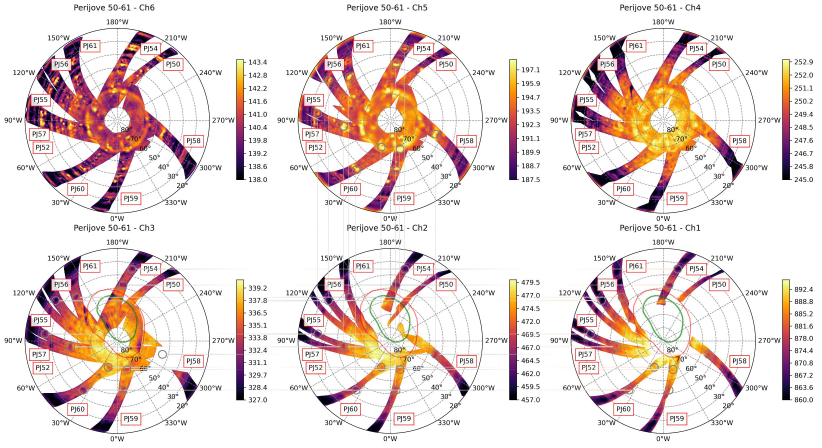}
}%
\caption{Polar projections of MWR data for PJ50 (April 2023) through 61 (May 2024).  Auroral and synchrotron contamination has been removed from channels 1 and 2, but the auroral oval \cite<green,>[]{17nichols} and Io footprints \cite<red,>[]{03grodent} are still shown to indicate auroral-affected regions.  Grey circles (linked by faint lines) are added to guide the eye, showing how microwave-bright spots in channel 5 (1.6 bars) \add{are} related to microwave-dark spots in channels 3 ($\sim14$ bars), 2 ($\sim35$ bars), and 1 ($\sim120$ bars). PJ51 and PJ53 are not shown as they would be overlain by later PJs, and measurements poleward of $80^\circ$N are omitted.  The $\Delta T_B$ version of this figure is shown in Fig. \ref{mwr-pj50-60resid}.  Versions without the annotation circles are available in our supporting data \cite{25fletcher_data}.}
\label{mwr-pj50-60}
\end{figure}

MWR observations before PJ40 had insufficient northern-hemisphere spatial resolution to reveal FFR-related contrasts in channels 2 and 1, and observations from PJ40-49 had longitudinal swaths too narrow to explore discrete features.  But after PJ50, the spatial resolution and swath width were sufficient to resolve features in channel 2 and (tentatively) channel 1, as shown in Fig. \ref{mwr-pj50-60}.  Despite the narrow longitudinal swaths, this collection of ten perijoves shows that the previously-identified microwave-dark spots in channel 3 (11.5 cm sounding $\sim14$ bars) are associated with microwave-dark features in channel 2 (24 cm sounding $\sim35$ bars) and even channel 1 (50 cm sounding $\sim120$ bars).  The weighting functions for these filters are broad, so their penetration depths are not known with high precision, but this demonstrates that the FFRs have deep roots, penetrating at least to the 50-100 bar level.  Such a depth was also reached by the cyclonic brown barge at $34.3^\circ$N (N2 domain) studied by \citeA{21bolton}.  Fig. \ref{mwr-pj50-60} also provides excellent coverage of the NPFB (with multiple microwave-bright FFRs in the belt), and indications of FFRs poleward of $70^\circ$N, all the way up \add{to} the latitudes of the circumpolar cyclones.

\section{Lightning within FFRs}  
\label{lightning}

In Section \ref{junocam}, we noted that the prominent FFR observed during PJ35 exhibited a discrete bright spot (Fig. \ref{junocam_morphology}), tentatively associated with a lightning flash near the centre of the FFR.  In Fig. \ref{ffrmap}, we indicated the latitudes of visible lightning observed by Juno's Stellar Reference Unit (SRU) between PJ11 and PJ17 \cite{20becker}.  In particular, shallow lightning flashes (red for $p<2$ bars, blue for 2-4 bars) are found exclusively in cyclonic belts where the FFRs are most common, whereas some of the deeper flashes (where half-power widths were not measured) are also present in anticyclonic zones.  SRU observed a cluster of flashes in the NPFB.  MWR has previously suggested that lightning discharges are localised, with peaks near 45$^\circ$N (N4 domain), 56$^\circ$N (N5 domain), 68$^\circ$N (N7 domain in the NPFB) and 80$^\circ$N surrounding the circumpolar cyclones from the first eight perijoves \cite{18brown}.  Intriguingly, the N6 domain ($61.2-66.1^\circ$N) was a local minimum in lightning detections.  This is consistent with the histogram of FFR detections in this domain, where there are fewer FFRs identified than in any of the other cyclonic belts (Fig. \ref{ffrmap}).  The N6 domain has been referred to as a `bland zone' in descriptions of Jupiter's high northern \add{latitudes} \cite{23rogers_baa}, which is supported by the absence of lightning detection, and the apparent dearth of FFRs.  In this section, we extend the survey of MWR lightning to consider the full spatial distribution, rather than just the latitudinal distribution.

\subsection{Lightning Detection with MWR}

Lightning discharges within the Jovian clouds generate sferics (directly-propagating broadband pulses) that can be detected by the long-wavelength channels of MWR, particularly in the 50 cm (600 MHz) channel \cite{18brown}.  These low frequencies are not attenuated by cloud opacity, nor by ionospheric plasma, making channel 1 an excellent lightning detector.  The pulses manifest as spikes in the time series of $T_A$, and can be identified by fitting a spline \add{to the channel-1 time series as Juno spins}, and searching for  $\delta T_A$ deviations of more than $4\sigma$, where $\sigma$ is the noise level for the channel \cite{20oyafuso}.  Following \citeA{18brown}, the effective isotropic radiating power ($P_{iso}$, EIRP in Watts) for channel 1 ($\lambda=0.5$ m) can be estimated from the lightning $\delta T_A$ as follows:
\begin{equation}
P_{iso} = \frac{2kB\delta T_A(4\pi R)^2}{G_r \lambda^2}
\end{equation} 
Here $k$ is the Boltzmann constant, $R$ is the range to the 1-bar surface along the boresight vector, $G_r=19.77$ dB is the unitless maximum antenna gain, and $B=18$ MHz is the channel-1 receiver bandwidth.  

However, the MWR sampling time is finite (100 ms), significantly longer than the expected 0.3-16 ms duration of a lightning strike \cite<as measured by Juno's Waves \add{and SRU instruments,}>[]{20imai, 20becker, 23kolmasova}, so it is impossible to know how long the discharge lasted for, meaning that $P_{iso}$ could actually be 6 (i.e., 100/16) to 300 (i.e., 100/0.3) times larger than that calculated here. Furthermore, the channel-1 beam pattern is large ($21^\circ$), making mapping lightning locations rather challenging, particularly when Juno is far from the planet. Detections are made in a single MWR sample, so it is impossible to know where the lightning originated from within the antenna beam pattern.  If the discharge occurred at the boresight location (i.e., the location of maximum gain), then the EIRP calculated above represents the minimum power of the lightning flash \cite{18brown} - discharges at the edges of the gain pattern would need to be much stronger to result in the same $T_A$ pulse \cite{25wong_epsc}.   

\begin{figure}[!h]
\makebox[\textwidth]{%
\includegraphics[width=1.5\textwidth]{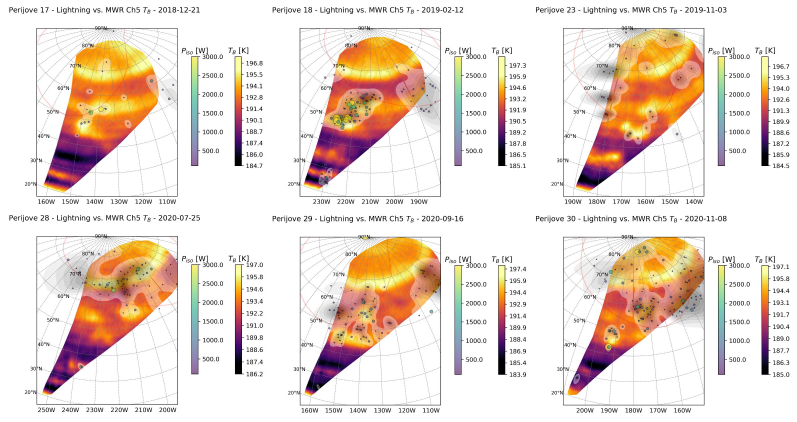}
}%
\caption{Orthographic projections comparing microwave $T_B$ (MWR channel 5, 3 cm, sounding 1.6 bars) to MWR lightning detections using two techniques:  assigning the $P_{iso}$ to the centre of the channel-1 boresight (coloured circles, with sizes scaled to the isotropic power); and antenna-weighted 2D histograms over the perijove (light grey contours, described in the main text).  Six representative perijoves are shown, all perijoves are available in the supporting data \cite{25fletcher_data}.  The locus of Io footprints is shown as a faint red oval, to avoid spurious features associated with the aurora. }
\label{mwr_lightning}
\end{figure}

We therefore map lightning distributions in two ways in Fig. \ref{mwr_lightning}, which shows six representative perijoves.  Firstly, we simply ascribe the minimum EIRP to the latitude and longitude of the boresight \cite{20oyafuso, 25brueshaber}, as shown by the circular points.  Secondly, we calculate a probability distribution for lightning locations, weighting each detection by the projected spatial coverage of the antenna pattern at the time of the detection \cite{18brown}.  We use the gain pattern from \citeA{17janssen}, using SPICE \cite<via SpiceyPy,>[]{20annex} to project the beam coverage for each detection, and then summing these overlapping detections during a perijove to generate a two-dimensional histogram.  These are shown as grey contours in Fig. \ref{mwr_lightning}, and indicate the uncertainty in the discharge locations.  We only display probability distributions and boresight detections for $P_{iso}>100$ W \cite<MWR previously measured minimum EIRP from 1.2-1800W,>[]{18brown}, emission angles $<65^\circ$, and spacecraft ranges $<3\times10^5$ km, focussing on stronger strikes on the planetary disc.

The six examples in Fig. \ref{mwr_lightning} demonstrate the challenge of comparing microwave brightness (channel 5, sounding 1.6 bars) with the location of lightning detections \cite<these are zoomed-in orthographic projections, equivalent polar projections are available in our supporting data,>[]{25fletcher_data}.  Some of the lightning detections are acquired when the MWR boresight is pointing towards the limb, such that we cannot identify their counterparts in the near-nadir maps of microwave brightness.  There is a compelling correspondence between lightning flashes and microwave hotspots in (i) the NPFB in the N7 domain; (ii) discrete FFRs in the N3-N6 domains; and (iii) some of the microwave-bright barges in the N1-N2 domains.  But this qualitative correspondence is not robust - there are many microwave-bright features that lack lightning; and many lightning detections that lack microwave counterparts.  Even when the correspondence is good, we lack the spatial resolution to tell \textit{where} in an FFR the lightning originates from - the small-scale cumulus clouds, or towards the edges of the stratiform lobes.  We find no correlation between the size of the $T_B$ anomaly and $P_{iso}$.

\subsection{Lightning Comparison to JunoCam}

\begin{figure}[!h]
\makebox[\textwidth]{%
\includegraphics[width=1.3\textwidth]{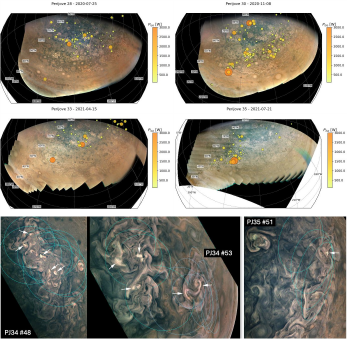}
}%
\caption{Top rows:  Four examples of MWR lightning detections ascribed to the channel-1 boresight, compared to orthographic projections of JunoCam maps for PJ28, 30, 33 and 35 (centred on $55^\circ$N).  This representative sample shows how the most prominent strikes (shown as darker orange and larger points) are co-located with folded filamentary regions seen by JunoCam. \add{Bottom row:  three examples (PJ34-35) overlaying MWR half-power beam widths (cyan) for lightning detections on top of high-resolution JunoCam imaging. All overlap FFRs containing bright white storms flanked by greenish haze (white arrows); further examples are provided in our Supplementary Material.}  }
\label{junocam-lightning}
\end{figure}

As JunoCam covers a broader longitudinal domain than the MWR brightness maps, Fig. \ref{junocam-lightning} compares the MWR boresight lightning detections to JunoCam orthographic projections for four representative perijoves.  Identification of the JunoCam cloud formations responsible for the MWR sferics is challenging because the lightning discharge may have occurred anywhere within the channel-1 beam, whose projected footprint size varies significantly as the spacecraft spins.  \remove{An initial survey of lightning clusters up to PJ24 (not shown), when Juno's closest-approach latitude was still near the equator ($24^\circ$N, December 2019), provided suggestive results:  of 16 identified clusters, all coincided with an active FFR (one in the NPFB, 5 in the N2 domain, the others in the N4 and N5 domains).}  

We initially focussed on perijoves PJ28 to PJ35 (perijove latitudes $25-30^\circ$N), when the spatial resolution at northern latitudes was improving, \add{and projected the MWR channel-1 half-power beam width (HPBW) at the moment of the lightning discharge onto Jupiter's 1-bar surface for comparison with JunoCam images.} To select for strong flashes with higher precision, we restricted our survey to clusters or pairs of discharges containing at least one with $P_{iso} > 2000$ W, close to the sub-spacecraft track so that the channel-1 footprint would be relatively small \add{(Figs. \ref{rogers_junocam1}-\ref{rogers_junocam4}, with three examples given in \mbox{Fig. \ref{junocam-lightning}}).}  Many, but not all, of the cloud features identified as FFRs are associated with clusters of boresight lightning detections in Fig. \ref{junocam-lightning}.   We found eight such clusters, from PJ28, PJ30 (two), PJ33, PJ34 (three), and PJ35.  For example, PJ28 shows clusters of flashes in FFRs near 65-70$^\circ$N in the N7 domain, PJ30 features numerous detections in FFRs near $55^\circ$N and $70^\circ$N.  All eight clusters overlapped an FFR, and in seven of them, there was a bright white storm within or on the edge of the cluster.  The exception, from PJ35, straddled an FFR with bright white cloud strips just outside the cluster.  Note that from PJ33 to PJ35, MWR consistently detected lightning flashes in the $55^\circ$N, $270^\circ$W region of the N5 domain, suggesting activity in a single location lasting for at least $\sim100$ days, even though the morphology of the FFRs evolved substantially in that time.  \add{Another smaller grouping observed during PJ35 near $50^\circ$N, $290^\circ$W was the location of the FFR shown in \mbox{Fig. \ref{junocam_morphology}} where JunoCam also detected an optical lightning flash.}


\add{For a more definitive survey of strong flashes, from PJ6 to PJ40, we included only lightning detections with $P_{iso}>1000$ W and emission angle $<65^\circ$, i.e., not far from the sub-spacecraft track, and then selected those with compact footprints (HPBW diameter $<\sim10^\circ$ latitude), and overlaid them onto the JunoCam maps (e.g., \ref{appendix3}, Figs. \ref{rogers_junocam1}-\ref{rogers_junocam4}). There were 113 such flashes (12 single, the others in 21 pairs or clusters), across 18 perijoves. The other PJs had no flashes meeting the criteria, or no JunoCam map. There were 20 flashes in the N7 domain (NPFB), 47 in N5, 31 in N4, 11 in N2 and 4 in N1.  We found that all the footprints included or overlapped a FFR. For 102 of the bursts (90\%), the FFR included a bright white patch or strip, within or on the edge of the footprint.} All these FFRs had bright embedded cumulus-like features on the stratiform clouds; almost all of them contained a very bright white convective storm, and exhibited some greenish haze nearby.  \add{For 10 of the remaining 11 flashes, there was a bright storm nearby, outside the footprint by up to $\sim0.4\times$ its radius (which does not exclude the storm as source, given the probability distribution of the beam footprint).  Only one footprint was further from the nearest bright storm (although it did overlap a weaker white strip: Fig. \ref{rogers_junocam4}, PJ11, N2 domain). }

A clear pattern emerged from this survey:  the great majority of strong lightning discharges in Jupiter's northern mid-latitudes are associated with the active FFRs containing bright white clouds, and are usually accompanied by a greenish haze.  However, not all FFRs are associated with lightning activity.  Maybe the discharges are sporadic in nature, so that the Juno PJ was encountering the FFR during a quiescent period?  Or does lightning activity indicate a particular phase during the storm lifetime, with younger storms being more convectively active?  As each PJ represents a snapshot, we cannot distinguish between these possibilities.   Although MWR lacks the spatial resolution to pinpoint where within the FFR the discharges occur,  we might expect them to be associated with the small-scale cumulus-like clouds seen to be popping up through the broader stratiform layer.  On Earth lightning is also observed originating in stratiform clouds connected to convective squall lines \cite{05carey}.  It is notable that we do not detect any lightning associated with anticyclonic vortices, or darker quiescent regions. 



\subsection{Lightning Distribution with Latitude}

\begin{figure}[!h]
\makebox[\textwidth]{%
\includegraphics[width=1.4\textwidth]{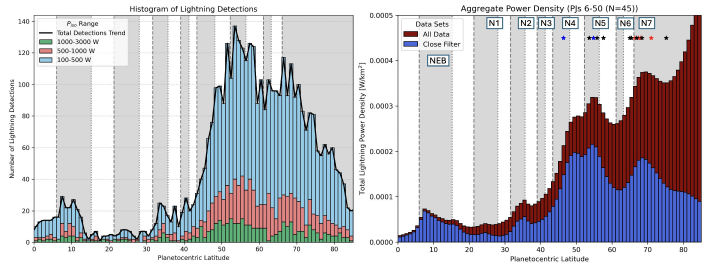}
}%
\caption{Latitudinal distribution of MWR lightning detections for PJ6-50. \textit{Left:} histogram of boresight latitudes of lightning detections, counting only detections with emission angles less than $65^\circ$ and spacecraft ranges less than $2.4\times10^5$ km, and splitting into bins depending on $P_{iso}$. \add{\textit{Right: } Total power density (W/km$^2$) summed over all perijoves for each latitude, accounting for the channel-1 antenna pattern projected onto the 1-bar surface.  Red bars use the same filtering as the left histogram, blue bars only include detections when Juno was closer than $1.0\times10^5$ km, limiting detections to the hemisphere nearest to perijove.  Note that the increased power density towards the pole is removed when this more stringent filter is applied, so should be treated with caution.  In both panels, }eastward (westward) jets are indicated by vertical dashed (dotted) lines, and cyclonic belts are shaded, from the NEB on the left to the N7 domain (polar region) on the right.  Stars indicate latitudes of lightning flash detection by Juno's Stellar Reference Unit between PJ11-17 \cite{20becker} - red for shallow $p<2$ bar lightning; blue for 2-4 bar lightning; and black where no half-width was measured.   
}
\label{lightning-hist}
\end{figure}

The latitudinal trend of lightning activity detected by MWR during the early phases of Juno's mission has been discussed at length elsewhere \cite{18brown, 20becker, 20guillot_ammonia}.  Here we extend this analysis using the MWR detections gathered from PJ6-50 to demonstrate the frequency of lightning detections as a function of latitude in Fig. \ref{lightning-hist}.  \add{The left-hand histogram} uses the minimum $P_{iso}$ assuming that the flash is located at the boresight, and therefore does not account for the \add{broad} antenna gain pattern of channel 1.  We split the histogram according to $P_{iso}$, to determine whether focussing on only the strongest detected signals produced a different distribution.  This panel confirms the trends observed by \citeA{18brown}: namely, the broad peak in detections between $45^\circ$N and $80^\circ$N compared to other latitudes.  \add{Next, we use the 2D histograms calculated by projecting the channel-1 antenna-weighted beam onto Jupiter's 1-bar surface for each lightning detection, multiplying by the minimum $P_{iso}$, and dividing by the area to estimate a 2D power density in W/km$^2$.  An example of this calculation for a single perijove is shown in Fig. \ref{lightning_density}.  We then averaged the 2D power density into 1-degree latitude bins, and summed over all perijoves to create the histogram on the right of \mbox{Fig. \ref{lightning-hist},} using two filters for distance:  one including data acquired within $2.4\times10^5$ km of Jupiter, and a more stringent filter for data within $1\times10^5$ km (removing data points acquired far from perijove).} The optical flashes detected by the SRU \cite{20becker} are shown for comparison.

Moving northwards from the equator, there are local maxima \add{in detection frequency and power density} in the North Equatorial Belt (NEB), but then few lightning detections until we reach the N2 domain (and brown barges in the North North Temperate Belt).  Neither N3 nor N4 show local maxima in lightning activity, but a broad peak is encountered in the N5 domain (poleward of $52.3^\circ$N), the N6 domain is a local minimum (consistent with it being considered as a `bland zone'), before the detection frequency \add{and power density} increase again in the N7 domain, peaking at the location of the NPFB ($66-70^\circ$N), \remove{with another local maximum near $74^\circ$N,} before declining towards the north pole.  There is a minor difference from \citeA{18brown}, who considered PJ1-8, in that we do not see their peak in the N4 domain (near $45^\circ$N), but all other qualitative details are similar.  Comparing to the histogram of microwave-bright spots in Fig. \ref{ffrmap}, the correspondence between local maxima in lightning, and local maxima in FFR detections, lends weight to the connection between FFRs and lightning activity.  However, the lightning detections appears to peak in the N5 and N7 domains, whereas the \textit{number} of FFR hotspots is similar here to the other cyclonic domains - the high lightning frequency of the N5/N7 domains therefore cannot be explained simply by the number of FFRs, lightning instead appears to be more frequent in FFRs found in N5 and N7.

\section{Discussion}
\label{discussion}

Juno has revealed that the chaotic, multi-lobed and filamentary structures of FFRs evident at Jupiter's cloud-tops (JunoCam and JIRAM) are just the top-most manifestation of structures that extend deep into the Jovian troposphere.  \remove{The turbulent appearance of the cloud-tops gives the impression of baroclinic instability within the weather layer:  the necessary Charney-Stern criterion is violated in the centre of each domain, near the westward jets \mbox{\cite{06read_jup}}}.  The microwave brightness displays an inversion with depth, whereby the structures are microwave-bright at wavelengths sounding $p<5$ bar, and microwave-dark for wavelengths sounding $p>10$ bars, with the transition likely associated with the increased stability (from molecular weight changes and latent heat release) near the water-condensation level at 6-7 bars.  Such an inversion has been previously identified in Jupiter's belt/zone structure \cite{17ingersoll, 20oyafuso, 21fletcher, 21duer}, and in a subset of vortices:  the Great Red Spot, an anticyclone near $16.7^\circ$N, and a cyclonic barge in the N2 domain near $34.3^\circ$N \cite{21bolton}.  The transitional layer was nicknamed the `jovicline' \cite{21fletcher}, and the inversion appears to be a typical feature of the majority of FFRs identified in the northern hemisphere.  Similar stable layers, which could appear as inversions, have been identified in numerical simulations of moist convection in giant planets \cite{25ge}.

Juno's MWR instrument has proven to be an excellent FFR detector, confirming that FFRs are preferentially found in cyclonic belts just poleward of the prograde jets \cite{79ingersoll, 22rogers}, and in particular in a prominent North Polar Filamentary Belt (NPFB) in the N7 domain, poleward of the last clear prograde jet at $66^\circ$N.  This NPFB marks the transition from organised zonal structure of the mid-latitudes, into the turbulent domain of the polar domain.  Furthermore, lightning sferics observed by MWR are clustered in the vicinity of the FFRs, suggesting that convection within FFRs is the dominant source of the enhanced lightning in northern-hemisphere cyclonic belts observed in previous studies \cite{18brown} and by Juno's Stellar Reference Unit \cite{20becker}.  However, FFR statistics in Fig. \ref{ffrmap} show the most detections in the N2, N4, N5 and N7 (NPFB) domains (with minor peaks in N3 and N6), whereas sferic detection is maximum in the N5 domain and over a broad range from $45-80^\circ$N.   We cannot rule out a sampling bias, but this might indicate a preference for FFRs that are more lightning-active in the N5 domain than any other region, confirming suggestions from Voyager \cite{91magalhaes}, Galileo \cite{99little}, and New Horizons \cite{07baines} that this domain has the most optically-active lightning sources.  The origin of this peak remains unclear, but might be related to changes in the availability of moisture (i.e., water) as a function of latitude (see Section \ref{energy_transport}).  In any case, the oft-cited ``increase in lightning near to the pole'' should be corrected, as FFRs in the $45-80^\circ$N domain (and particularly the N5 and N7 domains) appear to be the biggest contributors to the lightning budget in the northern hemisphere. 


\subsection{Vertical Structure of FFRs}

\begin{figure}[!h]
\makebox[\textwidth]{%
    \includegraphics[width=1.3\textwidth]{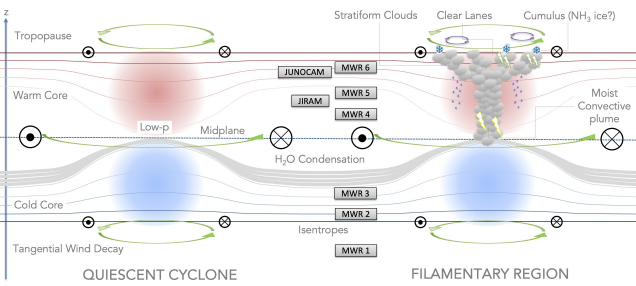}
}%

\caption{Schematic of a possible vertical cross-section through a Jovian cyclone, both in a quiescent phase (left), and when convective outbursts have generated an FFR (right). Deformations of the isentropes are shown by the pinched lines, curving towards the mid-plane where the highest tangential velocities will be found (green arrows, with circled dots indicating flow out of the page, circled crosses indicating flow into the page).  These velocities tentatively `de-spin' (i.e., get slower) with altitude and depth, thermal wind balance implies warm (light red) and cold (light blue) anomalies above/below the midplane, respectively.  We caution that we are ignoring changes to molecular weight in this scenario, and that the deep circulation shown here is not constrained.  The stable layer (water condensation, or the jovicline) is shown by the thinning of the grey shaded region near the vortex centre. This thinning may increase the availability of moisture, increase the stability, and lead to a build-up of CAPE that is released by the moist convective plume (grey clouds), with associated deep lightning (yellow edges), shallow lightning (green edges), stratiform lobed clouds, embedded cumulus, small-scale eddies (purple arrows), clear filamentary lanes, and potential precipitation.  Approximate depths sounded by JunoCam, JIRAM, and MWR are indicated.  Anticyclones would have a lens-shaped cross-section, with the thermal anomalies reversed in height. }
\label{ffr_cartoon}
\end{figure}

How might we explain the co-location of lightning, multi-lobed stratiform clouds with embedded cumulus, and the vertical inversion in microwave brightness?  Fig. \ref{ffr_cartoon} presents a simple schematic of an FFR as a cyclonic feature, with winds circulating around a low-pressure centre.  Moist convective plume activity is known to be associated with regions of cyclonic shear \cite<i.e., the Jovian belts,>[]{99little, 00gierasch}, but there are also numerous observations of outbursts within individual cyclones.  Voyager 2 captured plumes erupting in a southern-hemisphere cyclone in the S2 domain near $39^\circ$S \cite{79smith}; ground-based observations of `revival' plumes and outbreaks in the South Equatorial Belt (SEB) have been associated with cyclones \cite{17fletcher_seb, 19depater_alma}; convective plumes were observed erupting within cyclonic features in the South Temperate Belt (S1 domain) in 2018 \cite<the STB Ghost,>[]{20inurrigarro} and 2020 \cite<Clyde's Spot,>[]{22hueso, 22inurrigarro}.  Clyde's Spot was a small cyclone at $30.8^\circ$S that first displayed a double-lobed structure two days after the outbreak in May 2020, and slowly expanded to become an FFR, and then a turbulent sector within the STB a year later.

\citeA{89dowling_dps} explained the connection between convective outbursts and Jovian cyclones by stating that geostrophy in vortices cause vertical deformations to potential-temperature surfaces (isentropes, or `material surfaces').  Cyclones (low-pressure centres) would have a depression of isentropes in upper layers (the downward pull resisted by the stably-stratified upper troposphere) and a rise of isentropes in deeper layers. Closely spaced isentropes represent stronger static stability, so the `thinning' of isentrope spacing near the midplane of a cyclone represents a local stable layer.  \add{Similar insights were gained from isentropic potential vorticity mapping within terrestrial mid-latitude vortices \mbox{\cite{85hoskins, 86thorpe}}}.  These isentrope deformations are shown graphically in \add{Fig. 6 of} \citeA{22inurrigarro}, and are represented qualitatively in Fig. \ref{ffr_cartoon}. Convection may be more strongly inhibited across this stable layer, allowing a build-up of convective available potential energy (CAPE). The accumulated CAPE could then trigger moist convection due to vertical displacement across the stable layer, or due to horizontal transport as found to be important in low-latitude simulations of moist convective activity \cite{25sankar}.  Conversely, anticyclones (high-pressure centres) would have an upward bulge of isentropes in upper layers, and a downward bulge in deeper layers, with a `lens' of well-separated isentropes in the midplane.  The weakly stratified lens would be less effective at inhibiting convection and accumulating CAPE. Both cyclonic and anticyclonic vortices are `pancake-shaped,' broader in horizontal extent than they are in the vertical due to the effects of stratification \cite{20lemasquerier}, and are embedded and drifting within the neutrally stratified atmospheric layers between the tropopause and the deeper convective zones.  They are shallower than the deep winds of the zonal jets \cite{18kaspi}. 

A consequence of the deformation of isentropes is that vortices display temperature and density anomalies that invert with depth:  a cyclone should exhibit maximum tangential velocities at its midplane (where temperature anomalies vanish) and are expected to despin (i.e., become slower) with height and with depth, such that the thermal-wind relation for geostrophic vortices requires a warm, positive buoyancy anomaly in the top half, and a cold anomaly in the bottom half \cite<e.g.,>[]{13marcus, 14palotai, 20lemasquerier, 23read}.  The opposite would be true for anticyclones like the Great Red Spot \cite<as recently computed via modelling by>[]{24zhang}, and \citeA{21parisi} showed that the gravity field over the GRS could be produced by a positive mass concentration at upper levels, and negative mass concentration (i.e., positive buoyancy, warmer temperatures) at lower levels, consistent with MWR observations \cite{21bolton}.   We caution that we are ignoring contrasts in molecular weight (and/or ammonia gas) in this simple scenario, where NH$_3$ and H$_2$O gradients could affect the estimated windshear \cite{21fletcher}.  These temperature/composition contrasts are \textit{primary circulation} features with no implied upwelling or subsidence, \add{so in the absence of strong convective processes, stratiform cloud layers should} condense where it is cold (tops of anticyclones) and evaporate/sublimate where it is warm (tops of cyclones).  Examples of the latter include brown cyclonic barges \cite{21bolton} and pale cyclones \cite{22hueso} observed at Jupiter's mid-latitudes, although these are uncommon at high latitudes.   

\subsection{FFRs and Moist Convection}

At low- and mid-latitudes, quiescent cyclonic circulations are the most common locations for new convective outbreaks that expand into turbulent phenomena like FFRs \cite{79smith, 17fletcher_seb, 19depater_alma, 20inurrigarro, 22hueso}.  By analogy, we propose that high-latitude FFRs are cyclonic features where enhanced moist convection has injected significant cloud material into a cyclone.  This may be a simplification, as the enhanced turbulence and mixing across the boundary of the cyclone may pull moisture-rich air into the FFR horizontally \cite{25sankar}, resulting in convection and cumulus clouds at the edges of the FFR, rather than the centre.  In either case this is conjecture:  we have not witnessed the `birth' of a high-latitude FFR with Juno, but this does explain why FFRs are dominated by thick cloud (blocking 5-$\mu$m flux observed by JIRAM, and creating the high, reflective clouds observed by JunoCam).  Two factors cause a `thinning' of the barrier to convection at the midplane (e.g., the deformation of the isentropes at the low-pressure centre):  the cyclonic nature of the belt itself \cite{16thomson, 20fletcher}, intensified by the cyclonic nature of the embedded cyclone \cite{89dowling_dps, 22inurrigarro}.  These cyclonic belts could be suffused with cyclone-induced thinned stable layers, permitting moist convective outbursts.

The detection of lightning sferics clustered in and around these FFRs, and evidence that many lightning flashes occur in the very bright storms within them, is suggestive of water-driven convection rooted in the water cloud, with charge separation in the mixed liquid-ice region of the water clouds resulting in a form of cloud-to-cloud lightning, as on Earth.  As shown in Fig. \ref{ffr_cartoon}, these rising plumes reach the high static stability of the upper troposphere and spread laterally like outflows creating terrestrial thunderstorm anvils, producing the stratiform lobes (and intervening gaps), with secondary features like whorls/eddies and embedded cumulus (overshooting convection?) being embedded within the stratiform layers.  Strong precipitation \cite<ammonia snow, liquid water, ice, and potentially mixtures known as `mushballs,'>[]{20guillot_mushball} might be expected below these clouds.

\subsection{Lifetime of FFRs}

The `birth' of a high-latitude FFR has never been observed, due to the lack of sustained high-resolution imaging.  A 7.5-year survey of cyclonic features in the SSTB \cite<S2 domain,>[]{24rogers_epsc} suggested that FFRs in this domain have lifetimes ranging from 4 months to at least 8 years. New FFRs appeared about once a year, usually in low-contrast regions, sometimes in long pale oblong circulations. In the few examples where convective outbreaks were seen to initiate an FFR, they have been small and only occasionally bright in the 0.89 $\mu$m methane band, thus much less energetic than examples recorded in the STB \cite{20inurrigarro, 22hueso}.  It is also possible that FFRs arise inconspicuously in an eddy in a nondescript cyclonic region.  A young FFR in the SSTB is usually small and expands, though they do not grow indefinitely; their size and apparent activity can fluctuate, without any progressive evolution.   

Assessments of FFR lifetimes in the northern mid-high latitudes appear to be restricted only by our observational limitations.  Cassini's colour movie of Jupiter \cite{03porco} spanned 10 days (2000-Oct-31 to 2000-Nov-09), and revealed 4 FFRs in the N2 domain, and 8 FFRs in the N4 domain, and they all persisted throughout the movie.  Ground-based tracking, coupled with morphological identification from JunoCam, showed that some FFRs in the N4 domain and some in the NPFB last for at least 1 to 2 months, and one in N4 for 6 months \cite{23rogers_baa}. Comparisons of successive JunoCam maps at high latitudes suggest that FFRs can expand, shrink, split, or merge, so some may persist for a long time despite being unrecognisable from one Juno orbit to the next.



\subsection{Ammonia or Temperature Contrasts?}

Microwave contrasts observed by MWR have been typically assumed to be driven by ammonia gas \cite{17li}, with enhanced abundances creating microwave-dark signatures.  An example is a model of the cyclonic barge near $34.3^\circ$N \cite{21bolton}, which would need an excess of 20 ppm near 10 bar to explain the microwave-dark anomaly.  This would require explaining gaseous enhancements within cyclones at depths where condensation processes are unimportant, possibly due to evaporation of NH$_3$-rich precipitates falling from above \cite{20guillot_ammonia}.  Thermal contrasts provide an \add{alternative} explanation, as these can be linked to the despinning of vortices above/below their midplanes, although we stress that this despinning has never been directly measured. Attempts to break this ammonia-temperature degeneracy have been presented at the equator \cite{24li} and poles \cite{24hu_agu}, but this remains a significant challenge.  

\citeA{25biagiotti} presented JIRAM 2-5 $\mu$m spectroscopy of an FFR at $40^\circ$N (N3 domain), acquired during Juno's first perijove (August 2016).  They find (based on a 2.95-$\mu$m absorption feature and a shift in the 2.6-2.8 $\mu$m reflection peak) evidence of pure NH$_3$ ice located within what appears to be fresh, high-albedo clouds at the centre of the FFR. In addition, they find that the FFR exhibits high-altitude aerosols with smaller radii than in the surrounding atmosphere, alongside elevated ammonia relative humidity.   All of these are consistent with powerful upwelling within the cyclone, but how would this appear to MWR?  If all the ammonia vapour were removed by condensation in the top-half of the cyclone, we would produce the microwave-bright detections in shallow-sounding channels.  But excess NH$_3$ gas from upwelling would darken the microwave spectrum, so there must be physical warming to (i) enable sublimation of the clouds to create the high humidity observed by JIRAM; and (ii) counteract the darkening effect of the excess vapour, producing the microwave-bright features near 1-5 bars (MWR channels 4 and 5).  Some of this warming could originate from the hypothetical warm anomaly in the top half of cyclones, but latent heat release from water condensation may also contribute, making the FFRs appear microwave bright.   We note that the situation for channel 6 (0.65 bars) is more complex as it sounds altitudes above the ammonia condensation layer.  Here there is a buffering effect between temperature and ammonia concentration that reduces the sensitivity to contrasts:  cold spots would cause NH$_3$ condensation; warm spots would cause NH$_3$ sublimation, with the effects counterbalancing one another in the microwave brightness.  However, \citeA{20wong}, using ground-based mid-IR observations, showed that southern-hemisphere FFRs were bright at 10.8 $\mu$m (sounding $\sim0.5$ bar), supporting the suggestion that FFRs are warmer than their surroundings in the upper troposphere.

\subsection{Energy Transport at High Latitudes}
\label{energy_transport}

The absence of significant equator-to-pole contrasts in upper tropospheric temperatures has led to suggestions that Jupiter's outward-directed internal heat flux emerges preferentially at high latitudes, to make up for the sunlight deficit at the pole compared to the equator \cite{78ingersoll, 84pirraglia, 04ingersoll}.  The Galileo and Juno-derived lightning distributions have been cited as evidence supporting this claim \cite{99little, 18brown}.  On Earth, static stability generally increases towards the poles, so that lightning and moist convection are most prevalent at the tropics.  On Jupiter, the reverse may be occurring \cite<e.g.,>[]{00allison}, with weaker stability at high latitudes resulting in the increased prevalence of stormy FFRs and lightning, which provide a strong indication that convective activity is greatest in the $45-80^\circ$N region, and particularly the N5 domain.

In addition to the increased heat flux and weaker static stability at high latitudes, the availability of condensibles - particularly water - might be playing a role.  JIRAM spectroscopy \cite{20grassi} has previously suggested that water near 5 bars is enhanced within cyclonic shear zones, particularly in the N5 domain.  As cyclones drift in longitude, deforming deep isentropes upwards, they tap into the deeper reservoir of water to relinquish the stored potential energy.  We might consider them as an `eddy pump' (see below) mixing these water-rich layers upwards.   In their numerical simulations of convective outbursts in cyclones, \citeA{22inurrigarro} indicated that sustained moist convection was limited by the availability of water within the closed cyclones, and that the different water availability affected whether cyclones manifested as cloud-free barges or cloudy FFRs.  The strength (i.e., initial vorticity) and vertical extension of the cyclone must also play a role in the vigour of FFR outbursts.   These wandering cyclonic FFRs therefore provide the means for the internal heat flux to escape via moist convection, warming the upper troposphere so that it is microwave-bright.  The vigour of the convection (and lightning) results from a combination of the strength of the circulation, the availability of water, and (potentially) the increased internal heat flux at high latitudes.  FFRs are therefore an important component of Jupiter's high-latitude energy balance, but the latitudinal variation of deep water and static stability remains to be determined.

\subsection{Comparison to Saturn}

Although Saturn's atmosphere exhibits less visible contrast than Jupiter, Voyager and Cassini showed that vortices and storms are still evident at certain latitudes \cite{81smith, 06vasavada}, most notably two `storm alleys' near $\pm33-39^\circ$ (planetocentric), near the peak of the first westward jet ($36^\circ$) in each hemisphere \cite{11garcia}.   \citeA{05porco} present a time series of a 2004 convective storm (the `dragon storm,' their Fig. 5), \citeA{07dyudina} and \citeA{19fischer} show similar time series for spiralling convective storms in 2006 and 2008, respectively.  Both share morphological similarities with Jupiter's FFRs:  bulbous multi-lobed reflective clouds, dark filaments, and lightning-induced electrostatic discharges.  However, some of Saturn's cloud outflows appeared to exhibit anticyclonic circulation, and spawned anticyclonic vortices in their wakes \cite{05porco, 18sromovsky}, consistent with the numerical simulations of \citeA{22heimpel} that suggest `lower latitudes favour shallow anticyclonic vortices that form due to upward and divergent flow.'  At higher latitudes, Saturn does seem to exhibit a connection between cyclones and moist convection:  \citeA{23gunnarson} observed multiple convective storms over a period of years within a previously-quiescent cyclone at $50^\circ$N, bearing striking similarities to Jupiter's FFRs.  And a 2018 storm at $65^\circ$N was traced back to a cyclone \cite{20sanchez}. 

\subsection{Analogies to Oceanic Eddies}

We started Section \ref{intro} by noting the qualitative similarity of Jupiter's FFRs to mesoscale eddies in Earth's oceans, with swirls, spirals and filaments visualised through phytoplankton blooms, temperature, and salinity contrasts.  Our oceans are `seas of eddies,' from mesoscale eddies a few hundred km in scale that can last from weeks to a year \cite{24zhang_ocean}, to submesoscale eddies with scales up to a few tens of kilometres that last for tens of days \cite{23taylor}.  Both are associated with anomalies in sea temperature and surface height.  Our detection of an inversion in microwave brightness is a strong indication of stratification in Jupiter's weather layer (maybe due to latent heat effects, or changes in molecular weight), making an analogy to Earth's oceans appropriate.  Just as for the deformation of isentropes discussed for Jupiter, Earth's ocean eddies also deform material surfaces \cite{15mcgillicuddy}. In `surface eddies,' isopycnals (surfaces of constant density) and the thermocline (region of rapid temperature change) are bent upwards in cyclonic eddies, bringing cold, salty, nutrient-rich waters upwards (promoting phytoplankton growth), and downwards in anticyclonic eddies \cite{17chang, 22purkiani}.  Cyclonic eddies are sometimes referred to as `eddy pumps,' being responsible for bringing deep waters up to the surface.  These `surface eddies' have their maximum horizontal velocities near the ocean surface, and can be compared to the bottom-half of the vortices seen on Jupiter, with cyclones being cold-core, and anticyclones being warm-core down to a few hundred metres below the surface \cite{23ni}, and some energetic eddies can penetrate deep towards the ocean floor \cite{16zhang_ocean}.

A sub-category of oceanic eddies known as `subsurface,' `depth-intensified' or `intrathermocline' eddies (ITEs) may be more analogous to the cyclonic FFRs seen on Jupiter, as they cause both upward and downward deformations of material surfaces, with the maximum-velocity midplane between the `seasonal' and `permanent' thermoclines.  Anticyclonic ITEs exhibit a lens-like structure, with shallow isopycnals rising upwards, and the deep thermocline/pycnocline (analogous to our \textit{jovicline}) displaced downwards.  For example, an extra-large anticyclonic ITE observed in the Pacific \cite{17nan} exhibited a shallow ($\sim100$ m) cool anomaly and a deep ($\sim450$ m) warm anomaly.  Cyclonic ITEs are the opposite, and result in pinching or thinning of isopycnals at the midplane, sometimes being referred to as a `cyclonic thinny' \cite{15mcgillicuddy}, and bearing most resemblance to the isentrope deformations in Fig. \ref{ffr_cartoon}.  Numerous cyclonic eddies are observed in the northwestern Mediterranean, including some with the `pinched' ITE structure, and others characterised as cold-core surface cyclonic eddies \cite{16bosse}.  A cyclonic ITE in the Arabian Sea \cite{20demarez} showed a warm anomaly atop a cooler anomaly, similar to our suggested FFR structure revealed by Juno.  Both the surface and subsurface eddies account for a significant amount of the ocean's kinetic energy and heat transport.  Maybe Jupiter's FFRs have a similar dominance on the energetics of the high-latitude atmosphere, but it is clear that terrestrial oceanography has much to contribute to the exploration of Jupiter.




\section{Conclusion}
\label{conclusion}

Juno's multi-wavelength capabilities permit a step-change in our characterisation of sprawling, chaotic, and turbulent cloud formations at Jupiter's mid-to-high latitudes, known collectively as `Folded Filamentary Regions,' or FFRs.  We have shown that these deep-rooted cyclonic features dominate the meteorology and lightning activity of Jupiter's high latitudes, including a prominent `North Polar Filamentary Belt' (poleward of the prograde jet at $66^\circ$N), which marks the transition from the organised structure of belts and zones, to the chaotic polar domain.  Our conclusions are summarised as follows:

\begin{itemize}

\item \textbf{Cloud-top appearance:} Visible light (JunoCam) and 5-$\mu$m imaging (JIRAM) reveal the top-level appearance of structures that extend deep beneath the clouds.  FFRs display an exceptional variety of swirling masses of overlapping cloud patterns:  white aerosols in multiple stratiform lobes that cast shadows on deeper clouds and block deep 5-$\mu$m radiance (diffuse 5-$\mu$m glow characterises the northern hemisphere poleward of $40^\circ$N), separated by turbulent, visibly dark filaments (gaps in the clouds that are bright at 5 $\mu$m).  Smaller discrete cloud-free whorls/eddies, and the densest clusters of cumulus (or cumulonimbus) clouds, referred to as bright white storms, are embedded in the stratiform layer, appearing bright near 890 nm, indicating that they reach into the upper troposphere.  The cumulus clouds are tentatively associated with optical lightning (a flash was observed in the centre of an FFR during PJ35), and with an unusual green haze whose composition remains unknown.  Anticyclonic white ovals also appear dark and cloudy at 5 $\mu$m, but display less internal structure than the FFRs.  The turbulent structures evolve too rapidly to be tracked over consecutive perijoves, so the evolution and longevity of the FFRs remain poorly understood.

\item \textbf{Inversion in microwave brightness:} We show that FFRs, and thus the majority of meteorological features in the northern hemisphere, are microwave-bright at wavelengths sounding $p<5$ bar, and microwave-dark at wavelengths sounding $p>10$ bar, with the transition expected to be in the water-condensation region (6-7 bar).  This property is exhibited by both cloudy FFRs in the N3 through N7 domains, and cloud-free brown barges in the N2-N3 domains.  The microwave-dark anomalies are visible in Juno's deepest-sounding wavelength (50 cm, sounding $p\sim100$ bars), revealing that these cyclonic structures extend deep into the troposphere.  Nevertheless, the broad horizontal extent of FFRs remains an order of magnitude larger than their vertical extent, suggesting these to be `pancake-like' vortices.  The microwave brightness at shallow depths is consistent with a warm anomaly at altitudes above the vortex midplane (and a corresponding cool anomaly at greater depths), implying a decay of wind strength with both height and depth from the midplane (which exists somewhere between 5-14 bars, between MWR channels 3 and 4).  Although the microwave brightness could be associated with a local depletion in ammonia gas opacity, this appears inconsistent with recent JIRAM spectroscopic analysis \cite{25biagiotti}, such that the FFR anomalies are likely to be due to a \textit{combination} of temperature and ammonia contrasts.

\item \textbf{FFR locations:}  Via comparison of JunoCam and MWR maps, Juno's microwave radiometer has proven to be an excellent tool for studying the distribution of FFRs, finding them to be most common poleward of prograde jets (i.e., within cyclonic belts) in the N3 through N7 domains, particularly forming the microwave-bright band that marks the transition from belt/zone structure to the polar domain (i.e., the north polar filamentary belt, matching a similar southern belt in the S6 domain).  Conversely, there was a relative absence of FFRs in the N6 domain ($61-66^\circ$N), consistent with the impression of it being a `bland zone.'  

\item \textbf{FFRs as lightning sources:}  Many (but not all) of the microwave-bright FFRs observed by MWR and JunoCam are associated with lightning sferics observed at 50 cm.  In some cases, sferic detections cluster around active FFRs, but we cannot tell whether these originate in the stratiform layers, the core of the vortex, or the embedded cumulus clouds (bright white storms).  It is possible that these bright white storms, often flanked by greenish haze, are the predominant thunderstorms. We confirm that MWR detects a broad peak in lightning activity between $45^\circ$N and $80^\circ$N compared to other latitudes.  The latitudinal distribution shows that neither the N3 nor N4 domains show local maxima in lightning activity.  A broad peak is encountered in the N5 domain (poleward of $52.3^\circ$N), whereas the N6 domain is a local minimum (the `bland zone').  The detection frequency increases again at the location of the NPFB ($66-70^\circ$N), before declining towards the north pole.  This is consistent with the distribution of FFRs in cyclonic belts, and with optical detections of lightning by the SRU \cite{20becker}.  However, the high lightning frequency of the N5/N7 domains cannot be explained simply by the spatial density of FFRs (which is similar to the N3 and N4 domains), so lightning may be more active in FFRs found in N5 and N7 domains.  This excess lightning may be related to local maxima in the 5-bar water abundance identified by \citeA{20grassi} for the N5 domain.  We conclude that FFRs are the dominant source of lightning to explain the distribution observed by \citeA{18brown}.  However, we note that some of the FFRs observed by JunoCam and MWR exhibited no lightning, suggesting that discharges are sporadic in nature, and maybe lightning only occurs during distinct phases of the storm evolution.

\item \textbf{FFRs as eddy pumps:}. By analogy to the cyclonic thinning (deformation of isopycnals and the thermocline) in Earth's oceanic surface and subsurface eddies, we showed a qualitative schematic of the vertical structure of an FFR, with warm anomalies above the midplane, and cool anomalies below.  The `pinching' of isentropes and thinning of the stable layer near the low-pressure midplane, which intensifies a stable layer near the water condensation level, could enable accumulation of CAPE below cyclones. The release of this stored energy in moist convective plumes (marked by lightning) would then inject the top-half of the cyclone with turbulent, chaotic, and rapidly evolving cloud layers.  Thus the cyclones promote convective outbursts, and the release of latent heat, as a key component of Jupiter's escaping heat flux at high latitudes.  We speculate that differences between mid-latitude barges (cloud-free but prone to sporadic outbursts) and high-latitude FFRs (cloudy) could be explained by the availability of water to power convective outbursts, suggesting more water (or lower static stability) in Jupiter's high-latitude belts (particularly the N5 and N7 domains).

\end{itemize}

Juno's orbital geometry can only provide snapshots of this enormously complex `sea of eddies,' leaving many questions unresolved.  Are there cyclonic precursors visible before an eruption, as there was for cyclones in the S1 domain \cite{22hueso} and S2 domain \cite{79smith}?  Are the convective outbursts and lightning continuous, or do storms evolve through discrete phases, exhausting their supply of deep water?  How do multiple FFRs interact with one another, and do they show latitudinal drifts across belt/zone boundaries, as has been observed for some small anticyclones \cite{24morales_dps}? Why do some cyclones exhibit outbursts, but not all?  What happens when an FFR is exhausted, how long do FFRs last at different latitudes, and are there cyclonic remnants that can have a second life as a new storm?  A time series over days and weeks is required to explore the life cycle of FFRs.  The rapid changes in the morphology of FFRs from one PJ to the next, coupled with the regular detection of lightning, suggests that some level of moist convection is being sustained over weeks and months, indicating that the cyclonic structures are tapping into a significant reservoir of moisture as they drift with the zonal winds.  Why are anticyclones apparently shallower than the cyclones, do these deep roots depend on latitude, and will Jupiter's circumpolar and polar cyclones be similar to FFRs?  

Juno has characterised the structure and energetics of Jupiter's filamentary storms, but long-term monitoring from future spacecraft (e.g., JUICE), or ground-based facilities with sufficient resolution to observe the polar meteorology (e.g., observatories like ELT), are needed to make progress in understanding their life cycle.  Finally, FFR-like features have been observed to erupt from Saturnian cyclones, and we wonder whether similar convective features could exist on Uranus and Neptune, to be explored by the next generation of orbital missions.
%

%
%

\section*{Open Research Section}

Juno observations are available through the Planetary Data System Atmospheres Node \cite{pds_juno}, with links to the specific calibrated MWR data \cite{pds_junomwr}, JIRAM data \cite{pds_junojiram}, and JunoCam data \cite{pds_junocam}.  Data for individual figures presented in this manuscript are available via \citeA{25fletcher_data}.

\section*{Conflict of Interest Statement}

\add{The authors have no conflicts of interest to disclose.}


\acknowledgments
Fletcher is a Juno Participating Scientist supported by STFC Consolidated Grant reference ST/W00089X/1 and Small Grant reference UKRI1205.  Wong received support from the NASA Juno Participating Scientist program (80NSSC19K1265); Wong and Sankar received support from the New Frontiers Data Analysis Program (80NSSC25K0362). For the purpose of open access, the author has applied a Creative Commons Attribution (CC BY) licence to the Author Accepted Manuscript version arising from this submission.  A portion of this work used the ALICE high performance computing facility at the University of Leicester.  Some of this research was carried out at the Jet Propulsion Laboratory, California Institute of Technology, under a contract with the National Aeronautics and Space Administration (80NM0018D0004).  The JIRAM project was funded by the Italian Space Agency (ASI).

We are immensely grateful to the members of the Juno team responsible for planning, executing, and processing the datasets presented in this paper since 2016. We thank Tim Dowling, Phil Marcus, and Daphne Lemasquerier for illuminating discussions on the nature of Jovian vortices.  We thank Kevin M. Gill for sharing his JunoCam compositions for Figures 1 and 2.  \add{Finally, we thank two anonymous reviewers for their careful reviews of this manuscript.}

\appendix

\section{FFRs in the Southern Hemisphere}
\label{appendix1}

Although this study focusses on the northern hemisphere to enable coverage by the microwave radiometer, we note that both JunoCam \cite{22rogers} and JIRAM were able to study multiple FFRs in the high southern latitudes.  We adopt the same nomenclature for the southern bands as those in the north, namely splitting into domains of homogenised potential vorticity (PV), separated by prograde jets serving as mixing barriers.  These include the S1 domain (including the South Temperate Belt and Zone, poleward of $23.7^\circ$S), S2 domain ($32.2^\circ$S), S3 ($39.2^\circ$S), S4 ($48.6^\circ$S), S5 ($58^\circ$S) and S6 ($64^\circ$S).

\begin{figure*}[!t] 
\makebox[\textwidth]{%
    \includegraphics[width=1.3\textwidth]{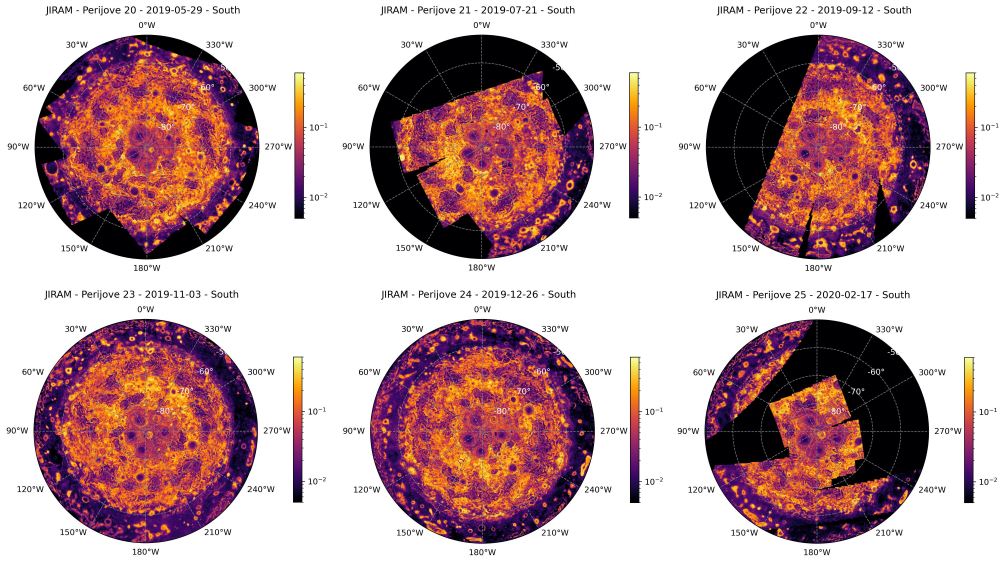}
}%
\caption{JIRAM 5-$\mu$m maps of Jupiter's south pole, displayed in azimuthal equidistant projection poleward of $50^\circ$S.  Six maps are shown in sequence between PJ20 and PJ25 (May 2019 to February 2020), and FFRs are shown as multi-lobed filamentary dark objects against the diffuse background glow, distinguished from the regular oval shape of the dark anticyclones.  The circumpolar cyclones (CPCs) were undergoing perturbations from a wavenumber-5 to a wavenumber-6 feature during this interval.  }
\label{jiram-south}
\end{figure*}

The most complete JIRAM spatial coverage across consecutive perijoves was acquired between PJ20 and PJ25 (May 2019 to February 2020), shown in Fig. \ref{jiram-south}.  These confirm that southern FFRs resemble their northern counterparts, with multi-lobed and filamentary structures appearing dark (i.e., high aerosol opacity) against the diffuse 5-$\mu$m flow.  It was initially hoped that this complete coverage would enable tracking of the same feature over the $\sim50$-day intervals between perijoves, but blink comparisons between these maps suggest such tracking is impossible.  The features evolve over timescales of a few days, such that complete and uninterrupted views on 2-3 day timescales would be needed to observe the changes and motions of these chaotic structures.  

Fig. \ref{jiram-south} shows that FFRs appear most common in a ``south polar filamentary belt" (SPFB in the S6 domain) between 65$^\circ$S and 70$^\circ$S, a counterpart to the NPFB discussed in the main text.  These belts represent the transition from the mid-latitude banded structure into the polar domain, and the 5-bar water abundance derived from JIRAM \cite{20grassi} shows a significant local maximum here (their Fig. 6).  Finally, the diffuse 5-$\mu$m glow appears to be more tightly constrained poleward of $58^\circ$S (S5 domain) than in the north (where the boundary in Fig. 4 (main article) is closer to $40^\circ$N), suggesting a difference in the cloud opacity between the two hemispheres.  This asymmetry in deep cloud opacity will be the topic of a future study.
\newline

\section{Global Views from MWR}
\label{appendix2}

\begin{figure*}[!t] 
\makebox[\textwidth]{%
\noindent\includegraphics[width=1.2\textwidth]{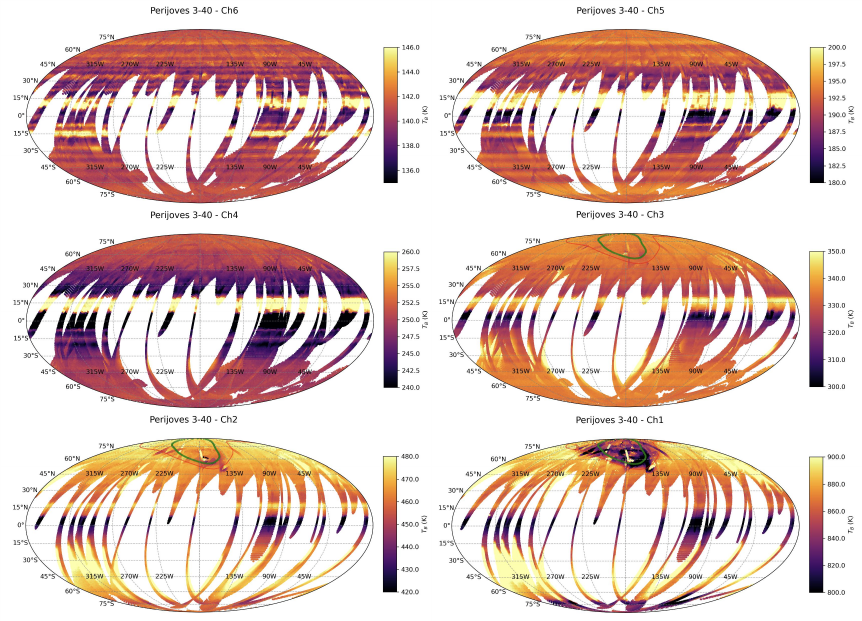}
}%
\caption{Mollweide projection of MWR $T_B$ observations between PJ3 (December 2016) and PJ40 (February 2022), showing the spatial coverage of the dataset in each filter, and revealing Jupiter's banded pattern.  Emission angles have been restricted to $<65^\circ$, but there has been no filtering of synchrotron emission (showing as bright patches, particularly at southern mid-latitudes) or the cold auroral electrons (visible as dark patches in channels 1 and 2, interior to the locus of Io footprints in red, and close to the main auroral oval in green).  Scale bars have been adjusted to best show mid-latitude contrasts, meaning that bright emission from the NEB is saturated at the top of the colour bars.  Wavelengths are shown from the shallow-sounding channel 6 (1.4 cm sounding 0.65 bars), channel 5 (3.0 cm sounding 1.6 bars), channel 4 (5.75 cm sounding 5 bars), channel 3 (11.5 cm sounding 14 bars), channel 2 (24 cm sound 35 bars) and the deep-sounding channel 1 in the bottom right (50 cm sounding $\sim120$ bars). }
\label{global-mwr-mollweide}
\end{figure*}

\begin{figure*}[!t] 
\makebox[\textwidth]{%
 \noindent\includegraphics[width=1.3\textwidth]{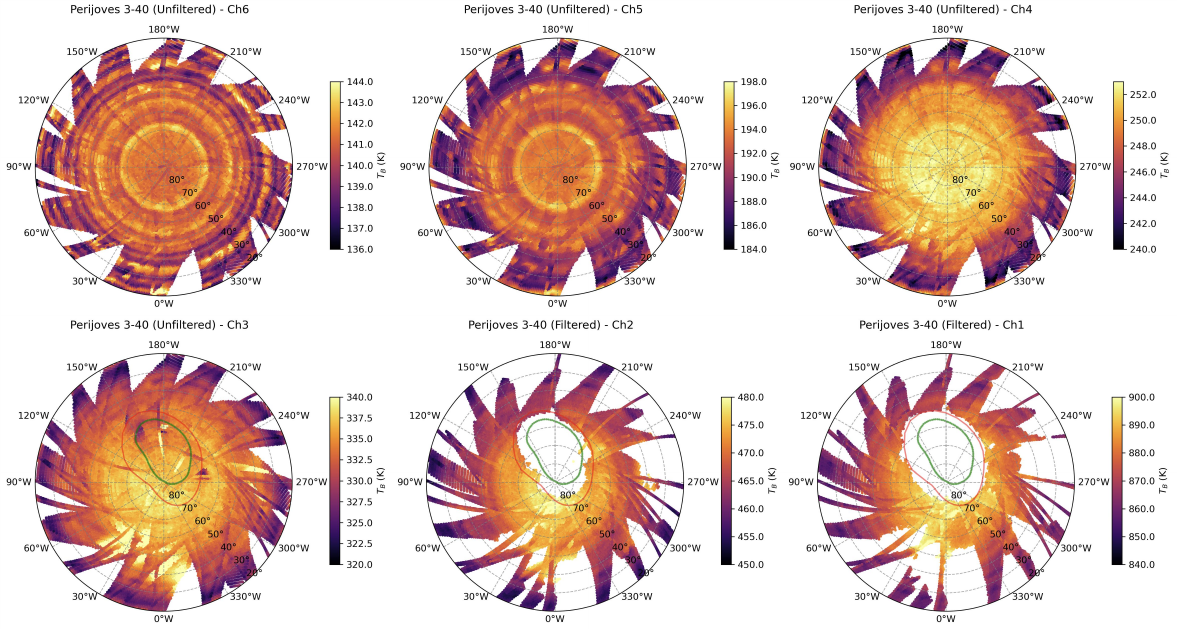}
}%
\caption{Azimuthal equidistant north polar projection of the MWR $T_B$ data shown in Fig. \ref{global-mwr-mollweide}, but with the synchrotron and auroral contamination filtered from each perijove for channels 1 and 2, which explains the lack of coverage interior to the locus of Io footprints (red oval).  Note that the spatial resolution across the north polar region ($>80^\circ$) before February 2022 was insufficient to resolve details of Jupiter's circumpolar cyclones, so they are not shown in this figure. }
\label{global-mwr-polar}
\end{figure*}

In the main text we split the MWR observations into `quartets,' groups that were acquired on adjacent perijoves to focus on the brightness contrasts associated with the FFRs.  In Fig. \ref{global-mwr-mollweide} we show a global Mollweide projection of the MWR data, overlapping observations (emission angles $<65^\circ$) from PJ3 to PJ40.  As Jupiter exhibits significant atmospheric variability over the 6 years of the data, small-scale contrasts should be ignored, but the figure shows the MWR spatial coverage and the overall banded pattern in each of the six channels.  Contamination from synchrotron (high emission at the top of the colour bar for each channel) and the auroral electrons (dark absorption at high latitudes) are retained in Fig. \ref{global-mwr-mollweide}, but in the FFR analysis these are filtered out for every perijove.  This is shown in Fig. \ref{global-mwr-polar}, which displays the same data in an azimuthal equidistant plot, but with auroral and synchrotron emission removed from channels 1 and 2.  Note that the spatial resolution over the polar domain was not yet sufficient to resolve contrasts associated with Jupiter's circumpolar cyclones, which will be the topic of future work \cite{24hu_agu}.

\clearpage

 \begin{figure*}[!t] 
\makebox[\textwidth]{%
 \noindent\includegraphics[width=1.2\textwidth]{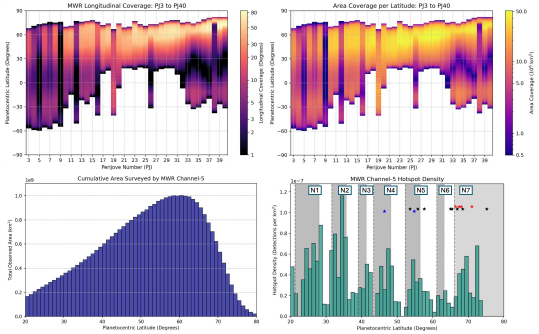}
}%
\caption{Calculation of MWR channel-5 hotspot density in the main text.  We first estimate the longitudinal swath for each perijove as a function of latitude (top left), and convert this into the spatial area of the 1-degree latitude bin (top right, accounting for the surface curvature on an oblate spheroid).  The cumulative sum of hotspot detections is then divided by the cumulative sum of area surveyed by MWR (bottom left) to calculate a hotspot density (bottom right). Eastward (westward) jets are indicated by vertical dashed (dotted) lines, and cyclonic belts are shaded, from the NEB on the left to the N7 domain (polar region) on the right.  Stars indicate latitudes of lightning flash detection by Juno's Stellar Reference Unit between PJ11-17 \cite{20becker} - red for shallow $p<2$ bar lightning; blue for 2-4 bar lightning; and black where no half-width was measured.  }
\label{hotspot_density}
\end{figure*}

\begin{figure*}[!t] 
\makebox[\textwidth]{%
 \noindent\includegraphics[width=1.3\textwidth]{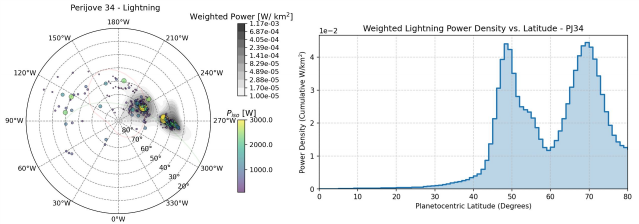}
}%
\caption{Example of lightning density calculation for a single perijove.  In the left panel, coloured dots on the polar projection show the minimum power $P_{iso}$ assuming the lightning discharge is located at the boresight of the channel-1 antenna. We project the channel-1 beam onto Jupiter's 1-bar surface, and use the gain pattern to calculate the dimensionless probability distribution for the lightning location.  This is multiplied by $P_{iso}$, and divided by the area covered by the beam, to create the 2D power density contours in grey.  The green line indicates the sub-spacecraft track, the red line indicates the locations of the Io footprint \cite{03grodent}.  We then calculate the histogram of the 2D power density as a function of latitude to create the plot on the right, showing the peaks in lightning density.  These are summed over all perijoves to create the final histogram in Fig. \ref{lightning-hist} of the main text.}
\label{lightning_density}
\end{figure*}

\begin{figure*}[!h]
\makebox[\textwidth]{%
\noindent\includegraphics[width=1.3\textwidth]{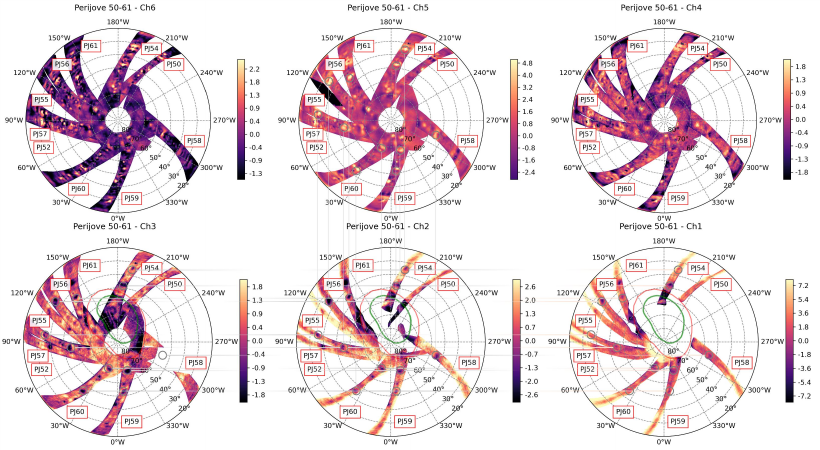}
}%
\caption{Polar projections of MWR $\Delta T_B$ for PJ50 (April 2023) through 61 (May 2024).  Auroral and synchrotron contamination has been removed from channels 1 and 2, but the auroral oval \cite<green,>[]{17nichols} and Io footprints \cite<red,>[]{03grodent} are still shown to indicate auroral-affected regions.  Grey circles (linked by faint lines) are added to guide the eye, showing how microwave-bright spots in channel 5 (1.6 bars) is related to microwave-dark spots in channels 3 ($\sim14$ bars), 2 ($\sim35$ bars), and 1 ($\sim120$ bars). PJ51 and PJ53 are not shown as they would be overlain by later PJs, and measurements poleward of $80^\circ$N are omitted.}
\label{mwr-pj50-60resid}
\end{figure*}

\section{JunoCam and MWR Lightning}
\label{appendix3}

\begin{figure*}
\makebox[\textwidth]{%
\noindent\includegraphics[width=1.2\textwidth]{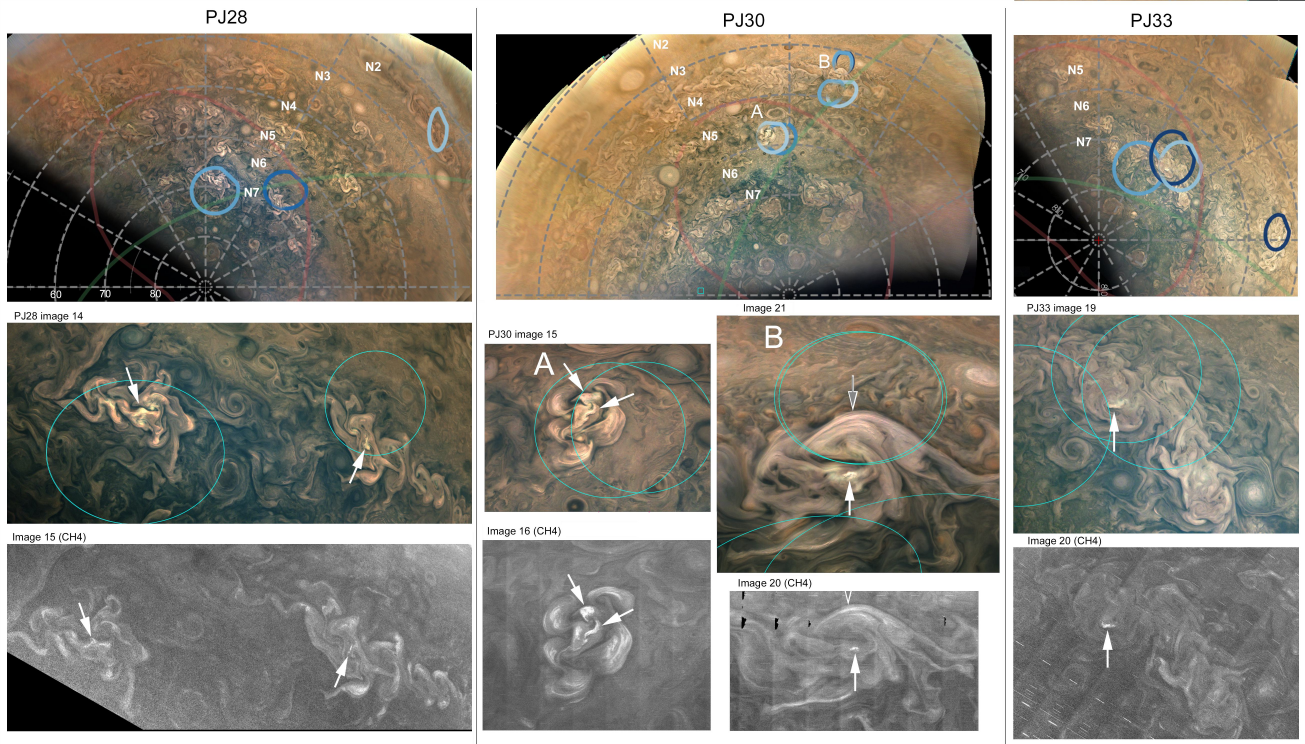}
}%
\caption{Active FFRs imaged by JunoCam that coincide with clusters of lightning discharges detected by MWR, at PJ28, 30, and 33 (compared to Fig. \ref{junocam-lightning} in the main text).  Blue and cyan mark the half-power beam width (HPBW) of the MWR channel 1 at the time of individual strong discharges. All overlap FFRs containing bright white storms flanked by greenish haze (white arrows), one of which is within or on the edge of the lightning footprint, except for PJ30 FFR (B) where a similar storm is just outside the footprints, but there is a dense white strip of elevated clouds within two footprints. \textit{Top row:} North polar projection maps (10 pixels/degree, enlarged $\times2$). The pale green curve marks the sub-spacecraft track; the pale red curve marks the Io magnetic footprint. \textit{Second row:} RGB images at full resolution, with the MWR lightning footprints (cyan) overlaid approximately. Scales vary because the spacecraft altitude is decreasing during this stage of each flyby, and some views are oblique. \textit{Third row:} Methane-band images. Labels (A) and (B) indicate individual FFRs, to reduce ambiguity from the top to bottom rows. 
}
\label{rogers_junocam1}
\end{figure*}

\begin{figure*}
\makebox[\textwidth]{%
\noindent\includegraphics[width=1.2\textwidth]{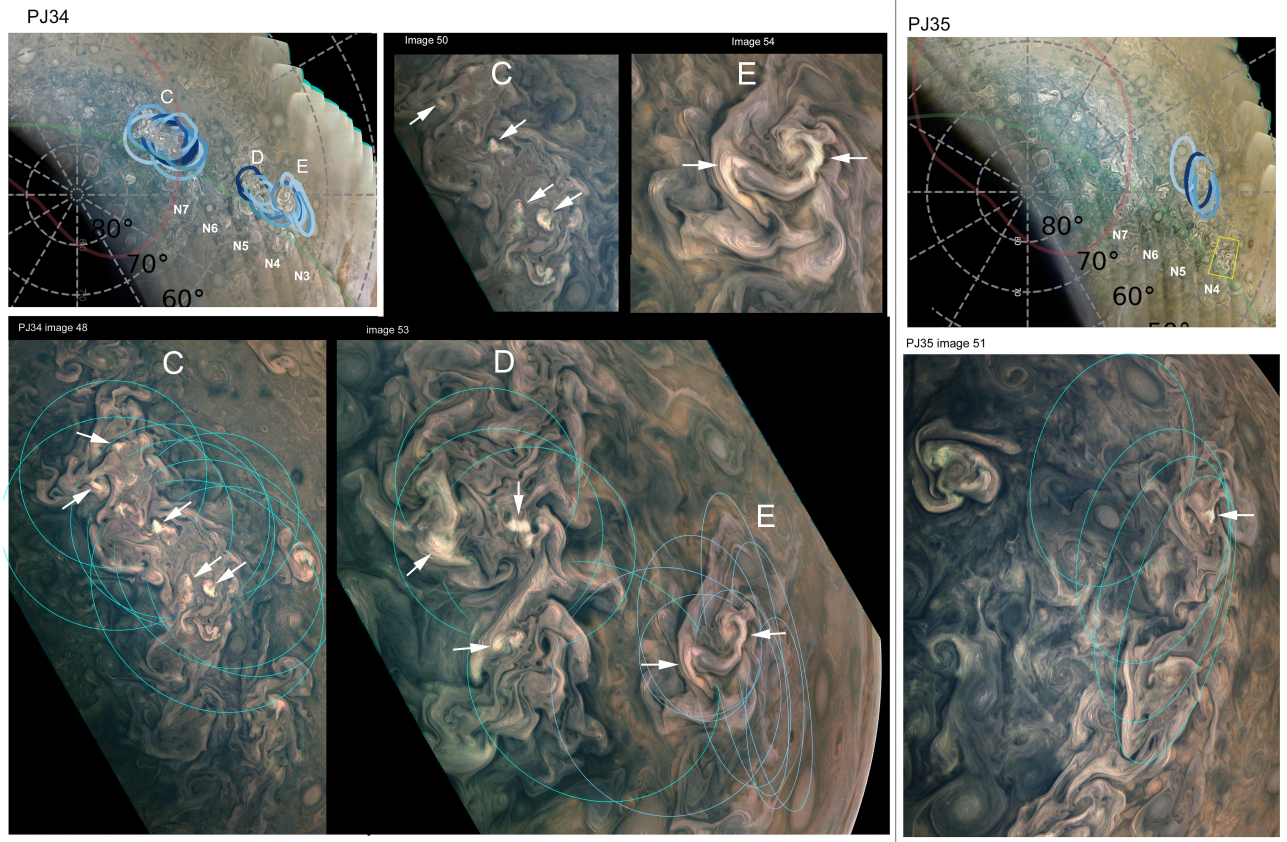}
}%
\caption{Active FFRs imaged by JunoCam that coincide with clusters of lightning discharges detected by MWR, at PJ34 and 35.  Blue and cyan ovals mark the MWR half-power beam widths for individual strong discharges. All overlap FFRs containing bright white storms flanked by greenish haze (white arrows), one of which is within or on the edge of the lightning footprints, except for one footprint at PJ35 where the storm is just outside it. \textit{Top row:} North polar projection maps (10 pixels/degree, enlarged $\times2$). For PJ35, the yellow box indicates the FFR shown in Fig. \ref{junocam_morphology} of the main text, from which MWR detected only a few weak discharges (Fig. \ref{junocam-lightning}). \textit{Second row:} RGB images at full resolution.  For PJ34, labels (C) to (E) indicate individual FFRs, to reduce ambiguity from the top to bottom rows, and images 50 \& 54 show closer views of two of them.  The PJ34 images show three particularly active FFRs producing numerous strong flashes, in domains N7 (C), N5 (D), and N4 (E).
}
\label{rogers_junocam2}
\end{figure*}

\begin{figure*}
\makebox[\textwidth]{%
\noindent\includegraphics[width=1.2\textwidth]{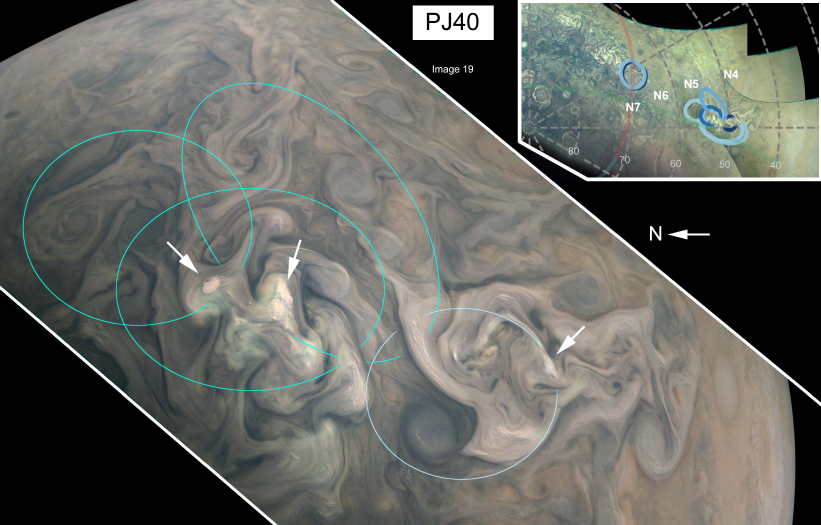}
}%
\caption{A pair of active FFRs imaged by JunoCam that coincide with a cluster of lightning discharges detected by MWR at PJ40. Ovals in shades of blue or cyan mark the HPBW of MWR channel 1 during individual strong discharges. All include one of the bright white storms flanked by greenish haze (white arrows), here viewed at particularly high resolution.  \textit{Top right:} North polar projection map (10 pixels/degree, enlarged $\times2$).  \textit{Main image:} RGB image at full resolution, with the MWR footprints overlaid approximately (omitting two which overlapped all three storms). 
}
\label{rogers_junocam3}
\end{figure*}

\begin{figure*}
\makebox[\textwidth]{%
\noindent\includegraphics[width=1.2\textwidth]{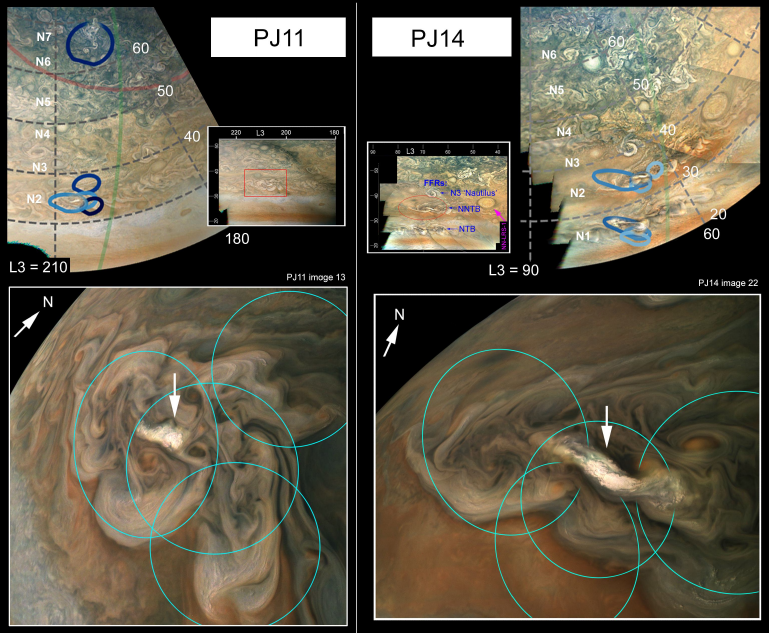}
}%
\caption{Active FFRs imaged by JunoCam in the N2 domain (NNTB), coinciding with clusters of lightning discharges detected by MWR channel 1 at PJ11 and PJ14.  Blue and cyan ovals provide the projected MWR channel-1 HPBW at the time of individual strong discharges.  \textit{Top row:} North polar projection maps (10 pixels/degree, enlarged $\times2$) and cylindrical maps (10 pixels/degree).  \textit{Bottom row:}  RGB images at full resolution, with the footprints overlaid approximately. These views were oblique, with the limb at upper left.  Each FFR contains a bright white storm flanked by greenish haze (white arrows). In all but one case the storm is within or just outside the lightning footprint, except for the upper right footprint at PJ11 which is the only one in this study that is not close to such a storm, although it does overlap a much weaker white cloud strip.
}
\label{rogers_junocam4}
\end{figure*}

Figs. \ref{rogers_junocam1} to \ref{rogers_junocam4} present high-resolution views of FFRs from JunoCam, and match these to MWR lightning discharge detections.  The half-power beam widths (HPBW) of MWR channel 1 were projected onto the 1-bar surface of Jupiter at the time of the detection of a lightning discharge (a spike in the MWR time series).  This reveals individual lightning detections, as well as clusters of activity that can be compared to high-resolution views from JunoCam.  These figures reveal that each FFR contains a bright white storm flanked by greenish haze (white arrows). In all but one case the storm is within or just outside the lightning footprint.  Image processing and map projection was by G. Eichst\"{a}dt, with post-processing by J.H. Rogers. Colours and intensities have been adjusted for best visibility, retaining the high contrast of the bright white storms. Some of the bright white storms were saturated in the raw images.

\clearpage
\bibliography{references.bib}

\begin{thebibliography}{}

\bibitem [\protect \citeauthoryear {%
{Adriani}%
\ \protect \BOthers {.}}{%
{Adriani}%
\ \protect \BOthers {.}}{%
{\protect \APACyear {2017}}%
}]{%
17adriani}
\APACinsertmetastar {%
17adriani}%
\begin{APACrefauthors}%
{Adriani}, A.%
, {Filacchione}, G.%
, {Di Iorio}, T.%
, {Turrini}, D.%
, {Noschese}, R.%
, {Cicchetti}, A.%
\BDBL {}{Olivieri}, A.%
\end{APACrefauthors}%
\unskip\
\newblock
\APACrefYearMonthDay{2017}{{\APACmonth{11}}}{}.
\newblock
{\BBOQ}\APACrefatitle {{JIRAM, the Jovian Infrared Auroral Mapper}} {{JIRAM,
  the Jovian Infrared Auroral Mapper}}.{\BBCQ}
\newblock
\APACjournalVolNumPages{Space Sci. Rev.}{213}{}{393-446}.
\newblock
\begin{APACrefDOI} \doi{10.1007/s11214-014-0094-y} \end{APACrefDOI}
\PrintBackRefs{\CurrentBib}

\bibitem [\protect \citeauthoryear {%
{Adriani}%
\ \protect \BOthers {.}}{%
{Adriani}%
\ \protect \BOthers {.}}{%
{\protect \APACyear {2018}}%
}]{%
18adriani_jiram}
\APACinsertmetastar {%
18adriani_jiram}%
\begin{APACrefauthors}%
{Adriani}, A.%
, {Mura}, A.%
, {Orton}, G.%
, {Hansen}, C.%
, {Altieri}, F.%
, {Moriconi}, M\BPBI L.%
\BDBL {}{Amoroso}, M.%
\end{APACrefauthors}%
\unskip\
\newblock
\APACrefYearMonthDay{2018}{{\APACmonth{03}}}{}.
\newblock
{\BBOQ}\APACrefatitle {{Clusters of cyclones encircling
  Jupiter{\textquoteright}s poles}} {{Clusters of cyclones encircling
  Jupiter{\textquoteright}s poles}}.{\BBCQ}
\newblock
\APACjournalVolNumPages{Nature}{555}{7695}{216-219}.
\newblock
\begin{APACrefDOI} \doi{10.1038/nature25491} \end{APACrefDOI}
\PrintBackRefs{\CurrentBib}

\bibitem [\protect \citeauthoryear {%
{Allison}%
}{%
{Allison}%
}{%
{\protect \APACyear {2000}}%
}]{%
00allison}
\APACinsertmetastar {%
00allison}%
\begin{APACrefauthors}%
{Allison}, M.%
\end{APACrefauthors}%
\unskip\
\newblock
\APACrefYearMonthDay{2000}{{\APACmonth{06}}}{}.
\newblock
{\BBOQ}\APACrefatitle {{A similarity model for the windy jovian thermocline}}
  {{A similarity model for the windy jovian thermocline}}.{\BBCQ}
\newblock
\APACjournalVolNumPages{Planetary and Space Science}{48}{7-8}{753-774}.
\newblock
\begin{APACrefDOI} \doi{10.1016/S0032-0633(00)00032-5} \end{APACrefDOI}
\PrintBackRefs{\CurrentBib}

\bibitem [\protect \citeauthoryear {%
{Annex}%
\ \protect \BOthers {.}}{%
{Annex}%
\ \protect \BOthers {.}}{%
{\protect \APACyear {2020}}%
}]{%
20annex}
\APACinsertmetastar {%
20annex}%
\begin{APACrefauthors}%
{Annex}, A.%
, {Pearson}, B.%
, {Seignovert}, B.%
, {Carcich}, B.%
, {Eichhorn}, H.%
, {Mapel}, J.%
\BDBL {}{Murakami}, S\BHBI y.%
\end{APACrefauthors}%
\unskip\
\newblock
\APACrefYearMonthDay{2020}{{\APACmonth{02}}}{}.
\newblock
{\BBOQ}\APACrefatitle {{SpiceyPy: a Pythonic Wrapper for the SPICE Toolkit}}
  {{SpiceyPy: a Pythonic Wrapper for the SPICE Toolkit}}.{\BBCQ}
\newblock
\APACjournalVolNumPages{The Journal of Open Source Software}{5}{46}{2050}.
\newblock
\begin{APACrefDOI} \doi{10.21105/joss.02050} \end{APACrefDOI}
\PrintBackRefs{\CurrentBib}

\bibitem [\protect \citeauthoryear {%
{Baines}%
\ \protect \BOthers {.}}{%
{Baines}%
\ \protect \BOthers {.}}{%
{\protect \APACyear {2007}}%
}]{%
07baines}
\APACinsertmetastar {%
07baines}%
\begin{APACrefauthors}%
{Baines}, K\BPBI H.%
, {Simon-Miller}, A\BPBI A.%
, {Orton}, G\BPBI S.%
, {Weaver}, H\BPBI A.%
, {Lunsford}, A.%
, {Momary}, T\BPBI W.%
\BDBL {}{Ressler}, M\BPBI E.%
\end{APACrefauthors}%
\unskip\
\newblock
\APACrefYearMonthDay{2007}{{\APACmonth{10}}}{}.
\newblock
{\BBOQ}\APACrefatitle {{Polar Lightning and Decadal-Scale Cloud Variability on
  Jupiter}} {{Polar Lightning and Decadal-Scale Cloud Variability on
  Jupiter}}.{\BBCQ}
\newblock
\APACjournalVolNumPages{Science}{318}{}{226-228}.
\newblock
\begin{APACrefDOI} \doi{10.1126/science.1147912} \end{APACrefDOI}
\PrintBackRefs{\CurrentBib}

\bibitem [\protect \citeauthoryear {%
{Bardet}%
\ \protect \BOthers {.}}{%
{Bardet}%
\ \protect \BOthers {.}}{%
{\protect \APACyear {2024}}%
}]{%
24bardet}
\APACinsertmetastar {%
24bardet}%
\begin{APACrefauthors}%
{Bardet}, D.%
, {Donnelly}, P\BPBI T.%
, {Fletcher}, L\BPBI N.%
, {Antu{\~n}ano}, A.%
, {Roman}, M\BPBI T.%
, {Sinclair}, J\BPBI A.%
\BDBL {}{Harkett}, J.%
\end{APACrefauthors}%
\unskip\
\newblock
\APACrefYearMonthDay{2024}{{\APACmonth{02}}}{}.
\newblock
{\BBOQ}\APACrefatitle {{Investigating Thermal Contrasts Between Jupiter's
  Belts, Zones, and Polar Vortices With VLT/VISIR}} {{Investigating Thermal
  Contrasts Between Jupiter's Belts, Zones, and Polar Vortices With
  VLT/VISIR}}.{\BBCQ}
\newblock
\APACjournalVolNumPages{Journal of Geophysical Research
  (Planets)}{129}{2}{e2023JE007902}.
\newblock
\begin{APACrefDOI} \doi{10.1029/2023JE007902} \end{APACrefDOI}
\PrintBackRefs{\CurrentBib}

\bibitem [\protect \citeauthoryear {%
{Becker}%
\ \protect \BOthers {.}}{%
{Becker}%
\ \protect \BOthers {.}}{%
{\protect \APACyear {2020}}%
}]{%
20becker}
\APACinsertmetastar {%
20becker}%
\begin{APACrefauthors}%
{Becker}, H\BPBI N.%
, {Alexander}, J\BPBI W.%
, {Atreya}, S\BPBI K.%
, {Bolton}, S\BPBI J.%
, {Brennan}, M\BPBI J.%
, {Brown}, S\BPBI T.%
\BDBL {}{Steffes}, P\BPBI G.%
\end{APACrefauthors}%
\unskip\
\newblock
\APACrefYearMonthDay{2020}{{\APACmonth{08}}}{}.
\newblock
{\BBOQ}\APACrefatitle {{Small lightning flashes from shallow electrical storms
  on Jupiter}} {{Small lightning flashes from shallow electrical storms on
  Jupiter}}.{\BBCQ}
\newblock
\APACjournalVolNumPages{Nature}{584}{7819}{55-58}.
\newblock
\begin{APACrefDOI} \doi{10.1038/s41586-020-2532-1} \end{APACrefDOI}
\PrintBackRefs{\CurrentBib}

\bibitem [\protect \citeauthoryear {%
{Bhattacharya}%
\ \protect \BOthers {.}}{%
{Bhattacharya}%
\ \protect \BOthers {.}}{%
{\protect \APACyear {2025}}%
}]{%
25bhattacharya}
\APACinsertmetastar {%
25bhattacharya}%
\begin{APACrefauthors}%
{Bhattacharya}, A.%
, {Waite}, J\BPBI H.%
, {Levin}, S\BPBI M.%
, {Oyafuso}, F\BPBI A.%
, {Steffes}, P\BPBI G.%
, {Lu}, Y.%
\BDBL {}{Bolton}, S\BPBI J.%
\end{APACrefauthors}%
\unskip\
\newblock
\APACrefYearMonthDay{2025}{{\APACmonth{02}}}{}.
\newblock
{\BBOQ}\APACrefatitle {{Jupiter's Auroral Ionosphere: Juno Microwave Radiometer
  Observations of Energetic Electron Precipitation Events}} {{Jupiter's Auroral
  Ionosphere: Juno Microwave Radiometer Observations of Energetic Electron
  Precipitation Events}}.{\BBCQ}
\newblock
\APACjournalVolNumPages{Journal of Geophysical Research (Space
  Physics)}{130}{2}{2024JA033431}.
\newblock
\begin{APACrefDOI} \doi{10.1029/2024JA033431} \end{APACrefDOI}
\PrintBackRefs{\CurrentBib}

\bibitem [\protect \citeauthoryear {%
{Biagiotti}%
\ \protect \BOthers {.}}{%
{Biagiotti}%
\ \protect \BOthers {.}}{%
{\protect \APACyear {2025}}%
}]{%
25biagiotti}
\APACinsertmetastar {%
25biagiotti}%
\begin{APACrefauthors}%
{Biagiotti}, F.%
, {Grassi}, D.%
, {Liuzzi}, G.%
, {Villanueva}, G.%
, {Piccioni}, G.%
, {Guillot}, T.%
\BDBL {}{Bolton}, S.%
\end{APACrefauthors}%
\unskip\
\newblock
\APACrefYearMonthDay{2025}{{\APACmonth{04}}}{}.
\newblock
{\BBOQ}\APACrefatitle {{Evidence of pure ammonia clouds in Jupiter's Northern
  Temperate domain from Juno/JIRAM infrared spectral data}} {{Evidence of pure
  ammonia clouds in Jupiter's Northern Temperate domain from Juno/JIRAM
  infrared spectral data}}.{\BBCQ}
\newblock
\APACjournalVolNumPages{Monthly Notices of the Royal Astronomical
  Society}{538}{3}{1535-1564}.
\newblock
\begin{APACrefDOI} \doi{10.1093/mnras/staf381} \end{APACrefDOI}
\PrintBackRefs{\CurrentBib}

\bibitem [\protect \citeauthoryear {%
{Bolton}%
\ \protect \BOthers {.}}{%
{Bolton}%
\ \protect \BOthers {.}}{%
{\protect \APACyear {2021}}%
}]{%
21bolton}
\APACinsertmetastar {%
21bolton}%
\begin{APACrefauthors}%
{Bolton}, S\BPBI J.%
, {Levin}, S\BPBI M.%
, {Guillot}, T.%
, {Li}, C.%
, {Kaspi}, Y.%
, {Orton}, G.%
\BDBL {}{Zhang}, Z.%
\end{APACrefauthors}%
\unskip\
\newblock
\APACrefYearMonthDay{2021}{{\APACmonth{11}}}{}.
\newblock
{\BBOQ}\APACrefatitle {{Microwave observations reveal the deep extent and
  structure of Jupiter{\textquoteright}s atmospheric vortices}} {{Microwave
  observations reveal the deep extent and structure of
  Jupiter{\textquoteright}s atmospheric vortices}}.{\BBCQ}
\newblock
\APACjournalVolNumPages{Science}{374}{6570}{968-972}.
\newblock
\begin{APACrefDOI} \doi{10.1126/science.abf1015} \end{APACrefDOI}
\PrintBackRefs{\CurrentBib}

\bibitem [\protect \citeauthoryear {%
{Borucki}%
\ \BBA {} {Magalhaes}%
}{%
{Borucki}%
\ \BBA {} {Magalhaes}%
}{%
{\protect \APACyear {1992}}%
}]{%
92borucki}
\APACinsertmetastar {%
92borucki}%
\begin{APACrefauthors}%
{Borucki}, W\BPBI J.%
\BCBT {}\ \BBA {} {Magalhaes}, J\BPBI A.%
\end{APACrefauthors}%
\unskip\
\newblock
\APACrefYearMonthDay{1992}{{\APACmonth{03}}}{}.
\newblock
{\BBOQ}\APACrefatitle {{Analysis of Voyager 2 images of Jovian lightning}}
  {{Analysis of Voyager 2 images of Jovian lightning}}.{\BBCQ}
\newblock
\APACjournalVolNumPages{Icarus}{96}{}{1-14}.
\newblock
\begin{APACrefDOI} \doi{10.1016/0019-1035(92)90002-O} \end{APACrefDOI}
\PrintBackRefs{\CurrentBib}

\bibitem [\protect \citeauthoryear {%
Bosse%
\ \protect \BOthers {.}}{%
Bosse%
\ \protect \BOthers {.}}{%
{\protect \APACyear {2016}}%
}]{%
16bosse}
\APACinsertmetastar {%
16bosse}%
\begin{APACrefauthors}%
Bosse, A.%
, Testor, P.%
, Houpert, L.%
, Damien, P.%
, Prieur, L.%
, Hayes, D.%
\BDBL {}Mortier, L.%
\end{APACrefauthors}%
\unskip\
\newblock
\APACrefYearMonthDay{2016}{}{}.
\newblock
{\BBOQ}\APACrefatitle {Scales and dynamics of Submesoscale Coherent Vortices
  formed by deep convection in the northwestern Mediterranean Sea} {Scales and
  dynamics of submesoscale coherent vortices formed by deep convection in the
  northwestern mediterranean sea}.{\BBCQ}
\newblock
\APACjournalVolNumPages{Journal of Geophysical Research:
  Oceans}{121}{10}{7716-7742}.
\newblock
\begin{APACrefURL}
  \url{https://agupubs.onlinelibrary.wiley.com/doi/abs/10.1002/2016JC012144}
  \end{APACrefURL}
\newblock
\begin{APACrefDOI} \doi{https://doi.org/10.1002/2016JC012144} \end{APACrefDOI}
\PrintBackRefs{\CurrentBib}

\bibitem [\protect \citeauthoryear {%
{Brown}%
\ \protect \BOthers {.}}{%
{Brown}%
\ \protect \BOthers {.}}{%
{\protect \APACyear {2018}}%
}]{%
18brown}
\APACinsertmetastar {%
18brown}%
\begin{APACrefauthors}%
{Brown}, S.%
, {Janssen}, M.%
, {Adumitroaie}, V.%
, {Atreya}, S.%
, {Bolton}, S.%
, {Gulkis}, S.%
\BDBL {}{Connerney}, J.%
\end{APACrefauthors}%
\unskip\
\newblock
\APACrefYearMonthDay{2018}{Jun}{}.
\newblock
{\BBOQ}\APACrefatitle {{Prevalent lightning sferics at 600 megahertz near
  Jupiter's poles}} {{Prevalent lightning sferics at 600 megahertz near
  Jupiter's poles}}.{\BBCQ}
\newblock
\APACjournalVolNumPages{Nature}{558}{}{87-90}.
\newblock
\begin{APACrefDOI} \doi{10.1038/s41586-018-0156-5} \end{APACrefDOI}
\PrintBackRefs{\CurrentBib}

\bibitem [\protect \citeauthoryear {%
{Brueshaber}%
\ \protect \BOthers {.}}{%
{Brueshaber}%
\ \protect \BOthers {.}}{%
{\protect \APACyear {2025}}%
}]{%
25brueshaber}
\APACinsertmetastar {%
25brueshaber}%
\begin{APACrefauthors}%
{Brueshaber}, S\BPBI R.%
, {Zhang}, Z.%
, {Rogers}, J\BPBI H.%
, {Eichst{\"a}dt}, G.%
, {Orton}, G\BPBI S.%
, {Grassi}, D.%
\BDBL {}{Bolton}, S.%
\end{APACrefauthors}%
\unskip\
\newblock
\APACrefYearMonthDay{2025}{{\APACmonth{05}}}{}.
\newblock
{\BBOQ}\APACrefatitle {{Multi-instrument sounding of a Jovian thunderstorm from
  Juno}} {{Multi-instrument sounding of a Jovian thunderstorm from
  Juno}}.{\BBCQ}
\newblock
\APACjournalVolNumPages{{Icarus}}{432}{}{116465}.
\newblock
\begin{APACrefDOI} \doi{10.1016/j.icarus.2025.116465} \end{APACrefDOI}
\PrintBackRefs{\CurrentBib}

\bibitem [\protect \citeauthoryear {%
Carey%
, Murphy%
, McCormick%
\BCBL {}\ \BBA {} Demetriades%
}{%
Carey%
\ \protect \BOthers {.}}{%
{\protect \APACyear {2005}}%
}]{%
05carey}
\APACinsertmetastar {%
05carey}%
\begin{APACrefauthors}%
Carey, L\BPBI D.%
, Murphy, M\BPBI J.%
, McCormick, T\BPBI L.%
\BCBL {}\ \BBA {} Demetriades, N\BPBI W\BPBI S.%
\end{APACrefauthors}%
\unskip\
\newblock
\APACrefYearMonthDay{2005}{}{}.
\newblock
{\BBOQ}\APACrefatitle {Lightning location relative to storm structure in a
  leading-line, trailing-stratiform mesoscale convective system} {Lightning
  location relative to storm structure in a leading-line, trailing-stratiform
  mesoscale convective system}.{\BBCQ}
\newblock
\APACjournalVolNumPages{Journal of Geophysical Research:
  Atmospheres}{110}{D3}{}.
\newblock
\begin{APACrefURL}
  \url{https://agupubs.onlinelibrary.wiley.com/doi/abs/10.1029/2003JD004371}
  \end{APACrefURL}
\newblock
\begin{APACrefDOI} \doi{https://doi.org/10.1029/2003JD004371} \end{APACrefDOI}
\PrintBackRefs{\CurrentBib}

\bibitem [\protect \citeauthoryear {%
Chang%
, Miyazawa%
, Oey%
, Kodaira%
\BCBL {}\ \BBA {} Huang%
}{%
Chang%
\ \protect \BOthers {.}}{%
{\protect \APACyear {2017}}%
}]{%
17chang}
\APACinsertmetastar {%
17chang}%
\begin{APACrefauthors}%
Chang, Y\BHBI L.%
, Miyazawa, Y.%
, Oey, L\BHBI Y.%
, Kodaira, T.%
\BCBL {}\ \BBA {} Huang, S.%
\end{APACrefauthors}%
\unskip\
\newblock
\APACrefYearMonthDay{2017}{}{}.
\newblock
{\BBOQ}\APACrefatitle {The formation processes of phytoplankton growth and
  decline in mesoscale eddies in the western N orth P acific O cean} {The
  formation processes of phytoplankton growth and decline in mesoscale eddies
  in the western n orth p acific o cean}.{\BBCQ}
\newblock
\APACjournalVolNumPages{Journal of Geophysical Research:
  Oceans}{122}{5}{4444--4455}.
\PrintBackRefs{\CurrentBib}

\bibitem [\protect \citeauthoryear {%
{de Pater}%
\ \protect \BOthers {.}}{%
{de Pater}%
\ \protect \BOthers {.}}{%
{\protect \APACyear {2019}}%
}]{%
19depater_alma}
\APACinsertmetastar {%
19depater_alma}%
\begin{APACrefauthors}%
{de Pater}, I.%
, {Sault}, R\BPBI J.%
, {Moeckel}, C.%
, {Moullet}, A.%
, {Wong}, M\BPBI H.%
, {Goullaud}, C.%
\BDBL {}{Villard}, E.%
\end{APACrefauthors}%
\unskip\
\newblock
\APACrefYearMonthDay{2019}{Oct}{}.
\newblock
{\BBOQ}\APACrefatitle {{First ALMA Millimeter-wavelength Maps of Jupiter, with
  a Multiwavelength Study of Convection}} {{First ALMA Millimeter-wavelength
  Maps of Jupiter, with a Multiwavelength Study of Convection}}.{\BBCQ}
\newblock
\APACjournalVolNumPages{Astronomical Journal}{158}{4}{139}.
\newblock
\begin{APACrefDOI} \doi{10.3847/1538-3881/ab3643} \end{APACrefDOI}
\PrintBackRefs{\CurrentBib}

\bibitem [\protect \citeauthoryear {%
{de Pater}%
\ \protect \BOthers {.}}{%
{de Pater}%
\ \protect \BOthers {.}}{%
{\protect \APACyear {2010}}%
}]{%
10depater_jup}
\APACinsertmetastar {%
10depater_jup}%
\begin{APACrefauthors}%
{de Pater}, I.%
, {Wong}, M\BPBI H.%
, {Marcus}, P.%
, {Luszcz-Cook}, S.%
, {{\'A}d{\'a}mkovics}, M.%
, {Conrad}, A.%
\BDBL {}{Go}, C.%
\end{APACrefauthors}%
\unskip\
\newblock
\APACrefYearMonthDay{2010}{{\APACmonth{12}}}{}.
\newblock
{\BBOQ}\APACrefatitle {{Persistent rings in and around Jupiter's anticyclones -
  Observations and theory}} {{Persistent rings in and around Jupiter's
  anticyclones - Observations and theory}}.{\BBCQ}
\newblock
\APACjournalVolNumPages{{Icarus}}{210}{}{742-762}.
\newblock
\begin{APACrefDOI} \doi{10.1016/j.icarus.2010.07.027} \end{APACrefDOI}
\PrintBackRefs{\CurrentBib}

\bibitem [\protect \citeauthoryear {%
de Marez%
, Carton%
, Corr{\'e}ard%
, L'H{\'e}garet%
\BCBL {}\ \BBA {} Morvan%
}{%
de Marez%
\ \protect \BOthers {.}}{%
{\protect \APACyear {2020}}%
}]{%
20demarez}
\APACinsertmetastar {%
20demarez}%
\begin{APACrefauthors}%
de Marez, C.%
, Carton, X.%
, Corr{\'e}ard, S.%
, L'H{\'e}garet, P.%
\BCBL {}\ \BBA {} Morvan, M.%
\end{APACrefauthors}%
\unskip\
\newblock
\APACrefYearMonthDay{2020}{}{}.
\newblock
{\BBOQ}\APACrefatitle {Observations of a Deep Submesoscale Cyclonic Vortex in
  the Arabian Sea} {Observations of a deep submesoscale cyclonic vortex in the
  arabian sea}.{\BBCQ}
\newblock
\APACjournalVolNumPages{Geophysical Research Letters}{47}{13}{e2020GL087881}.
\newblock
\begin{APACrefURL}
  \url{https://agupubs.onlinelibrary.wiley.com/doi/abs/10.1029/2020GL087881}
  \end{APACrefURL}
\newblock
\APACrefnote{e2020GL087881 10.1029/2020GL087881}
\newblock
\begin{APACrefDOI} \doi{https://doi.org/10.1029/2020GL087881} \end{APACrefDOI}
\PrintBackRefs{\CurrentBib}

\bibitem [\protect \citeauthoryear {%
{Dowling}%
\ \BBA {} {Gierasch}%
}{%
{Dowling}%
\ \BBA {} {Gierasch}%
}{%
{\protect \APACyear {1989}}%
}]{%
89dowling_dps}
\APACinsertmetastar {%
89dowling_dps}%
\begin{APACrefauthors}%
{Dowling}, T\BPBI E.%
\BCBT {}\ \BBA {} {Gierasch}, P\BPBI J.%
\end{APACrefauthors}%
\unskip\
\newblock
\APACrefYearMonthDay{1989}{{\APACmonth{06}}}{}.
\newblock
{\BBOQ}\APACrefatitle {{Cyclones and Moist Convection on Jovian Planets}}
  {{Cyclones and Moist Convection on Jovian Planets}}.{\BBCQ}
\newblock
\BIn{} \APACrefbtitle {Bulletin of the American Astronomical Society} {Bulletin
  of the american astronomical society}\ (\BVOL~21, \BPG~946).
\PrintBackRefs{\CurrentBib}

\bibitem [\protect \citeauthoryear {%
{Duer}%
\ \protect \BOthers {.}}{%
{Duer}%
\ \protect \BOthers {.}}{%
{\protect \APACyear {2021}}%
}]{%
21duer}
\APACinsertmetastar {%
21duer}%
\begin{APACrefauthors}%
{Duer}, K.%
, {Gavriel}, N.%
, {Galanti}, E.%
, {Kaspi}, Y.%
, {Fletcher}, L\BPBI N.%
, {Guillot}, T.%
\BDBL {}{Waite}, J\BPBI H.%
\end{APACrefauthors}%
\unskip\
\newblock
\APACrefYearMonthDay{2021}{{\APACmonth{12}}}{}.
\newblock
{\BBOQ}\APACrefatitle {{Evidence for Multiple Ferrel-Like Cells on Jupiter}}
  {{Evidence for Multiple Ferrel-Like Cells on Jupiter}}.{\BBCQ}
\newblock
\APACjournalVolNumPages{Geophys. Res. Lett.}{48}{23}{e95651}.
\newblock
\begin{APACrefDOI} \doi{10.1029/2021GL095651} \end{APACrefDOI}
\PrintBackRefs{\CurrentBib}

\bibitem [\protect \citeauthoryear {%
Dyudina%
\ \protect \BOthers {.}}{%
Dyudina%
\ \protect \BOthers {.}}{%
{\protect \APACyear {2007}}%
}]{%
07dyudina}
\APACinsertmetastar {%
07dyudina}%
\begin{APACrefauthors}%
Dyudina, U.%
, Ingersoll, A.%
, Ewald, S.%
, Porco, C.%
, Fischer, G.%
, Kurth, W.%
\BDBL {}Ferrier, J.%
\end{APACrefauthors}%
\unskip\
\newblock
\APACrefYearMonthDay{2007}{}{}.
\newblock
{\BBOQ}\APACrefatitle {{Lightning storms on Saturn observed by Cassini ISS and
  RPWS during 2004--2006}} {{Lightning storms on Saturn observed by Cassini ISS
  and RPWS during 2004--2006}}.{\BBCQ}
\newblock
\APACjournalVolNumPages{Icarus}{190}{2}{545--555}.
\PrintBackRefs{\CurrentBib}

\bibitem [\protect \citeauthoryear {%
{Dyudina}%
\ \protect \BOthers {.}}{%
{Dyudina}%
\ \protect \BOthers {.}}{%
{\protect \APACyear {2004}}%
}]{%
04dyudina}
\APACinsertmetastar {%
04dyudina}%
\begin{APACrefauthors}%
{Dyudina}, U\BPBI A.%
, {Del Genio}, A\BPBI D.%
, {Ingersoll}, A\BPBI P.%
, {Porco}, C\BPBI C.%
, {West}, R\BPBI A.%
, {Vasavada}, A\BPBI R.%
\BCBL {}\ \BBA {} {Barbara}, J\BPBI M.%
\end{APACrefauthors}%
\unskip\
\newblock
\APACrefYearMonthDay{2004}{{\APACmonth{11}}}{}.
\newblock
{\BBOQ}\APACrefatitle {{Lightning on Jupiter observed in the
  H$_{{\ensuremath{\alpha}}}$ line by the Cassini imaging science subsystem}}
  {{Lightning on Jupiter observed in the H$_{{\ensuremath{\alpha}}}$ line by
  the Cassini imaging science subsystem}}.{\BBCQ}
\newblock
\APACjournalVolNumPages{{Icarus}}{172}{1}{24-36}.
\newblock
\begin{APACrefDOI} \doi{10.1016/j.icarus.2004.07.014} \end{APACrefDOI}
\PrintBackRefs{\CurrentBib}

\bibitem [\protect \citeauthoryear {%
{Fischer}%
\ \protect \BOthers {.}}{%
{Fischer}%
\ \protect \BOthers {.}}{%
{\protect \APACyear {2019}}%
}]{%
19fischer}
\APACinsertmetastar {%
19fischer}%
\begin{APACrefauthors}%
{Fischer}, G.%
, {Pagaran}, J\BPBI A.%
, {Zarka}, P.%
, {Delcroix}, M.%
, {Dyudina}, U\BPBI A.%
, {Kurth}, W\BPBI S.%
\BCBL {}\ \BBA {} {Gurnett}, D\BPBI A.%
\end{APACrefauthors}%
\unskip\
\newblock
\APACrefYearMonthDay{2019}{{\APACmonth{01}}}{}.
\newblock
{\BBOQ}\APACrefatitle {{Analysis of a long-lived, two-cell lightning storm on
  Saturn}} {{Analysis of a long-lived, two-cell lightning storm on
  Saturn}}.{\BBCQ}
\newblock
\APACjournalVolNumPages{Astron. Astrophys.}{621}{}{A113}.
\newblock
\begin{APACrefDOI} \doi{10.1051/0004-6361/201833014} \end{APACrefDOI}
\PrintBackRefs{\CurrentBib}

\bibitem [\protect \citeauthoryear {%
{Fletcher}%
}{%
{Fletcher}%
}{%
{\protect \APACyear {2025}}%
}]{%
25fletcher_data}
\APACinsertmetastar {%
25fletcher_data}%
\begin{APACrefauthors}%
{Fletcher}, L\BPBI N.%
\end{APACrefauthors}%
\unskip\
\newblock
\APACrefYearMonthDay{2025}{}{}.
\newblock
\APACrefbtitle {{Supporting Data for Juno Folded Filamentary Region Study
  [Data]}.} {{Supporting Data for Juno Folded Filamentary Region Study
  [Data]}.}
\newblock
\begin{APACrefURL}
  \url{https://github.com/leighfletcher/Juno-Folded-Filamentary-Regions}
  \end{APACrefURL}
\newblock
\begin{APACrefDOI} \doi{10.5281/zenodo.15755740} \end{APACrefDOI}
\PrintBackRefs{\CurrentBib}

\bibitem [\protect \citeauthoryear {%
{Fletcher}%
\ \protect \BOthers {.}}{%
{Fletcher}%
\ \protect \BOthers {.}}{%
{\protect \APACyear {2016}}%
}]{%
16fletcher_texes}
\APACinsertmetastar {%
16fletcher_texes}%
\begin{APACrefauthors}%
{Fletcher}, L\BPBI N.%
, {Greathouse}, T\BPBI K.%
, {Orton}, G\BPBI S.%
, {Sinclair}, J\BPBI A.%
, {Giles}, R\BPBI S.%
, {Irwin}, P\BPBI G\BPBI J.%
\BCBL {}\ \BBA {} {Encrenaz}, T.%
\end{APACrefauthors}%
\unskip\
\newblock
\APACrefYearMonthDay{2016}{{\APACmonth{11}}}{}.
\newblock
{\BBOQ}\APACrefatitle {{Mid-infrared mapping of Jupiter's temperatures, aerosol
  opacity and chemical distributions with IRTF/TEXES}} {{Mid-infrared mapping
  of Jupiter's temperatures, aerosol opacity and chemical distributions with
  IRTF/TEXES}}.{\BBCQ}
\newblock
\APACjournalVolNumPages{Icarus}{278}{}{128-161}.
\newblock
\begin{APACrefDOI} \doi{10.1016/j.icarus.2016.06.008} \end{APACrefDOI}
\PrintBackRefs{\CurrentBib}

\bibitem [\protect \citeauthoryear {%
{Fletcher}%
\ \protect \BOthers {.}}{%
{Fletcher}%
\ \protect \BOthers {.}}{%
{\protect \APACyear {2018}}%
}]{%
18fletcher_waves}
\APACinsertmetastar {%
18fletcher_waves}%
\begin{APACrefauthors}%
{Fletcher}, L\BPBI N.%
, {Melin}, H.%
, {Adriani}, A.%
, {Simon}, A\BPBI A.%
, {Sanchez-Lavega}, A.%
, {Donnelly}, P\BPBI T.%
\BDBL {}{Sindoni}, G.%
\end{APACrefauthors}%
\unskip\
\newblock
\APACrefYearMonthDay{2018}{{\APACmonth{08}}}{}.
\newblock
{\BBOQ}\APACrefatitle {{Jupiter's Mesoscale Waves Observed at 5
  {\ensuremath{\mu}}m by Ground-based Observations and Juno JIRAM}} {{Jupiter's
  Mesoscale Waves Observed at 5 {\ensuremath{\mu}}m by Ground-based
  Observations and Juno JIRAM}}.{\BBCQ}
\newblock
\APACjournalVolNumPages{Astronomical Journal}{156}{2}{67}.
\newblock
\begin{APACrefDOI} \doi{10.3847/1538-3881/aace02} \end{APACrefDOI}
\PrintBackRefs{\CurrentBib}

\bibitem [\protect \citeauthoryear {%
{Fletcher}%
\ \protect \BOthers {.}}{%
{Fletcher}%
\ \protect \BOthers {.}}{%
{\protect \APACyear {2020}}%
}]{%
20fletcher}
\APACinsertmetastar {%
20fletcher}%
\begin{APACrefauthors}%
{Fletcher}, L\BPBI N.%
, {Orton}, G\BPBI S.%
, {Greathouse}, T\BPBI K.%
, {Rogers}, J\BPBI H.%
, {Zhang}, Z.%
, {Oyafuso}, F\BPBI A.%
\BDBL {}{Adriani}, A.%
\end{APACrefauthors}%
\unskip\
\newblock
\APACrefYearMonthDay{2020}{{\APACmonth{08}}}{}.
\newblock
{\BBOQ}\APACrefatitle {{Jupiter's Equatorial Plumes and Hot Spots: Spectral
  Mapping from Gemini/TEXES and Juno/MWR}} {{Jupiter's Equatorial Plumes and
  Hot Spots: Spectral Mapping from Gemini/TEXES and Juno/MWR}}.{\BBCQ}
\newblock
\APACjournalVolNumPages{Journal of Geophysical Research
  (Planets)}{125}{8}{e06399}.
\newblock
\begin{APACrefDOI} \doi{10.1029/2020JE006399} \end{APACrefDOI}
\PrintBackRefs{\CurrentBib}

\bibitem [\protect \citeauthoryear {%
{Fletcher}%
\ \protect \BOthers {.}}{%
{Fletcher}%
\ \protect \BOthers {.}}{%
{\protect \APACyear {2017}}%
}]{%
17fletcher_seb}
\APACinsertmetastar {%
17fletcher_seb}%
\begin{APACrefauthors}%
{Fletcher}, L\BPBI N.%
, {Orton}, G\BPBI S.%
, {Rogers}, J\BPBI H.%
, {Giles}, R\BPBI S.%
, {Payne}, A\BPBI V.%
, {Irwin}, P\BPBI G\BPBI J.%
\BCBL {}\ \BBA {} {Vedovato}, M.%
\end{APACrefauthors}%
\unskip\
\newblock
\APACrefYearMonthDay{2017}{{\APACmonth{04}}}{}.
\newblock
{\BBOQ}\APACrefatitle {{Moist convection and the 2010-2011 revival of Jupiter's
  South Equatorial Belt}} {{Moist convection and the 2010-2011 revival of
  Jupiter's South Equatorial Belt}}.{\BBCQ}
\newblock
\APACjournalVolNumPages{{Icarus}}{286}{}{94-117}.
\newblock
\begin{APACrefDOI} \doi{10.1016/j.icarus.2017.01.001} \end{APACrefDOI}
\PrintBackRefs{\CurrentBib}

\bibitem [\protect \citeauthoryear {%
{Fletcher}%
\ \protect \BOthers {.}}{%
{Fletcher}%
\ \protect \BOthers {.}}{%
{\protect \APACyear {2022}}%
}]{%
22fletcher_epsc}
\APACinsertmetastar {%
22fletcher_epsc}%
\begin{APACrefauthors}%
{Fletcher}, L\BPBI N.%
, {Oyafuso}, F.%
, {Orton}, G.%
, {Zhang}, Z.%
, {Brueshaber}, S.%
, {Wong}, M.%
\BDBL {}{Brown}, S.%
\end{APACrefauthors}%
\unskip\
\newblock
\APACrefYearMonthDay{2022}{{\APACmonth{09}}}{}.
\newblock
{\BBOQ}\APACrefatitle {{Juno Characterisation of Cyclonic ``Folded Filamentary
  Regions'' within Jupiter's Polar Domains}} {{Juno Characterisation of
  Cyclonic ``Folded Filamentary Regions'' within Jupiter's Polar
  Domains}}.{\BBCQ}
\newblock
\BIn{} \APACrefbtitle {European Planetary Science Congress} {European planetary
  science congress}\ (\BPG~EPSC2022-475).
\newblock
\begin{APACrefDOI} \doi{10.5194/epsc2022-475} \end{APACrefDOI}
\PrintBackRefs{\CurrentBib}

\bibitem [\protect \citeauthoryear {%
{Fletcher}%
\ \protect \BOthers {.}}{%
{Fletcher}%
\ \protect \BOthers {.}}{%
{\protect \APACyear {2021}}%
}]{%
21fletcher}
\APACinsertmetastar {%
21fletcher}%
\begin{APACrefauthors}%
{Fletcher}, L\BPBI N.%
, {Oyafuso}, F\BPBI A.%
, {Allison}, M.%
, {Ingersoll}, A.%
, {Li}, L.%
, {Kaspi}, Y.%
\BDBL {}{Bolton}, S.%
\end{APACrefauthors}%
\unskip\
\newblock
\APACrefYearMonthDay{2021}{{\APACmonth{10}}}{}.
\newblock
{\BBOQ}\APACrefatitle {{Jupiter's Temperate Belt/Zone Contrasts Revealed at
  Depth by Juno Microwave Observations}} {{Jupiter's Temperate Belt/Zone
  Contrasts Revealed at Depth by Juno Microwave Observations}}.{\BBCQ}
\newblock
\APACjournalVolNumPages{Journal of Geophysical Research
  (Planets)}{126}{10}{e06858}.
\newblock
\begin{APACrefDOI} \doi{10.1029/2021JE006858} \end{APACrefDOI}
\PrintBackRefs{\CurrentBib}

\bibitem [\protect \citeauthoryear {%
{Garc{\'{\i}}a-Melendo}%
, {P{\'e}rez-Hoyos}%
, {S{\'a}nchez-Lavega}%
\BCBL {}\ \BBA {} {Hueso}%
}{%
{Garc{\'{\i}}a-Melendo}%
\ \protect \BOthers {.}}{%
{\protect \APACyear {2011}}%
}]{%
11garcia}
\APACinsertmetastar {%
11garcia}%
\begin{APACrefauthors}%
{Garc{\'{\i}}a-Melendo}, E.%
, {P{\'e}rez-Hoyos}, S.%
, {S{\'a}nchez-Lavega}, A.%
\BCBL {}\ \BBA {} {Hueso}, R.%
\end{APACrefauthors}%
\unskip\
\newblock
\APACrefYearMonthDay{2011}{{\APACmonth{09}}}{}.
\newblock
{\BBOQ}\APACrefatitle {{Saturn's zonal wind profile in 2004-2009 from Cassini
  ISS images and its long-term variability}} {{Saturn's zonal wind profile in
  2004-2009 from Cassini ISS images and its long-term variability}}.{\BBCQ}
\newblock
\APACjournalVolNumPages{Icarus}{215}{}{62-74}.
\newblock
\begin{APACrefDOI} \doi{10.1016/j.icarus.2011.07.005} \end{APACrefDOI}
\PrintBackRefs{\CurrentBib}

\bibitem [\protect \citeauthoryear {%
{Ge}%
, {Li}%
, {Zhang}%
, {Ingersoll}%
\BCBL {}\ \BBA {} {Chen}%
}{%
{Ge}%
\ \protect \BOthers {.}}{%
{\protect \APACyear {2025}}%
}]{%
25ge}
\APACinsertmetastar {%
25ge}%
\begin{APACrefauthors}%
{Ge}, H.%
, {Li}, C.%
, {Zhang}, X.%
, {Ingersoll}, A\BPBI P.%
\BCBL {}\ \BBA {} {Chen}, S.%
\end{APACrefauthors}%
\unskip\
\newblock
\APACrefYearMonthDay{2025}{{\APACmonth{09}}}{}.
\newblock
{\BBOQ}\APACrefatitle {{Nonuniform water distribution in Jupiter's
  midlatitudes: Influence of precipitation and planetary rotation}}
  {{Nonuniform water distribution in Jupiter's midlatitudes: Influence of
  precipitation and planetary rotation}}.{\BBCQ}
\newblock
\APACjournalVolNumPages{Proceedings of the National Academy of
  Science}{122}{41}{e2419087122}.
\newblock
\begin{APACrefDOI} \doi{10.1073/pnas.2419087122} \end{APACrefDOI}
\PrintBackRefs{\CurrentBib}

\bibitem [\protect \citeauthoryear {%
{Gehrels}%
}{%
{Gehrels}%
}{%
{\protect \APACyear {1976}}%
}]{%
76gehrels}
\APACinsertmetastar {%
76gehrels}%
\begin{APACrefauthors}%
{Gehrels}, T.%
\end{APACrefauthors}%
\unskip\
\newblock
\APACrefYearMonthDay{1976}{{\APACmonth{01}}}{}.
\newblock
{\BBOQ}\APACrefatitle {{The results of the imaging photopolarimeter on Pioneers
  10 and 11}} {{The results of the imaging photopolarimeter on Pioneers 10 and
  11}}.{\BBCQ}
\newblock
\BIn{} T.~{Gehrels}\ \BBA {} S.~{Matthews}\ (\BEDS), \APACrefbtitle {IAU
  Colloq. 30: Jupiter: Studies of the Interior, Atmosp here, Magnetosphere and
  Satellites} {Iau colloq. 30: Jupiter: Studies of the interior, atmosp here,
  magnetosphere and satellites}\ (\BPG~531-563).
\PrintBackRefs{\CurrentBib}

\bibitem [\protect \citeauthoryear {%
{Gierasch}%
\ \protect \BOthers {.}}{%
{Gierasch}%
\ \protect \BOthers {.}}{%
{\protect \APACyear {2000}}%
}]{%
00gierasch}
\APACinsertmetastar {%
00gierasch}%
\begin{APACrefauthors}%
{Gierasch}, P\BPBI J.%
, {Ingersoll}, A\BPBI P.%
, {Banfield}, D.%
, {Ewald}, S\BPBI P.%
, {Helfenstein}, P.%
, {Simon-Miller}, A.%
\BDBL {}{Galileo Imaging Team}%
\end{APACrefauthors}%
\unskip\
\newblock
\APACrefYearMonthDay{2000}{{\APACmonth{02}}}{}.
\newblock
{\BBOQ}\APACrefatitle {{Observation of moist convection in Jupiter's
  atmosphere}} {{Observation of moist convection in Jupiter's
  atmosphere}}.{\BBCQ}
\newblock
\APACjournalVolNumPages{Nature}{403}{}{628-630}.
\newblock
\begin{APACrefDOI} \doi{10.1038/35001017} \end{APACrefDOI}
\PrintBackRefs{\CurrentBib}

\bibitem [\protect \citeauthoryear {%
{Giles}%
\ \protect \BOthers {.}}{%
{Giles}%
\ \protect \BOthers {.}}{%
{\protect \APACyear {2020}}%
}]{%
20giles}
\APACinsertmetastar {%
20giles}%
\begin{APACrefauthors}%
{Giles}, R\BPBI S.%
, {Greathouse}, T\BPBI K.%
, {Bonfond}, B.%
, {Gladstone}, G\BPBI R.%
, {Kammer}, J\BPBI A.%
, {Hue}, V.%
\BDBL {}{Levin}, S\BPBI M.%
\end{APACrefauthors}%
\unskip\
\newblock
\APACrefYearMonthDay{2020}{{\APACmonth{11}}}{}.
\newblock
{\BBOQ}\APACrefatitle {{Possible Transient Luminous Events Observed in
  Jupiter's Upper Atmosphere}} {{Possible Transient Luminous Events Observed in
  Jupiter's Upper Atmosphere}}.{\BBCQ}
\newblock
\APACjournalVolNumPages{Journal of Geophysical Research
  (Planets)}{125}{11}{e06659}.
\newblock
\begin{APACrefDOI} \doi{10.1029/2020JE006659} \end{APACrefDOI}
\PrintBackRefs{\CurrentBib}

\bibitem [\protect \citeauthoryear {%
{Grassi}%
\ \protect \BOthers {.}}{%
{Grassi}%
\ \protect \BOthers {.}}{%
{\protect \APACyear {2020}}%
}]{%
20grassi}
\APACinsertmetastar {%
20grassi}%
\begin{APACrefauthors}%
{Grassi}, D.%
, {Adriani}, A.%
, {Mura}, A.%
, {Atreya}, S\BPBI K.%
, {Fletcher}, L\BPBI N.%
, {Lunine}, J\BPBI I.%
\BDBL {}{Turrini}, D.%
\end{APACrefauthors}%
\unskip\
\newblock
\APACrefYearMonthDay{2020}{{\APACmonth{04}}}{}.
\newblock
{\BBOQ}\APACrefatitle {{On the Spatial Distribution of Minor Species in
  Jupiter's Troposphere as Inferred From Juno JIRAM Data}} {{On the Spatial
  Distribution of Minor Species in Jupiter's Troposphere as Inferred From Juno
  JIRAM Data}}.{\BBCQ}
\newblock
\APACjournalVolNumPages{Journal of Geophysical Research
  (Planets)}{125}{4}{e06206}.
\newblock
\begin{APACrefDOI} \doi{10.1029/2019JE006206} \end{APACrefDOI}
\PrintBackRefs{\CurrentBib}

\bibitem [\protect \citeauthoryear {%
{Grodent}%
\ \protect \BOthers {.}}{%
{Grodent}%
\ \protect \BOthers {.}}{%
{\protect \APACyear {2003}}%
}]{%
03grodent}
\APACinsertmetastar {%
03grodent}%
\begin{APACrefauthors}%
{Grodent}, D.%
, {Clarke}, J\BPBI T.%
, {Waite}, J\BPBI H.%
, {Cowley}, S\BPBI W\BPBI H.%
, {G{\'e}Rard}, J\BPBI C.%
\BCBL {}\ \BBA {} {Kim}, J.%
\end{APACrefauthors}%
\unskip\
\newblock
\APACrefYearMonthDay{2003}{{\APACmonth{10}}}{}.
\newblock
{\BBOQ}\APACrefatitle {{Jupiter's polar auroral emissions}} {{Jupiter's polar
  auroral emissions}}.{\BBCQ}
\newblock
\APACjournalVolNumPages{Journal of Geophysical Research (Space
  Physics)}{108}{A10}{1366}.
\newblock
\begin{APACrefDOI} \doi{10.1029/2003JA010017} \end{APACrefDOI}
\PrintBackRefs{\CurrentBib}

\bibitem [\protect \citeauthoryear {%
{Guillot}%
\ \protect \BOthers {.}}{%
{Guillot}%
\ \protect \BOthers {.}}{%
{\protect \APACyear {2024}}%
}]{%
24guillot_egu}
\APACinsertmetastar {%
24guillot_egu}%
\begin{APACrefauthors}%
{Guillot}, T.%
, {Biagiotti}, F.%
, {Davide}, G.%
, {Mike}, W.%
, {Leigh}, F.%
, {Glenn}, O.%
\BDBL {}{Bolton}, S.%
\end{APACrefauthors}%
\unskip\
\newblock
\APACrefYearMonthDay{2024}{{\APACmonth{04}}}{}.
\newblock
{\BBOQ}\APACrefatitle {{How high are Jupiter's clouds? From high-resolution
  JunoCam images to a multi-wavelength analysis}} {{How high are Jupiter's
  clouds? From high-resolution JunoCam images to a multi-wavelength
  analysis}}.{\BBCQ}
\newblock
\BIn{} \APACrefbtitle {EGU General Assembly Conference Abstracts} {Egu general
  assembly conference abstracts}\ (\BPG~17351).
\newblock
\begin{APACrefDOI} \doi{10.5194/egusphere-egu24-17351} \end{APACrefDOI}
\PrintBackRefs{\CurrentBib}

\bibitem [\protect \citeauthoryear {%
{Guillot}%
, {Li}%
\BCBL {}\ \protect \BOthers {.}}{%
{Guillot}%
, {Li}%
\BCBL {}\ \protect \BOthers {.}}{%
{\protect \APACyear {2020}}%
}]{%
20guillot_ammonia}
\APACinsertmetastar {%
20guillot_ammonia}%
\begin{APACrefauthors}%
{Guillot}, T.%
, {Li}, C.%
, {Bolton}, S\BPBI J.%
, {Brown}, S\BPBI T.%
, {Ingersoll}, A\BPBI P.%
, {Janssen}, M\BPBI A.%
\BDBL {}{Stevenson}, D\BPBI J.%
\end{APACrefauthors}%
\unskip\
\newblock
\APACrefYearMonthDay{2020}{{\APACmonth{08}}}{}.
\newblock
{\BBOQ}\APACrefatitle {{Storms and the Depletion of Ammonia in Jupiter: II.
  Explaining the Juno Observations}} {{Storms and the Depletion of Ammonia in
  Jupiter: II. Explaining the Juno Observations}}.{\BBCQ}
\newblock
\APACjournalVolNumPages{Journal of Geophysical Research
  (Planets)}{125}{8}{e06404}.
\newblock
\begin{APACrefDOI} \doi{10.1029/2020JE006404} \end{APACrefDOI}
\PrintBackRefs{\CurrentBib}

\bibitem [\protect \citeauthoryear {%
{Guillot}%
, {Stevenson}%
, {Atreya}%
, {Bolton}%
\BCBL {}\ \BBA {} {Becker}%
}{%
{Guillot}%
, {Stevenson}%
\BCBL {}\ \protect \BOthers {.}}{%
{\protect \APACyear {2020}}%
}]{%
20guillot_mushball}
\APACinsertmetastar {%
20guillot_mushball}%
\begin{APACrefauthors}%
{Guillot}, T.%
, {Stevenson}, D\BPBI J.%
, {Atreya}, S\BPBI K.%
, {Bolton}, S\BPBI J.%
\BCBL {}\ \BBA {} {Becker}, H\BPBI N.%
\end{APACrefauthors}%
\unskip\
\newblock
\APACrefYearMonthDay{2020}{{\APACmonth{08}}}{}.
\newblock
{\BBOQ}\APACrefatitle {{Storms and the Depletion of Ammonia in Jupiter: I.
  Microphysics of ``Mushballs''}} {{Storms and the Depletion of Ammonia in
  Jupiter: I. Microphysics of ``Mushballs''}}.{\BBCQ}
\newblock
\APACjournalVolNumPages{Journal of Geophysical Research
  (Planets)}{125}{8}{e06403}.
\newblock
\begin{APACrefDOI} \doi{10.1029/2020JE006403} \end{APACrefDOI}
\PrintBackRefs{\CurrentBib}

\bibitem [\protect \citeauthoryear {%
{Gunnarson}%
\ \protect \BOthers {.}}{%
{Gunnarson}%
\ \protect \BOthers {.}}{%
{\protect \APACyear {2023}}%
}]{%
23gunnarson}
\APACinsertmetastar {%
23gunnarson}%
\begin{APACrefauthors}%
{Gunnarson}, J\BPBI L.%
, {Sayanagi}, K\BPBI M.%
, {Fischer}, G.%
, {Barry}, T.%
, {Wesley}, A.%
, {Dyudina}, U\BPBI A.%
\BDBL {}{Ingersoll}, A\BPBI P.%
\end{APACrefauthors}%
\unskip\
\newblock
\APACrefYearMonthDay{2023}{{\APACmonth{01}}}{}.
\newblock
{\BBOQ}\APACrefatitle {{Multiple convective storms within a single cyclone on
  Saturn}} {{Multiple convective storms within a single cyclone on
  Saturn}}.{\BBCQ}
\newblock
\APACjournalVolNumPages{{Icarus}}{389}{}{115228}.
\newblock
\begin{APACrefDOI} \doi{10.1016/j.icarus.2022.115228} \end{APACrefDOI}
\PrintBackRefs{\CurrentBib}

\bibitem [\protect \citeauthoryear {%
{Hansen}%
, {Brueshaber}%
, {Orton}%
, {Momary}%
\BCBL {}\ \BBA {} {Bolton}%
}{%
{Hansen}%
\ \protect \BOthers {.}}{%
{\protect \APACyear {2019}}%
}]{%
19hansen_agu}
\APACinsertmetastar {%
19hansen_agu}%
\begin{APACrefauthors}%
{Hansen}, C\BPBI J.%
, {Brueshaber}, S.%
, {Orton}, G.%
, {Momary}, T.%
\BCBL {}\ \BBA {} {Bolton}, S\BPBI J.%
\end{APACrefauthors}%
\unskip\
\newblock
\APACrefYearMonthDay{2019}{{\APACmonth{12}}}{}.
\newblock
{\BBOQ}\APACrefatitle {{JunoCam Images of Castellanus Clouds on Jupiter}}
  {{JunoCam Images of Castellanus Clouds on Jupiter}}.{\BBCQ}
\newblock
\BIn{} \APACrefbtitle {AGU Fall Meeting Abstracts} {Agu fall meeting
  abstracts}\ (\BVOL\ 2019, \BPG~P44A-05).
\PrintBackRefs{\CurrentBib}

\bibitem [\protect \citeauthoryear {%
{Hansen}%
\ \protect \BOthers {.}}{%
{Hansen}%
\ \protect \BOthers {.}}{%
{\protect \APACyear {2017}}%
}]{%
17hansen}
\APACinsertmetastar {%
17hansen}%
\begin{APACrefauthors}%
{Hansen}, C\BPBI J.%
, {Caplinger}, M\BPBI A.%
, {Ingersoll}, A.%
, {Ravine}, M\BPBI A.%
, {Jensen}, E.%
, {Bolton}, S.%
\BCBL {}\ \BBA {} {Orton}, G.%
\end{APACrefauthors}%
\unskip\
\newblock
\APACrefYearMonthDay{2017}{Nov}{}.
\newblock
{\BBOQ}\APACrefatitle {{Junocam: Juno's Outreach Camera}} {{Junocam: Juno's
  Outreach Camera}}.{\BBCQ}
\newblock
\APACjournalVolNumPages{Space Science Reviews}{213}{1-4}{475-506}.
\newblock
\begin{APACrefDOI} \doi{10.1007/s11214-014-0079-x} \end{APACrefDOI}
\PrintBackRefs{\CurrentBib}

\bibitem [\protect \citeauthoryear {%
{Heimpel}%
, {Yadav}%
, {Featherstone}%
\BCBL {}\ \BBA {} {Aurnou}%
}{%
{Heimpel}%
\ \protect \BOthers {.}}{%
{\protect \APACyear {2022}}%
}]{%
22heimpel}
\APACinsertmetastar {%
22heimpel}%
\begin{APACrefauthors}%
{Heimpel}, M\BPBI H.%
, {Yadav}, R\BPBI K.%
, {Featherstone}, N\BPBI A.%
\BCBL {}\ \BBA {} {Aurnou}, J\BPBI M.%
\end{APACrefauthors}%
\unskip\
\newblock
\APACrefYearMonthDay{2022}{{\APACmonth{06}}}{}.
\newblock
{\BBOQ}\APACrefatitle {{Polar and mid-latitude vortices and zonal flows on
  Jupiter and Saturn}} {{Polar and mid-latitude vortices and zonal flows on
  Jupiter and Saturn}}.{\BBCQ}
\newblock
\APACjournalVolNumPages{Icarus}{379}{}{114942}.
\newblock
\begin{APACrefDOI} \doi{10.1016/j.icarus.2022.114942} \end{APACrefDOI}
\PrintBackRefs{\CurrentBib}

\bibitem [\protect \citeauthoryear {%
{Hodges}%
\ \protect \BOthers {.}}{%
{Hodges}%
\ \protect \BOthers {.}}{%
{\protect \APACyear {2020}}%
}]{%
20hodges}
\APACinsertmetastar {%
20hodges}%
\begin{APACrefauthors}%
{Hodges}, A.%
, {Steffes}, P.%
, {Bellotti}, A.%
, {Waite}, J\BPBI H.%
, {Brown}, S.%
, {Oyafuso}, F.%
\BDBL {}{Bolton}, S.%
\end{APACrefauthors}%
\unskip\
\newblock
\APACrefYearMonthDay{2020}{{\APACmonth{09}}}{}.
\newblock
{\BBOQ}\APACrefatitle {{Observations and Electron Density Retrievals of
  Jupiter's Discrete Auroral Arcs Using the Juno Microwave Radiometer}}
  {{Observations and Electron Density Retrievals of Jupiter's Discrete Auroral
  Arcs Using the Juno Microwave Radiometer}}.{\BBCQ}
\newblock
\APACjournalVolNumPages{Journal of Geophysical Research
  (Planets)}{125}{9}{e06293}.
\newblock
\begin{APACrefDOI} \doi{10.1029/2019JE006293} \end{APACrefDOI}
\PrintBackRefs{\CurrentBib}

\bibitem [\protect \citeauthoryear {%
Hoskins%
, McIntyre%
\BCBL {}\ \BBA {} Robertson%
}{%
Hoskins%
\ \protect \BOthers {.}}{%
{\protect \APACyear {1985}}%
}]{%
85hoskins}
\APACinsertmetastar {%
85hoskins}%
\begin{APACrefauthors}%
Hoskins, B\BPBI J.%
, McIntyre, M\BPBI E.%
\BCBL {}\ \BBA {} Robertson, A\BPBI W.%
\end{APACrefauthors}%
\unskip\
\newblock
\APACrefYearMonthDay{1985}{}{}.
\newblock
{\BBOQ}\APACrefatitle {On the use and significance of isentropic potential
  vorticity maps} {On the use and significance of isentropic potential
  vorticity maps}.{\BBCQ}
\newblock
\APACjournalVolNumPages{Quarterly Journal of the Royal Meteorological
  Society}{111}{470}{877-946}.
\newblock
\begin{APACrefURL}
  \url{https://rmets.onlinelibrary.wiley.com/doi/abs/10.1002/qj.49711147002}
  \end{APACrefURL}
\newblock
\begin{APACrefDOI} \doi{https://doi.org/10.1002/qj.49711147002}
  \end{APACrefDOI}
\PrintBackRefs{\CurrentBib}

\bibitem [\protect \citeauthoryear {%
{Hu}%
\ \protect \BOthers {.}}{%
{Hu}%
\ \protect \BOthers {.}}{%
{\protect \APACyear {2024}}%
}]{%
24hu_agu}
\APACinsertmetastar {%
24hu_agu}%
\begin{APACrefauthors}%
{Hu}, J.%
, {Li}, C.%
, {Zhang}, Z.%
, {Oyafuso}, F\BPBI A.%
, {Brown}, S\BPBI T.%
, {Orton}, G\BPBI S.%
\BDBL {}{Bolton}, S\BPBI J.%
\end{APACrefauthors}%
\unskip\
\newblock
\APACrefYearMonthDay{2024}{{\APACmonth{12}}}{}.
\newblock
{\BBOQ}\APACrefatitle {{Vertical structures of Jupiter circumpolar cyclones as
  retrieved from Juno/MWR polar observations}} {{Vertical structures of Jupiter
  circumpolar cyclones as retrieved from Juno/MWR polar observations}}.{\BBCQ}
\newblock
\BIn{} \APACrefbtitle {AGU Fall Meeting Abstracts} {Agu fall meeting
  abstracts}\ (\BVOL\ 2024, \BPG~P41G-2972).
\PrintBackRefs{\CurrentBib}

\bibitem [\protect \citeauthoryear {%
{Hueso}%
\ \protect \BOthers {.}}{%
{Hueso}%
\ \protect \BOthers {.}}{%
{\protect \APACyear {2022}}%
}]{%
22hueso}
\APACinsertmetastar {%
22hueso}%
\begin{APACrefauthors}%
{Hueso}, R.%
, {I{\~n}urrigarro}, P.%
, {S{\'a}nchez-Lavega}, A.%
, {Foster}, C\BPBI R.%
, {Rogers}, J\BPBI H.%
, {Orton}, G\BPBI S.%
\BDBL {}{Anguiano-Arteaga}, A.%
\end{APACrefauthors}%
\unskip\
\newblock
\APACrefYearMonthDay{2022}{{\APACmonth{07}}}{}.
\newblock
{\BBOQ}\APACrefatitle {{Convective storms in closed cyclones in Jupiter's South
  Temperate Belt: (I) observations}} {{Convective storms in closed cyclones in
  Jupiter's South Temperate Belt: (I) observations}}.{\BBCQ}
\newblock
\APACjournalVolNumPages{{Icarus}}{380}{}{114994}.
\newblock
\begin{APACrefDOI} \doi{10.1016/j.icarus.2022.114994} \end{APACrefDOI}
\PrintBackRefs{\CurrentBib}

\bibitem [\protect \citeauthoryear {%
{I{\~n}urrigarro}%
\ \protect \BOthers {.}}{%
{I{\~n}urrigarro}%
\ \protect \BOthers {.}}{%
{\protect \APACyear {2020}}%
}]{%
20inurrigarro}
\APACinsertmetastar {%
20inurrigarro}%
\begin{APACrefauthors}%
{I{\~n}urrigarro}, P.%
, {Hueso}, R.%
, {Legarreta}, J.%
, {S{\'a}nchez-Lavega}, A.%
, {Eichst{\"a}dt}, G.%
, {Rogers}, J\BPBI H.%
\BDBL {}{G{\'o}mez-Forrellad}, J\BPBI M.%
\end{APACrefauthors}%
\unskip\
\newblock
\APACrefYearMonthDay{2020}{{\APACmonth{01}}}{}.
\newblock
{\BBOQ}\APACrefatitle {{Observations and numerical modelling of a convective
  disturbance in a large-scale cyclone in Jupiter's South Temperate Belt}}
  {{Observations and numerical modelling of a convective disturbance in a
  large-scale cyclone in Jupiter's South Temperate Belt}}.{\BBCQ}
\newblock
\APACjournalVolNumPages{Icarus}{336}{}{113475}.
\newblock
\begin{APACrefDOI} \doi{10.1016/j.icarus.2019.113475} \end{APACrefDOI}
\PrintBackRefs{\CurrentBib}

\bibitem [\protect \citeauthoryear {%
{I{\~n}urrigarro}%
, {Hueso}%
, {S{\'a}nchez-Lavega}%
\BCBL {}\ \BBA {} {Legarreta}%
}{%
{I{\~n}urrigarro}%
\ \protect \BOthers {.}}{%
{\protect \APACyear {2022}}%
}]{%
22inurrigarro}
\APACinsertmetastar {%
22inurrigarro}%
\begin{APACrefauthors}%
{I{\~n}urrigarro}, P.%
, {Hueso}, R.%
, {S{\'a}nchez-Lavega}, A.%
\BCBL {}\ \BBA {} {Legarreta}, J.%
\end{APACrefauthors}%
\unskip\
\newblock
\APACrefYearMonthDay{2022}{{\APACmonth{11}}}{}.
\newblock
{\BBOQ}\APACrefatitle {{Convective storms in closed cyclones in Jupiter: (II)
  numerical modeling}} {{Convective storms in closed cyclones in Jupiter: (II)
  numerical modeling}}.{\BBCQ}
\newblock
\APACjournalVolNumPages{{Icarus}}{386}{}{115169}.
\newblock
\begin{APACrefDOI} \doi{10.1016/j.icarus.2022.115169} \end{APACrefDOI}
\PrintBackRefs{\CurrentBib}

\bibitem [\protect \citeauthoryear {%
{Imai}%
\ \protect \BOthers {.}}{%
{Imai}%
\ \protect \BOthers {.}}{%
{\protect \APACyear {2019}}%
}]{%
19imai}
\APACinsertmetastar {%
19imai}%
\begin{APACrefauthors}%
{Imai}, M.%
, {Kolma{\v{s}}ov{\'a}}, I.%
, {Kurth}, W\BPBI S.%
, {Santol{\'\i}k}, O.%
, {Hospodarsky}, G\BPBI B.%
, {Gurnett}, D\BPBI A.%
\BDBL {}{Levin}, S\BPBI M.%
\end{APACrefauthors}%
\unskip\
\newblock
\APACrefYearMonthDay{2019}{{\APACmonth{06}}}{}.
\newblock
{\BBOQ}\APACrefatitle {{Evidence for low density holes in Jupiter's
  ionosphere}} {{Evidence for low density holes in Jupiter's
  ionosphere}}.{\BBCQ}
\newblock
\APACjournalVolNumPages{Nature Communications}{10}{}{2751}.
\newblock
\begin{APACrefDOI} \doi{10.1038/s41467-019-10708-w} \end{APACrefDOI}
\PrintBackRefs{\CurrentBib}

\bibitem [\protect \citeauthoryear {%
{Imai}%
\ \protect \BOthers {.}}{%
{Imai}%
\ \protect \BOthers {.}}{%
{\protect \APACyear {2018}}%
}]{%
18imai}
\APACinsertmetastar {%
18imai}%
\begin{APACrefauthors}%
{Imai}, M.%
, {Santol{\'\i}k}, O.%
, {Brown}, S\BPBI T.%
, {Kolma{\v{s}}ov{\'a}}, I.%
, {Kurth}, W\BPBI S.%
, {Janssen}, M\BPBI A.%
\BDBL {}{Levin}, S\BPBI M.%
\end{APACrefauthors}%
\unskip\
\newblock
\APACrefYearMonthDay{2018}{{\APACmonth{08}}}{}.
\newblock
{\BBOQ}\APACrefatitle {{Jupiter Lightning-Induced Whistler and Sferic Events
  With Waves and MWR During Juno Perijoves}} {{Jupiter Lightning-Induced
  Whistler and Sferic Events With Waves and MWR During Juno Perijoves}}.{\BBCQ}
\newblock
\APACjournalVolNumPages{Geophys. Res. Lett.}{45}{15}{7268-7276}.
\newblock
\begin{APACrefDOI} \doi{10.1029/2018GL078864} \end{APACrefDOI}
\PrintBackRefs{\CurrentBib}

\bibitem [\protect \citeauthoryear {%
{Imai}%
\ \protect \BOthers {.}}{%
{Imai}%
\ \protect \BOthers {.}}{%
{\protect \APACyear {2020}}%
}]{%
20imai}
\APACinsertmetastar {%
20imai}%
\begin{APACrefauthors}%
{Imai}, M.%
, {Wong}, M\BPBI H.%
, {Kolma{\v{s}}ov{\'a}}, I.%
, {Brown}, S\BPBI T.%
, {Santol{\'\i}k}, O.%
, {Kurth}, W\BPBI S.%
\BDBL {}{Levin}, S\BPBI M.%
\end{APACrefauthors}%
\unskip\
\newblock
\APACrefYearMonthDay{2020}{{\APACmonth{08}}}{}.
\newblock
{\BBOQ}\APACrefatitle {{High-Spatiotemporal Resolution Observations of Jupiter
  Lightning-Induced Radio Pulses Associated With Sferics and Thunderstorms}}
  {{High-Spatiotemporal Resolution Observations of Jupiter Lightning-Induced
  Radio Pulses Associated With Sferics and Thunderstorms}}.{\BBCQ}
\newblock
\APACjournalVolNumPages{Geophysical Research Letters}{47}{15}{e88397}.
\newblock
\begin{APACrefDOI} \doi{10.1029/2020GL088397} \end{APACrefDOI}
\PrintBackRefs{\CurrentBib}

\bibitem [\protect \citeauthoryear {%
{Ingersoll}%
\ \protect \BOthers {.}}{%
{Ingersoll}%
\ \protect \BOthers {.}}{%
{\protect \APACyear {2017}}%
}]{%
17ingersoll}
\APACinsertmetastar {%
17ingersoll}%
\begin{APACrefauthors}%
{Ingersoll}, A\BPBI P.%
, {Adumitroaie}, V.%
, {Allison}, M\BPBI D.%
, {Atreya}, S.%
, {Bellotti}, A\BPBI A.%
, {Bolton}, S\BPBI J.%
\BDBL {}{Steffes}, P\BPBI G.%
\end{APACrefauthors}%
\unskip\
\newblock
\APACrefYearMonthDay{2017}{{\APACmonth{08}}}{}.
\newblock
{\BBOQ}\APACrefatitle {{Implications of the ammonia distribution on Jupiter
  from 1 to 100 bars as measured by the Juno microwave radiometer}}
  {{Implications of the ammonia distribution on Jupiter from 1 to 100 bars as
  measured by the Juno microwave radiometer}}.{\BBCQ}
\newblock
\APACjournalVolNumPages{Geophys. Res. Lett.}{44}{}{7676-7685}.
\newblock
\begin{APACrefDOI} \doi{10.1002/2017GL074277} \end{APACrefDOI}
\PrintBackRefs{\CurrentBib}

\bibitem [\protect \citeauthoryear {%
{Ingersoll}%
\ \protect \BOthers {.}}{%
{Ingersoll}%
\ \protect \BOthers {.}}{%
{\protect \APACyear {1979}}%
}]{%
79ingersoll}
\APACinsertmetastar {%
79ingersoll}%
\begin{APACrefauthors}%
{Ingersoll}, A\BPBI P.%
, {Beebe}, R\BPBI F.%
, {Collins}, S\BPBI A.%
, {Hunt}, G\BPBI E.%
, {Mitchell}, J\BPBI L.%
, {Muller}, P.%
\BDBL {}{Terrile}, R\BPBI J.%
\end{APACrefauthors}%
\unskip\
\newblock
\APACrefYearMonthDay{1979}{{\APACmonth{08}}}{}.
\newblock
{\BBOQ}\APACrefatitle {{Zonal velocity and texture in the jovian atmosphere
  inferred from Voyager images}} {{Zonal velocity and texture in the jovian
  atmosphere inferred from Voyager images}}.{\BBCQ}
\newblock
\APACjournalVolNumPages{Nature}{280}{5725}{773-775}.
\newblock
\begin{APACrefDOI} \doi{10.1038/280773a0} \end{APACrefDOI}
\PrintBackRefs{\CurrentBib}

\bibitem [\protect \citeauthoryear {%
{Ingersoll}%
\ \protect \BOthers {.}}{%
{Ingersoll}%
\ \protect \BOthers {.}}{%
{\protect \APACyear {2004}}%
}]{%
04ingersoll}
\APACinsertmetastar {%
04ingersoll}%
\begin{APACrefauthors}%
{Ingersoll}, A\BPBI P.%
, {Dowling}, T\BPBI E.%
, {Gierasch}, P\BPBI J.%
, {Orton}, G\BPBI S.%
, {Read}, P\BPBI L.%
, {S{\'a}nchez-Lavega}, A.%
\BDBL {}{Vasavada}, A\BPBI R.%
\end{APACrefauthors}%
\unskip\
\newblock
\APACrefYearMonthDay{2004}{}{}.
\newblock
{\BBOQ}\APACrefatitle {{Dynamics of Jupiter's atmosphere}} {{Dynamics of
  Jupiter's atmosphere}}.{\BBCQ}
\newblock
\BIn{} F.~{Bagenal}, T\BPBI E.~{Dowling}\BCBL {}\ \BBA {} W\BPBI B.~{McKinnon}\
  (\BEDS), (\BPG~105-128).
\newblock
\APACaddressPublisher{}{Jupiter.~The Planet, Satellites and Magnetosphere}.
\PrintBackRefs{\CurrentBib}

\bibitem [\protect \citeauthoryear {%
{Ingersoll}%
\ \BBA {} {Porco}%
}{%
{Ingersoll}%
\ \BBA {} {Porco}%
}{%
{\protect \APACyear {1978}}%
}]{%
78ingersoll}
\APACinsertmetastar {%
78ingersoll}%
\begin{APACrefauthors}%
{Ingersoll}, A\BPBI P.%
\BCBT {}\ \BBA {} {Porco}, C\BPBI C.%
\end{APACrefauthors}%
\unskip\
\newblock
\APACrefYearMonthDay{1978}{{\APACmonth{07}}}{}.
\newblock
{\BBOQ}\APACrefatitle {{Solar heating and internal heat flow on Jupiter}}
  {{Solar heating and internal heat flow on Jupiter}}.{\BBCQ}
\newblock
\APACjournalVolNumPages{Icarus}{35}{}{27-43}.
\newblock
\begin{APACrefDOI} \doi{10.1016/0019-1035(78)90058-1} \end{APACrefDOI}
\PrintBackRefs{\CurrentBib}

\bibitem [\protect \citeauthoryear {%
{Janssen}%
\ \protect \BOthers {.}}{%
{Janssen}%
\ \protect \BOthers {.}}{%
{\protect \APACyear {2017}}%
}]{%
17janssen}
\APACinsertmetastar {%
17janssen}%
\begin{APACrefauthors}%
{Janssen}, M\BPBI A.%
, {Oswald}, J\BPBI E.%
, {Brown}, S\BPBI T.%
, {Gulkis}, S.%
, {Levin}, S\BPBI M.%
, {Bolton}, S\BPBI J.%
\BDBL {}{Wang}, C\BPBI C.%
\end{APACrefauthors}%
\unskip\
\newblock
\APACrefYearMonthDay{2017}{{\APACmonth{11}}}{}.
\newblock
{\BBOQ}\APACrefatitle {{MWR: Microwave Radiometer for the Juno Mission to
  Jupiter}} {{MWR: Microwave Radiometer for the Juno Mission to
  Jupiter}}.{\BBCQ}
\newblock
\APACjournalVolNumPages{Space Sci. Rev.}{213}{}{139-185}.
\newblock
\begin{APACrefDOI} \doi{10.1007/s11214-017-0349-5} \end{APACrefDOI}
\PrintBackRefs{\CurrentBib}

\bibitem [\protect \citeauthoryear {%
{Juno Team}%
}{%
{Juno Team}%
}{%
{\protect \APACyear {2025}}%
}]{%
pds_juno}
\APACinsertmetastar {%
pds_juno}%
\begin{APACrefauthors}%
{Juno Team}.%
\end{APACrefauthors}%
\unskip\
\newblock
\APACrefYearMonthDay{2025}{}{}.
\newblock
\APACrefbtitle {{Juno Data Bundle - Planetary Data System [Data]}.} {{Juno Data
  Bundle - Planetary Data System [Data]}.}
\newblock
\begin{APACrefURL}
  \url{https://pds-atmospheres.nmsu.edu/data_and_services/atmospheres_data/JUNO/juno.html}
  \end{APACrefURL}
\PrintBackRefs{\CurrentBib}

\bibitem [\protect \citeauthoryear {%
{JunoCam}%
}{%
{JunoCam}%
}{%
{\protect \APACyear {2025}}%
}]{%
pds_junocam}
\APACinsertmetastar {%
pds_junocam}%
\begin{APACrefauthors}%
{JunoCam}.%
\end{APACrefauthors}%
\unskip\
\newblock
\APACrefYearMonthDay{2025}{}{}.
\newblock
\APACrefbtitle {{JunoCam Data - Planetary Data System [Data]}.} {{JunoCam Data
  - Planetary Data System [Data]}.}
\newblock
\begin{APACrefURL} \url{https://pds-imaging.jpl.nasa.gov/volumes/juno.html}
  \end{APACrefURL}
\PrintBackRefs{\CurrentBib}

\bibitem [\protect \citeauthoryear {%
{Juno/JIRAM}%
}{%
{Juno/JIRAM}%
}{%
{\protect \APACyear {2025}}%
}]{%
pds_junojiram}
\APACinsertmetastar {%
pds_junojiram}%
\begin{APACrefauthors}%
{Juno/JIRAM}.%
\end{APACrefauthors}%
\unskip\
\newblock
\APACrefYearMonthDay{2025}{}{}.
\newblock
\APACrefbtitle {{Juno JIRAM Data - Planetary Data System [Data]}.} {{Juno JIRAM
  Data - Planetary Data System [Data]}.}
\newblock
\begin{APACrefURL}
  \url{https://atmos.nmsu.edu/PDS/data/PDS4/juno_jiram_bundle/data_calibrated/}
  \end{APACrefURL}
\PrintBackRefs{\CurrentBib}

\bibitem [\protect \citeauthoryear {%
{Juno/MWR}%
}{%
{Juno/MWR}%
}{%
{\protect \APACyear {2025}}%
}]{%
pds_junomwr}
\APACinsertmetastar {%
pds_junomwr}%
\begin{APACrefauthors}%
{Juno/MWR}.%
\end{APACrefauthors}%
\unskip\
\newblock
\APACrefYearMonthDay{2025}{}{}.
\newblock
\APACrefbtitle {{Juno MWR Data - Planetary Data System [Data]}.} {{Juno MWR
  Data - Planetary Data System [Data]}.}
\newblock
\begin{APACrefURL} \url{https://pds-atmospheres.nmsu.edu/PDS/data/jnomwr_1100/}
  \end{APACrefURL}
\PrintBackRefs{\CurrentBib}

\bibitem [\protect \citeauthoryear {%
{Kaspi}%
\ \protect \BOthers {.}}{%
{Kaspi}%
\ \protect \BOthers {.}}{%
{\protect \APACyear {2018}}%
}]{%
18kaspi}
\APACinsertmetastar {%
18kaspi}%
\begin{APACrefauthors}%
{Kaspi}, Y.%
, {Galanti}, E.%
, {Hubbard}, W\BPBI B.%
, {Stevenson}, D\BPBI J.%
, {Bolton}, S\BPBI J.%
, {Iess}, L.%
\BDBL {}{Wahl}, S\BPBI M.%
\end{APACrefauthors}%
\unskip\
\newblock
\APACrefYearMonthDay{2018}{{\APACmonth{03}}}{}.
\newblock
{\BBOQ}\APACrefatitle {{Jupiter's atmospheric jet streams extend thousands of
  kilometres deep}} {{Jupiter's atmospheric jet streams extend thousands of
  kilometres deep}}.{\BBCQ}
\newblock
\APACjournalVolNumPages{{Nature}}{555}{}{223-226}.
\newblock
\begin{APACrefDOI} \doi{10.1038/nature25793} \end{APACrefDOI}
\PrintBackRefs{\CurrentBib}

\bibitem [\protect \citeauthoryear {%
{Kolma{\v{s}}ov{\'a}}%
\ \protect \BOthers {.}}{%
{Kolma{\v{s}}ov{\'a}}%
\ \protect \BOthers {.}}{%
{\protect \APACyear {2018}}%
}]{%
18kolmasova}
\APACinsertmetastar {%
18kolmasova}%
\begin{APACrefauthors}%
{Kolma{\v{s}}ov{\'a}}, I.%
, {Imai}, M.%
, {Santol{\'\i}k}, O.%
, {Kurth}, W\BPBI S.%
, {Hospodarsky}, G\BPBI B.%
, {Gurnett}, D\BPBI A.%
\BDBL {}{Bolton}, S\BPBI J.%
\end{APACrefauthors}%
\unskip\
\newblock
\APACrefYearMonthDay{2018}{{\APACmonth{06}}}{}.
\newblock
{\BBOQ}\APACrefatitle {{Discovery of rapid whistlers close to Jupiter implying
  lightning rates similar to those on Earth}} {{Discovery of rapid whistlers
  close to Jupiter implying lightning rates similar to those on Earth}}.{\BBCQ}
\newblock
\APACjournalVolNumPages{Nature Astronomy}{2}{}{544-548}.
\newblock
\begin{APACrefDOI} \doi{10.1038/s41550-018-0442-z} \end{APACrefDOI}
\PrintBackRefs{\CurrentBib}

\bibitem [\protect \citeauthoryear {%
{Kolma{\v{s}}ov{\'a}}%
\ \protect \BOthers {.}}{%
{Kolma{\v{s}}ov{\'a}}%
\ \protect \BOthers {.}}{%
{\protect \APACyear {2023}}%
}]{%
23kolmasova}
\APACinsertmetastar {%
23kolmasova}%
\begin{APACrefauthors}%
{Kolma{\v{s}}ov{\'a}}, I.%
, {Santol{\'\i}k}, O.%
, {Imai}, M.%
, {Kurth}, W\BPBI S.%
, {Hospodarsky}, G\BPBI B.%
, {Connerney}, J\BPBI E\BPBI P.%
\BDBL {}{L{\'a}n}, R.%
\end{APACrefauthors}%
\unskip\
\newblock
\APACrefYearMonthDay{2023}{{\APACmonth{05}}}{}.
\newblock
{\BBOQ}\APACrefatitle {{Lightning at Jupiter pulsates with a similar rhythm as
  in-cloud lightning at Earth}} {{Lightning at Jupiter pulsates with a similar
  rhythm as in-cloud lightning at Earth}}.{\BBCQ}
\newblock
\APACjournalVolNumPages{Nature Communications}{14}{}{2707}.
\newblock
\begin{APACrefDOI} \doi{10.1038/s41467-023-38351-6} \end{APACrefDOI}
\PrintBackRefs{\CurrentBib}

\bibitem [\protect \citeauthoryear {%
{Lemasquerier}%
, {Facchini}%
, {Favier}%
\BCBL {}\ \BBA {} {Le Bars}%
}{%
{Lemasquerier}%
\ \protect \BOthers {.}}{%
{\protect \APACyear {2020}}%
}]{%
20lemasquerier}
\APACinsertmetastar {%
20lemasquerier}%
\begin{APACrefauthors}%
{Lemasquerier}, D.%
, {Facchini}, G.%
, {Favier}, B.%
\BCBL {}\ \BBA {} {Le Bars}, M.%
\end{APACrefauthors}%
\unskip\
\newblock
\APACrefYearMonthDay{2020}{{\APACmonth{03}}}{}.
\newblock
{\BBOQ}\APACrefatitle {{Remote determination of the shape of Jupiter's vortices
  from laboratory experiments}} {{Remote determination of the shape of
  Jupiter's vortices from laboratory experiments}}.{\BBCQ}
\newblock
\APACjournalVolNumPages{Nature Physics}{16}{6}{695-700}.
\newblock
\begin{APACrefDOI} \doi{10.1038/s41567-020-0833-9} \end{APACrefDOI}
\PrintBackRefs{\CurrentBib}

\bibitem [\protect \citeauthoryear {%
{Li}%
\ \protect \BOthers {.}}{%
{Li}%
\ \protect \BOthers {.}}{%
{\protect \APACyear {2024}}%
}]{%
24li}
\APACinsertmetastar {%
24li}%
\begin{APACrefauthors}%
{Li}, C.%
, {Allison}, M.%
, {Atreya}, S.%
, {Brueshaber}, S.%
, {Fletcher}, L\BPBI N.%
, {Guillot}, T.%
\BDBL {}{Bolton}, S.%
\end{APACrefauthors}%
\unskip\
\newblock
\APACrefYearMonthDay{2024}{{\APACmonth{05}}}{}.
\newblock
{\BBOQ}\APACrefatitle {{Super-adiabatic temperature gradient at Jupiter's
  equatorial zone and implications for the water abundance}} {{Super-adiabatic
  temperature gradient at Jupiter's equatorial zone and implications for the
  water abundance}}.{\BBCQ}
\newblock
\APACjournalVolNumPages{{Icarus}}{414}{}{116028}.
\newblock
\begin{APACrefDOI} \doi{10.1016/j.icarus.2024.116028} \end{APACrefDOI}
\PrintBackRefs{\CurrentBib}

\bibitem [\protect \citeauthoryear {%
{Li}%
\ \protect \BOthers {.}}{%
{Li}%
\ \protect \BOthers {.}}{%
{\protect \APACyear {2017}}%
}]{%
17li}
\APACinsertmetastar {%
17li}%
\begin{APACrefauthors}%
{Li}, C.%
, {Ingersoll}, A.%
, {Janssen}, M.%
, {Levin}, S.%
, {Bolton}, S.%
, {Adumitroaie}, V.%
\BDBL {}{Williamson}, R.%
\end{APACrefauthors}%
\unskip\
\newblock
\APACrefYearMonthDay{2017}{{\APACmonth{06}}}{}.
\newblock
{\BBOQ}\APACrefatitle {{The distribution of ammonia on Jupiter from a
  preliminary inversion of Juno microwave radiometer data}} {{The distribution
  of ammonia on Jupiter from a preliminary inversion of Juno microwave
  radiometer data}}.{\BBCQ}
\newblock
\APACjournalVolNumPages{Geophys. Res. Lett.}{44}{}{5317-5325}.
\newblock
\begin{APACrefDOI} \doi{10.1002/2017GL073159} \end{APACrefDOI}
\PrintBackRefs{\CurrentBib}

\bibitem [\protect \citeauthoryear {%
{Little}%
\ \protect \BOthers {.}}{%
{Little}%
\ \protect \BOthers {.}}{%
{\protect \APACyear {1999}}%
}]{%
99little}
\APACinsertmetastar {%
99little}%
\begin{APACrefauthors}%
{Little}, B.%
, {Anger}, C\BPBI D.%
, {Ingersoll}, A\BPBI P.%
, {Vasavada}, A\BPBI R.%
, {Senske}, D\BPBI A.%
, {Breneman}, H\BPBI H.%
\BDBL {}{The Galileo SSI Team}%
\end{APACrefauthors}%
\unskip\
\newblock
\APACrefYearMonthDay{1999}{{\APACmonth{12}}}{}.
\newblock
{\BBOQ}\APACrefatitle {{Galileo Images of Lightning on Jupiter}} {{Galileo
  Images of Lightning on Jupiter}}.{\BBCQ}
\newblock
\APACjournalVolNumPages{Icarus}{142}{}{306-323}.
\newblock
\begin{APACrefDOI} \doi{10.1006/icar.1999.6195} \end{APACrefDOI}
\PrintBackRefs{\CurrentBib}

\bibitem [\protect \citeauthoryear {%
{Magalhaes}%
\ \BBA {} {Borucki}%
}{%
{Magalhaes}%
\ \BBA {} {Borucki}%
}{%
{\protect \APACyear {1991}}%
}]{%
91magalhaes}
\APACinsertmetastar {%
91magalhaes}%
\begin{APACrefauthors}%
{Magalhaes}, J\BPBI A.%
\BCBT {}\ \BBA {} {Borucki}, W\BPBI J.%
\end{APACrefauthors}%
\unskip\
\newblock
\APACrefYearMonthDay{1991}{{\APACmonth{01}}}{}.
\newblock
{\BBOQ}\APACrefatitle {{Spatial distribution of visible lightning on Jupiter}}
  {{Spatial distribution of visible lightning on Jupiter}}.{\BBCQ}
\newblock
\APACjournalVolNumPages{Nature}{349}{6307}{311-313}.
\newblock
\begin{APACrefDOI} \doi{10.1038/349311a0} \end{APACrefDOI}
\PrintBackRefs{\CurrentBib}

\bibitem [\protect \citeauthoryear {%
Marcus%
, Asay-Davis%
, Wong%
\BCBL {}\ \BBA {} De~Pater%
}{%
Marcus%
\ \protect \BOthers {.}}{%
{\protect \APACyear {2013}}%
}]{%
13marcus}
\APACinsertmetastar {%
13marcus}%
\begin{APACrefauthors}%
Marcus, P\BPBI S.%
, Asay-Davis, X.%
, Wong, M\BPBI H.%
\BCBL {}\ \BBA {} De~Pater, I.%
\end{APACrefauthors}%
\unskip\
\newblock
\APACrefYearMonthDay{2013}{}{}.
\newblock
{\BBOQ}\APACrefatitle {Jupiter's Red Oval BA: Dynamics, Color, and Relationship
  to Jovian Climate Change} {Jupiter's red oval ba: Dynamics, color, and
  relationship to jovian climate change}.{\BBCQ}
\newblock
\APACjournalVolNumPages{Journal of Heat Transfer}{135}{1}{011007}.
\PrintBackRefs{\CurrentBib}

\bibitem [\protect \citeauthoryear {%
{Marcus}%
\ \BBA {} {Shetty}%
}{%
{Marcus}%
\ \BBA {} {Shetty}%
}{%
{\protect \APACyear {2011}}%
}]{%
11marcus}
\APACinsertmetastar {%
11marcus}%
\begin{APACrefauthors}%
{Marcus}, P\BPBI S.%
\BCBT {}\ \BBA {} {Shetty}, S.%
\end{APACrefauthors}%
\unskip\
\newblock
\APACrefYearMonthDay{2011}{{\APACmonth{02}}}{}.
\newblock
{\BBOQ}\APACrefatitle {{Jupiter's zonal winds: are they bands of homogenized
  potential vorticity organized as a monotonic staircase?}} {{Jupiter's zonal
  winds: are they bands of homogenized potential vorticity organized as a
  monotonic staircase?}}{\BBCQ}
\newblock
\APACjournalVolNumPages{Philosophical Transactions of the Royal Society of
  London Series A}{369}{1937}{771-795}.
\newblock
\begin{APACrefDOI} \doi{10.1098/rsta.2010.0299} \end{APACrefDOI}
\PrintBackRefs{\CurrentBib}

\bibitem [\protect \citeauthoryear {%
McGillicuddy~Jr%
}{%
McGillicuddy~Jr%
}{%
{\protect \APACyear {2015}}%
}]{%
15mcgillicuddy}
\APACinsertmetastar {%
15mcgillicuddy}%
\begin{APACrefauthors}%
McGillicuddy~Jr, D\BPBI J.%
\end{APACrefauthors}%
\unskip\
\newblock
\APACrefYearMonthDay{2015}{}{}.
\newblock
{\BBOQ}\APACrefatitle {Formation of intrathermocline lenses by eddy--wind
  interaction} {Formation of intrathermocline lenses by eddy--wind
  interaction}.{\BBCQ}
\newblock
\APACjournalVolNumPages{Journal of Physical Oceanography}{45}{2}{606--612}.
\PrintBackRefs{\CurrentBib}

\bibitem [\protect \citeauthoryear {%
{Mitchell}%
\ \protect \BOthers {.}}{%
{Mitchell}%
\ \protect \BOthers {.}}{%
{\protect \APACyear {1979}}%
}]{%
79mitchell}
\APACinsertmetastar {%
79mitchell}%
\begin{APACrefauthors}%
{Mitchell}, J\BPBI L.%
, {Terrile}, R\BPBI J.%
, {Smith}, B\BPBI A.%
, {Muller}, J\BPBI P.%
, {Ingersoll}, A\BPBI P.%
, {Hunt}, G\BPBI E.%
\BDBL {}{Beebe}, R\BPBI F.%
\end{APACrefauthors}%
\unskip\
\newblock
\APACrefYearMonthDay{1979}{{\APACmonth{08}}}{}.
\newblock
{\BBOQ}\APACrefatitle {{Jovian cloud structure and velocity fields}} {{Jovian
  cloud structure and velocity fields}}.{\BBCQ}
\newblock
\APACjournalVolNumPages{Nature}{280}{5725}{776-778}.
\newblock
\begin{APACrefDOI} \doi{10.1038/280776a0} \end{APACrefDOI}
\PrintBackRefs{\CurrentBib}

\bibitem [\protect \citeauthoryear {%
{Moeckel}%
, {Ge}%
\BCBL {}\ \BBA {} {de Pater}%
}{%
{Moeckel}%
\ \protect \BOthers {.}}{%
{\protect \APACyear {2025}}%
}]{%
25moeckel}
\APACinsertmetastar {%
25moeckel}%
\begin{APACrefauthors}%
{Moeckel}, C.%
, {Ge}, H.%
\BCBL {}\ \BBA {} {de Pater}, I.%
\end{APACrefauthors}%
\unskip\
\newblock
\APACrefYearMonthDay{2025}{{\APACmonth{03}}}{}.
\newblock
{\BBOQ}\APACrefatitle {{Tempests in the troposphere: Mapping the impact of
  giant storms on Jupiter's deep atmosphere}} {{Tempests in the troposphere:
  Mapping the impact of giant storms on Jupiter's deep atmosphere}}.{\BBCQ}
\newblock
\APACjournalVolNumPages{Science Advances}{11}{13}{eado9779}.
\newblock
\begin{APACrefDOI} \doi{10.1126/sciadv.ado9779} \end{APACrefDOI}
\PrintBackRefs{\CurrentBib}

\bibitem [\protect \citeauthoryear {%
{Morales-Juberias}%
, {Li}%
, {Dowling}%
, {Mercuri}%
\BCBL {}\ \BBA {} {Simon}%
}{%
{Morales-Juberias}%
\ \protect \BOthers {.}}{%
{\protect \APACyear {2024}}%
}]{%
24morales_dps}
\APACinsertmetastar {%
24morales_dps}%
\begin{APACrefauthors}%
{Morales-Juberias}, R.%
, {Li}, L.%
, {Dowling}, T.%
, {Mercuri}, S.%
\BCBL {}\ \BBA {} {Simon}, A.%
\end{APACrefauthors}%
\unskip\
\newblock
\APACrefYearMonthDay{2024}{{\APACmonth{10}}}{}.
\newblock
{\BBOQ}\APACrefatitle {{Interactions between opposite signed vortices on
  adjacent domains in Jupiter.}} {{Interactions between opposite signed
  vortices on adjacent domains in Jupiter.}}{\BBCQ}
\newblock
\BIn{} \APACrefbtitle {56th Annual Meeting of the Division for Planetary
  Sciences} {56th annual meeting of the division for planetary sciences}\
  (\BVOL~56, \BPG~103.03).
\PrintBackRefs{\CurrentBib}

\bibitem [\protect \citeauthoryear {%
Nan%
, Yu%
, Wei%
, Ren%
\BCBL {}\ \BBA {} Fan%
}{%
Nan%
\ \protect \BOthers {.}}{%
{\protect \APACyear {2017}}%
}]{%
17nan}
\APACinsertmetastar {%
17nan}%
\begin{APACrefauthors}%
Nan, F.%
, Yu, F.%
, Wei, C.%
, Ren, Q.%
\BCBL {}\ \BBA {} Fan, C.%
\end{APACrefauthors}%
\unskip\
\newblock
\APACrefYearMonthDay{2017}{}{}.
\newblock
{\BBOQ}\APACrefatitle {Observations of an extra-large subsurface anticyclonic
  eddy in the northwestern Pacific subtropical gyre} {Observations of an
  extra-large subsurface anticyclonic eddy in the northwestern pacific
  subtropical gyre}.{\BBCQ}
\newblock
\APACjournalVolNumPages{Journal of Marine Science: Research \&
  Development}{7}{04}{234}.
\PrintBackRefs{\CurrentBib}

\bibitem [\protect \citeauthoryear {%
Ni%
, Zhai%
, Yang%
\BCBL {}\ \BBA {} Chen%
}{%
Ni%
\ \protect \BOthers {.}}{%
{\protect \APACyear {2023}}%
}]{%
23ni}
\APACinsertmetastar {%
23ni}%
\begin{APACrefauthors}%
Ni, Q.%
, Zhai, X.%
, Yang, Z.%
\BCBL {}\ \BBA {} Chen, D.%
\end{APACrefauthors}%
\unskip\
\newblock
\APACrefYearMonthDay{2023}{}{}.
\newblock
{\BBOQ}\APACrefatitle {Generation of cold anticyclonic eddies and warm cyclonic
  eddies in the tropical oceans} {Generation of cold anticyclonic eddies and
  warm cyclonic eddies in the tropical oceans}.{\BBCQ}
\newblock
\APACjournalVolNumPages{Journal of Physical Oceanography}{53}{6}{1485--1498}.
\PrintBackRefs{\CurrentBib}

\bibitem [\protect \citeauthoryear {%
{Nichols}%
\ \protect \BOthers {.}}{%
{Nichols}%
\ \protect \BOthers {.}}{%
{\protect \APACyear {2017}}%
}]{%
17nichols}
\APACinsertmetastar {%
17nichols}%
\begin{APACrefauthors}%
{Nichols}, J\BPBI D.%
, {Badman}, S\BPBI V.%
, {Bagenal}, F.%
, {Bolton}, S\BPBI J.%
, {Bonfond}, B.%
, {Bunce}, E\BPBI J.%
\BDBL {}{Yoshikawa}, I.%
\end{APACrefauthors}%
\unskip\
\newblock
\APACrefYearMonthDay{2017}{{\APACmonth{08}}}{}.
\newblock
{\BBOQ}\APACrefatitle {{Response of Jupiter's auroras to conditions in the
  interplanetary medium as measured by the Hubble Space Telescope and Juno}}
  {{Response of Jupiter's auroras to conditions in the interplanetary medium as
  measured by the Hubble Space Telescope and Juno}}.{\BBCQ}
\newblock
\APACjournalVolNumPages{Geophys. Res. Lett.}{44}{15}{7643-7652}.
\newblock
\begin{APACrefDOI} \doi{10.1002/2017GL073029} \end{APACrefDOI}
\PrintBackRefs{\CurrentBib}

\bibitem [\protect \citeauthoryear {%
{Orton}%
, {Hansen}%
\BCBL {}\ \protect \BOthers {.}}{%
{Orton}%
, {Hansen}%
\BCBL {}\ \protect \BOthers {.}}{%
{\protect \APACyear {2017}}%
}]{%
17orton_juno}
\APACinsertmetastar {%
17orton_juno}%
\begin{APACrefauthors}%
{Orton}, G\BPBI S.%
, {Hansen}, C.%
, {Caplinger}, M.%
, {Ravine}, M.%
, {Atreya}, S.%
, {Ingersoll}, A\BPBI P.%
\BDBL {}{Bolton}, S.%
\end{APACrefauthors}%
\unskip\
\newblock
\APACrefYearMonthDay{2017}{{\APACmonth{05}}}{}.
\newblock
{\BBOQ}\APACrefatitle {{The first close-up images of Jupiter's polar regions:
  Results from the Juno mission JunoCam instrument}} {{The first close-up
  images of Jupiter's polar regions: Results from the Juno mission JunoCam
  instrument}}.{\BBCQ}
\newblock
\APACjournalVolNumPages{Geophys. Res. Lett.}{44}{10}{4599-4606}.
\newblock
\begin{APACrefDOI} \doi{10.1002/2016GL072443} \end{APACrefDOI}
\PrintBackRefs{\CurrentBib}

\bibitem [\protect \citeauthoryear {%
{Orton}%
, {Momary}%
\BCBL {}\ \protect \BOthers {.}}{%
{Orton}%
, {Momary}%
\BCBL {}\ \protect \BOthers {.}}{%
{\protect \APACyear {2017}}%
}]{%
17orton}
\APACinsertmetastar {%
17orton}%
\begin{APACrefauthors}%
{Orton}, G\BPBI S.%
, {Momary}, T.%
, {Ingersoll}, A\BPBI P.%
, {Adriani}, A.%
, {Hansen}, C\BPBI J.%
, {Janssen}, M.%
\BDBL {}{Sindoni}, G.%
\end{APACrefauthors}%
\unskip\
\newblock
\APACrefYearMonthDay{2017}{{\APACmonth{05}}}{}.
\newblock
{\BBOQ}\APACrefatitle {{Multiple-wavelength sensing of Jupiter during the Juno
  mission's first perijove passage}} {{Multiple-wavelength sensing of Jupiter
  during the Juno mission's first perijove passage}}.{\BBCQ}
\newblock
\APACjournalVolNumPages{Geophys. Res. Lett.}{44}{}{4607-4614}.
\newblock
\begin{APACrefDOI} \doi{10.1002/2017GL073019} \end{APACrefDOI}
\PrintBackRefs{\CurrentBib}

\bibitem [\protect \citeauthoryear {%
{Oyafuso}%
\ \protect \BOthers {.}}{%
{Oyafuso}%
\ \protect \BOthers {.}}{%
{\protect \APACyear {2020}}%
}]{%
20oyafuso}
\APACinsertmetastar {%
20oyafuso}%
\begin{APACrefauthors}%
{Oyafuso}, F.%
, {Levin}, S.%
, {Orton}, G.%
, {Brown}, S\BPBI T.%
, {Adumitroaie}, V.%
, {Janssen}, M.%
\BDBL {}{Bolton}, S.%
\end{APACrefauthors}%
\unskip\
\newblock
\APACrefYearMonthDay{2020}{{\APACmonth{11}}}{}.
\newblock
{\BBOQ}\APACrefatitle {{Angular Dependence and Spatial Distribution of
  Jupiter's Centimeter-Wave Thermal Emission From Juno's Microwave Radiometer}}
  {{Angular Dependence and Spatial Distribution of Jupiter's Centimeter-Wave
  Thermal Emission From Juno's Microwave Radiometer}}.{\BBCQ}
\newblock
\APACjournalVolNumPages{Earth and Space Science}{7}{11}{e01254}.
\newblock
\begin{APACrefDOI} \doi{10.1029/2020EA001254} \end{APACrefDOI}
\PrintBackRefs{\CurrentBib}

\bibitem [\protect \citeauthoryear {%
{Palotai}%
, {Dowling}%
\BCBL {}\ \BBA {} {Fletcher}%
}{%
{Palotai}%
\ \protect \BOthers {.}}{%
{\protect \APACyear {2014}}%
}]{%
14palotai}
\APACinsertmetastar {%
14palotai}%
\begin{APACrefauthors}%
{Palotai}, C.%
, {Dowling}, T\BPBI E.%
\BCBL {}\ \BBA {} {Fletcher}, L\BPBI N.%
\end{APACrefauthors}%
\unskip\
\newblock
\APACrefYearMonthDay{2014}{}{}.
\newblock
{\BBOQ}\APACrefatitle {{3D Modeling of interactions between Jupiter's ammonia
  clouds and large anticyclones}} {{3D Modeling of interactions between
  Jupiter's ammonia clouds and large anticyclones}}.{\BBCQ}
\newblock
\APACjournalVolNumPages{Icarus}{232}{}{141-156}.
\PrintBackRefs{\CurrentBib}

\bibitem [\protect \citeauthoryear {%
{Parisi}%
\ \protect \BOthers {.}}{%
{Parisi}%
\ \protect \BOthers {.}}{%
{\protect \APACyear {2021}}%
}]{%
21parisi}
\APACinsertmetastar {%
21parisi}%
\begin{APACrefauthors}%
{Parisi}, M.%
, {Kaspi}, Y.%
, {Galanti}, E.%
, {Durante}, D.%
, {Bolton}, S\BPBI J.%
, {Levin}, S\BPBI M.%
\BDBL {}{Wong}, M\BPBI H.%
\end{APACrefauthors}%
\unskip\
\newblock
\APACrefYearMonthDay{2021}{{\APACmonth{11}}}{}.
\newblock
{\BBOQ}\APACrefatitle {{The depth of Jupiter{\textquoteright}s Great Red Spot
  constrained by Juno gravity overflights}} {{The depth of
  Jupiter{\textquoteright}s Great Red Spot constrained by Juno gravity
  overflights}}.{\BBCQ}
\newblock
\APACjournalVolNumPages{Science}{374}{6570}{964-968}.
\newblock
\begin{APACrefDOI} \doi{10.1126/science.abf1396} \end{APACrefDOI}
\PrintBackRefs{\CurrentBib}

\bibitem [\protect \citeauthoryear {%
{Pirraglia}%
}{%
{Pirraglia}%
}{%
{\protect \APACyear {1984}}%
}]{%
84pirraglia}
\APACinsertmetastar {%
84pirraglia}%
\begin{APACrefauthors}%
{Pirraglia}, J\BPBI A.%
\end{APACrefauthors}%
\unskip\
\newblock
\APACrefYearMonthDay{1984}{}{}.
\newblock
{\BBOQ}\APACrefatitle {{Meridional energy balance of Jupiter}} {{Meridional
  energy balance of Jupiter}}.{\BBCQ}
\newblock
\APACjournalVolNumPages{Icarus}{59}{}{169-176}.
\newblock
\begin{APACrefDOI} \doi{10.1016/0019-1035(84)90020-4} \end{APACrefDOI}
\PrintBackRefs{\CurrentBib}

\bibitem [\protect \citeauthoryear {%
{Porco}%
\ \protect \BOthers {.}}{%
{Porco}%
\ \protect \BOthers {.}}{%
{\protect \APACyear {2005}}%
}]{%
05porco}
\APACinsertmetastar {%
05porco}%
\begin{APACrefauthors}%
{Porco}, C\BPBI C.%
, {Baker}, E.%
, {Barbara}, J.%
, {Beurle}, K.%
, {Brahic}, A.%
, {Burns}, J\BPBI A.%
\BDBL {}{West}, R.%
\end{APACrefauthors}%
\unskip\
\newblock
\APACrefYearMonthDay{2005}{}{}.
\newblock
{\BBOQ}\APACrefatitle {{Cassini Imaging Science: Initial Results on Saturn's
  Atmosphere}} {{Cassini Imaging Science: Initial Results on Saturn's
  Atmosphere}}.{\BBCQ}
\newblock
\APACjournalVolNumPages{Science}{307}{}{1243-1247}.
\newblock
\begin{APACrefDOI} \doi{10.1126/science.1107691} \end{APACrefDOI}
\PrintBackRefs{\CurrentBib}

\bibitem [\protect \citeauthoryear {%
{Porco}%
\ \protect \BOthers {.}}{%
{Porco}%
\ \protect \BOthers {.}}{%
{\protect \APACyear {2003}}%
}]{%
03porco}
\APACinsertmetastar {%
03porco}%
\begin{APACrefauthors}%
{Porco}, C\BPBI C.%
, {West}, R\BPBI A.%
, {McEwen}, A.%
, {Del Genio}, A\BPBI D.%
, {Ingersoll}, A\BPBI P.%
, {Thomas}, P.%
\BDBL {}{Vasavada}, A\BPBI R.%
\end{APACrefauthors}%
\unskip\
\newblock
\APACrefYearMonthDay{2003}{{\APACmonth{03}}}{}.
\newblock
{\BBOQ}\APACrefatitle {{Cassini Imaging of Jupiter's Atmosphere, Satellites,
  and Rings}} {{Cassini Imaging of Jupiter's Atmosphere, Satellites, and
  Rings}}.{\BBCQ}
\newblock
\APACjournalVolNumPages{Science}{299}{}{1541-1547}.
\newblock
\begin{APACrefDOI} \doi{10.1126/science.1079462} \end{APACrefDOI}
\PrintBackRefs{\CurrentBib}

\bibitem [\protect \citeauthoryear {%
Purkiani%
\ \protect \BOthers {.}}{%
Purkiani%
\ \protect \BOthers {.}}{%
{\protect \APACyear {2022}}%
}]{%
22purkiani}
\APACinsertmetastar {%
22purkiani}%
\begin{APACrefauthors}%
Purkiani, K.%
, Haeckel, M.%
, Haalboom, S.%
, Schmidt, K.%
, Urban, P.%
, Gazis, I\BHBI Z.%
\BDBL {}Vink, A.%
\end{APACrefauthors}%
\unskip\
\newblock
\APACrefYearMonthDay{2022}{}{}.
\newblock
{\BBOQ}\APACrefatitle {Impact of a long-lived anticyclonic mesoscale eddy on
  seawater anomalies in the northeastern tropical Pacific Ocean: a composite
  analysis from hydrographic measurements, sea level altimetry data, and
  reanalysis model products} {Impact of a long-lived anticyclonic mesoscale
  eddy on seawater anomalies in the northeastern tropical pacific ocean: a
  composite analysis from hydrographic measurements, sea level altimetry data,
  and reanalysis model products}.{\BBCQ}
\newblock
\APACjournalVolNumPages{Ocean Science}{18}{4}{1163--1181}.
\newblock
\begin{APACrefURL} \url{https://os.copernicus.org/articles/18/1163/2022/}
  \end{APACrefURL}
\newblock
\begin{APACrefDOI} \doi{10.5194/os-18-1163-2022} \end{APACrefDOI}
\PrintBackRefs{\CurrentBib}

\bibitem [\protect \citeauthoryear {%
Read%
\ \protect \BOthers {.}}{%
Read%
\ \protect \BOthers {.}}{%
{\protect \APACyear {2006}}%
}]{%
06read_jup}
\APACinsertmetastar {%
06read_jup}%
\begin{APACrefauthors}%
Read, P.%
, Gierasch, P.%
, Conrath, B.%
, Simon-Miller, A.%
, Fouchet, T.%
\BCBL {}\ \BBA {} Yamazaki, Y.%
\end{APACrefauthors}%
\unskip\
\newblock
\APACrefYearMonthDay{2006}{}{}.
\newblock
{\BBOQ}\APACrefatitle {{Mapping potential-vorticity dynamics on Jupiter. I:
  Zonal-mean circulation from Cassini and Voyager 1 data}} {{Mapping
  potential-vorticity dynamics on Jupiter. I: Zonal-mean circulation from
  Cassini and Voyager 1 data}}.{\BBCQ}
\newblock
\APACjournalVolNumPages{Q. J. R. Meteorol. Soc.}{132}{}{1577-1603}.
\PrintBackRefs{\CurrentBib}

\bibitem [\protect \citeauthoryear {%
{Read}%
}{%
{Read}%
}{%
{\protect \APACyear {2023}}%
}]{%
23read}
\APACinsertmetastar {%
23read}%
\begin{APACrefauthors}%
{Read}, P\BPBI L.%
\end{APACrefauthors}%
\unskip\
\newblock
\APACrefYearMonthDay{2023}{{\APACmonth{10}}}{}.
\newblock
{\BBOQ}\APACrefatitle {{The Dynamics of Jupiter's and Saturn's Weather Layers:
  A Synthesis After Cassini and Juno}} {{The Dynamics of Jupiter's and Saturn's
  Weather Layers: A Synthesis After Cassini and Juno}}.{\BBCQ}
\newblock
\APACjournalVolNumPages{Annual Review of Fluid Mechanics}{56}{}{271-293}.
\newblock
\begin{APACrefDOI} \doi{10.1146/annurev-fluid-121021-040058} \end{APACrefDOI}
\PrintBackRefs{\CurrentBib}

\bibitem [\protect \citeauthoryear {%
{Rogers}%
, {Adamoli}%
\BCBL {}\ \BBA {} {Bullen}%
}{%
{Rogers}%
\ \protect \BOthers {.}}{%
{\protect \APACyear {2023}}%
}]{%
23rogers_baa}
\APACinsertmetastar {%
23rogers_baa}%
\begin{APACrefauthors}%
{Rogers}, J\BPBI H.%
, {Adamoli}, G.%
\BCBL {}\ \BBA {} {Bullen}, R.%
\end{APACrefauthors}%
\unskip\
\newblock
\APACrefYearMonthDay{2023}{}{}.
\newblock
\APACrefbtitle {{Jupiter in 2022/23, Report no.6: Final report on the high
  northern latitudes.}} {{Jupiter in 2022/23, Report no.6: Final report on the
  high northern latitudes.}}
\newblock
\begin{APACrefURL}
  \url{https://britastro.org/section_information_/jupiter-section-overview/jupiter-in-2022-23/jupiter-in-2022-23-report-no-6}
  \end{APACrefURL}
\PrintBackRefs{\CurrentBib}

\bibitem [\protect \citeauthoryear {%
{Rogers}%
, {Adamoli}%
\BCBL {}\ \protect \BOthers {.}}{%
{Rogers}%
, {Adamoli}%
\BCBL {}\ \protect \BOthers {.}}{%
{\protect \APACyear {2022}}%
}]{%
22rogers_epsc}
\APACinsertmetastar {%
22rogers_epsc}%
\begin{APACrefauthors}%
{Rogers}, J\BPBI H.%
, {Adamoli}, G.%
, {Hansen}, C.%
, {Eichst{\"a}dt}, G.%
, {Orton}, G.%
, {Momary}, T.%
\BDBL {}{Mettig}, H\BHBI J.%
\end{APACrefauthors}%
\unskip\
\newblock
\APACrefYearMonthDay{2022}{{\APACmonth{09}}}{}.
\newblock
{\BBOQ}\APACrefatitle {{Jupiter's high-latitude northern domains: Dynamics from
  Earth-based and JunoCam imaging}} {{Jupiter's high-latitude northern domains:
  Dynamics from Earth-based and JunoCam imaging}}.{\BBCQ}
\newblock
\BIn{} \APACrefbtitle {European Planetary Science Congress} {European planetary
  science congress}\ (\BPG~EPSC2022-16).
\newblock
\begin{APACrefDOI} \doi{10.5194/epsc2022-16} \end{APACrefDOI}
\PrintBackRefs{\CurrentBib}

\bibitem [\protect \citeauthoryear {%
{Rogers}%
, {Adamoli}%
, {Jacquesson}%
, {Vedovato}%
\BCBL {}\ \BBA {} {Mettig}%
}{%
{Rogers}%
\ \protect \BOthers {.}}{%
{\protect \APACyear {2017}}%
}]{%
17rogers_baa}
\APACinsertmetastar {%
17rogers_baa}%
\begin{APACrefauthors}%
{Rogers}, J\BPBI H.%
, {Adamoli}, G.%
, {Jacquesson}, M.%
, {Vedovato}, M.%
\BCBL {}\ \BBA {} {Mettig}, H\BPBI J.%
\end{APACrefauthors}%
\unskip\
\newblock
\APACrefYearMonthDay{2017}{}{}.
\newblock
\APACrefbtitle {{Jupiter's high northern latitudes: patterns and dynamics of
  the N3 to N6 domains}.} {{Jupiter's high northern latitudes: patterns and
  dynamics of the N3 to N6 domains}.}
\newblock
\begin{APACrefURL} \url{https://britastro.org/node/11328} \end{APACrefURL}
\PrintBackRefs{\CurrentBib}

\bibitem [\protect \citeauthoryear {%
{Rogers}%
, {Eichst{\"a}dt}%
\BCBL {}\ \protect \BOthers {.}}{%
{Rogers}%
, {Eichst{\"a}dt}%
\BCBL {}\ \protect \BOthers {.}}{%
{\protect \APACyear {2022}}%
}]{%
22rogers}
\APACinsertmetastar {%
22rogers}%
\begin{APACrefauthors}%
{Rogers}, J\BPBI H.%
, {Eichst{\"a}dt}, G.%
, {Hansen}, C\BPBI J.%
, {Orton}, G\BPBI S.%
, {Momary}, T.%
, {Casely}, A.%
\BDBL {}{Bolton}, S.%
\end{APACrefauthors}%
\unskip\
\newblock
\APACrefYearMonthDay{2022}{{\APACmonth{01}}}{}.
\newblock
{\BBOQ}\APACrefatitle {{Flow patterns of Jupiter's south polar region}} {{Flow
  patterns of Jupiter's south polar region}}.{\BBCQ}
\newblock
\APACjournalVolNumPages{Icarus}{372}{}{114742}.
\newblock
\begin{APACrefDOI} \doi{10.1016/j.icarus.2021.114742} \end{APACrefDOI}
\PrintBackRefs{\CurrentBib}

\bibitem [\protect \citeauthoryear {%
{Rogers}%
\ \protect \BOthers {.}}{%
{Rogers}%
\ \protect \BOthers {.}}{%
{\protect \APACyear {2024}}%
}]{%
24rogers_epsc}
\APACinsertmetastar {%
24rogers_epsc}%
\begin{APACrefauthors}%
{Rogers}, J\BPBI H.%
, {Hansen}, C.%
, {Eichst{\"a}dt}, G.%
, {Orton}, G.%
, {Momary}, T.%
, {Adamoli}, G.%
\BDBL {}{Mettig}, H\BHBI J.%
\end{APACrefauthors}%
\unskip\
\newblock
\APACrefYearMonthDay{2024}{{\APACmonth{09}}}{}.
\newblock
{\BBOQ}\APACrefatitle {{Longevity of cyclonic formations in Jupiter's S2 (South
  South Temperate) Domain}} {{Longevity of cyclonic formations in Jupiter's S2
  (South South Temperate) Domain}}.{\BBCQ}
\newblock
\BIn{} \APACrefbtitle {European Planetary Science Congress} {European planetary
  science congress}\ (\BPG~EPSC2024-378).
\newblock
\begin{APACrefDOI} \doi{10.5194/epsc2024-378} \end{APACrefDOI}
\PrintBackRefs{\CurrentBib}

\bibitem [\protect \citeauthoryear {%
{S{\'a}nchez-Lavega}%
\ \protect \BOthers {.}}{%
{S{\'a}nchez-Lavega}%
\ \protect \BOthers {.}}{%
{\protect \APACyear {2020}}%
}]{%
20sanchez}
\APACinsertmetastar {%
20sanchez}%
\begin{APACrefauthors}%
{S{\'a}nchez-Lavega}, A.%
, {Garc{\'\i}a-Melendo}, E.%
, {Legarreta}, J.%
, {Hueso}, R.%
, {del R{\'\i}o-Gaztelurrutia}, T.%
, {Sanz-Requena}, J\BPBI F.%
\BDBL {}{Ewald}, S.%
\end{APACrefauthors}%
\unskip\
\newblock
\APACrefYearMonthDay{2020}{{\APACmonth{02}}}{}.
\newblock
{\BBOQ}\APACrefatitle {{A complex storm system in Saturn's north polar
  atmosphere in 2018}} {{A complex storm system in Saturn's north polar
  atmosphere in 2018}}.{\BBCQ}
\newblock
\APACjournalVolNumPages{Nature Astronomy}{4}{}{180-187}.
\newblock
\begin{APACrefDOI} \doi{10.1038/s41550-019-0914-9} \end{APACrefDOI}
\PrintBackRefs{\CurrentBib}

\bibitem [\protect \citeauthoryear {%
{Sankar}%
\ \protect \BOthers {.}}{%
{Sankar}%
\ \protect \BOthers {.}}{%
{\protect \APACyear {2024}}%
}]{%
24sankar}
\APACinsertmetastar {%
24sankar}%
\begin{APACrefauthors}%
{Sankar}, R.%
, {Brueshaber}, S.%
, {Fortson}, L.%
, {Hansen-Koharcheck}, C.%
, {Lintott}, C.%
, {Mantha}, K.%
\BDBL {}{Orton}, G\BPBI S.%
\end{APACrefauthors}%
\unskip\
\newblock
\APACrefYearMonthDay{2024}{{\APACmonth{09}}}{}.
\newblock
{\BBOQ}\APACrefatitle {{Jovian Vortex Hunter: A Citizen Science Project to
  Study Jupiter's Vortices}} {{Jovian Vortex Hunter: A Citizen Science Project
  to Study Jupiter's Vortices}}.{\BBCQ}
\newblock
\APACjournalVolNumPages{Planetary Science Journal}{5}{9}{203}.
\newblock
\begin{APACrefDOI} \doi{10.3847/PSJ/ad6e75} \end{APACrefDOI}
\PrintBackRefs{\CurrentBib}

\bibitem [\protect \citeauthoryear {%
{Sankar}%
, {Wong}%
, {Palotai}%
\BCBL {}\ \BBA {} {Brueshaber}%
}{%
{Sankar}%
\ \protect \BOthers {.}}{%
{\protect \APACyear {2025}}%
}]{%
25sankar}
\APACinsertmetastar {%
25sankar}%
\begin{APACrefauthors}%
{Sankar}, R.%
, {Wong}, M\BPBI H.%
, {Palotai}, C.%
\BCBL {}\ \BBA {} {Brueshaber}, S.%
\end{APACrefauthors}%
\unskip\
\newblock
\APACrefYearMonthDay{2025}{{\APACmonth{05}}}{}.
\newblock
{\BBOQ}\APACrefatitle {{Wind Shear and the Role of Eddy Vapor Transport in
  Driving Water Convection on Jupiter}} {{Wind Shear and the Role of Eddy Vapor
  Transport in Driving Water Convection on Jupiter}}.{\BBCQ}
\newblock
\APACjournalVolNumPages{Planetary Science Journal}{6}{5}{109}.
\newblock
\begin{APACrefDOI} \doi{10.3847/PSJ/adcc21} \end{APACrefDOI}
\PrintBackRefs{\CurrentBib}

\bibitem [\protect \citeauthoryear {%
{Smith}%
\ \protect \BOthers {.}}{%
{Smith}%
\ \protect \BOthers {.}}{%
{\protect \APACyear {1981}}%
}]{%
81smith}
\APACinsertmetastar {%
81smith}%
\begin{APACrefauthors}%
{Smith}, B\BPBI A.%
, {Soderblom}, L.%
, {Beebe}, R\BPBI F.%
, {Boyce}, J\BPBI M.%
, {Briggs}, G.%
, {Bunker}, A.%
\BDBL {}{Suomi}, V\BPBI E.%
\end{APACrefauthors}%
\unskip\
\newblock
\APACrefYearMonthDay{1981}{{\APACmonth{04}}}{}.
\newblock
{\BBOQ}\APACrefatitle {{Encounter with Saturn - Voyager 1 imaging science
  results}} {{Encounter with Saturn - Voyager 1 imaging science
  results}}.{\BBCQ}
\newblock
\APACjournalVolNumPages{Science}{212}{}{163-191}.
\PrintBackRefs{\CurrentBib}

\bibitem [\protect \citeauthoryear {%
{Smith}%
\ \protect \BOthers {.}}{%
{Smith}%
\ \protect \BOthers {.}}{%
{\protect \APACyear {1979}}%
}]{%
79smith}
\APACinsertmetastar {%
79smith}%
\begin{APACrefauthors}%
{Smith}, B\BPBI A.%
, {Soderblom}, L\BPBI A.%
, {Johnson}, T\BPBI V.%
, {Ingersoll}, A\BPBI P.%
, {Collins}, S\BPBI A.%
, {Shoemaker}, E\BPBI M.%
\BDBL {}{Suomi}, V\BPBI E.%
\end{APACrefauthors}%
\unskip\
\newblock
\APACrefYearMonthDay{1979}{{\APACmonth{06}}}{}.
\newblock
{\BBOQ}\APACrefatitle {{The Jupiter system through the eyes of Voyager 1}}
  {{The Jupiter system through the eyes of Voyager 1}}.{\BBCQ}
\newblock
\APACjournalVolNumPages{Science}{204}{}{951-957}.
\newblock
\begin{APACrefDOI} \doi{10.1126/science.204.4396.951} \end{APACrefDOI}
\PrintBackRefs{\CurrentBib}

\bibitem [\protect \citeauthoryear {%
{Sromovsky}%
, {Karkoschka}%
, {Fry}%
, {de Pater}%
\BCBL {}\ \BBA {} {Hammel}%
}{%
{Sromovsky}%
\ \protect \BOthers {.}}{%
{\protect \APACyear {2019}}%
}]{%
18sromovsky}
\APACinsertmetastar {%
18sromovsky}%
\begin{APACrefauthors}%
{Sromovsky}, L\BPBI A.%
, {Karkoschka}, E.%
, {Fry}, P\BPBI M.%
, {de Pater}, I.%
\BCBL {}\ \BBA {} {Hammel}, H\BPBI B.%
\end{APACrefauthors}%
\unskip\
\newblock
\APACrefYearMonthDay{2019}{Jan}{}.
\newblock
{\BBOQ}\APACrefatitle {{The methane distribution and polar brightening on
  Uranus based on HST/STIS, Keck/NIRC2, and IRTF/SpeX observations through
  2015}} {{The methane distribution and polar brightening on Uranus based on
  HST/STIS, Keck/NIRC2, and IRTF/SpeX observations through 2015}}.{\BBCQ}
\newblock
\APACjournalVolNumPages{Icarus}{317}{}{266-306}.
\newblock
\begin{APACrefDOI} \doi{10.1016/j.icarus.2018.06.026} \end{APACrefDOI}
\PrintBackRefs{\CurrentBib}

\bibitem [\protect \citeauthoryear {%
Taylor%
\ \BBA {} Thompson%
}{%
Taylor%
\ \BBA {} Thompson%
}{%
{\protect \APACyear {2023}}%
}]{%
23taylor}
\APACinsertmetastar {%
23taylor}%
\begin{APACrefauthors}%
Taylor, J\BPBI R.%
\BCBT {}\ \BBA {} Thompson, A\BPBI F.%
\end{APACrefauthors}%
\unskip\
\newblock
\APACrefYearMonthDay{2023}{}{}.
\newblock
{\BBOQ}\APACrefatitle {Submesoscale dynamics in the upper ocean} {Submesoscale
  dynamics in the upper ocean}.{\BBCQ}
\newblock
\APACjournalVolNumPages{Annual Review of Fluid Mechanics}{55}{1}{103--127}.
\PrintBackRefs{\CurrentBib}

\bibitem [\protect \citeauthoryear {%
{Thomson}%
\ \BBA {} {McIntyre}%
}{%
{Thomson}%
\ \BBA {} {McIntyre}%
}{%
{\protect \APACyear {2016}}%
}]{%
16thomson}
\APACinsertmetastar {%
16thomson}%
\begin{APACrefauthors}%
{Thomson}, S\BPBI I.%
\BCBT {}\ \BBA {} {McIntyre}, M\BPBI E.%
\end{APACrefauthors}%
\unskip\
\newblock
\APACrefYearMonthDay{2016}{{\APACmonth{03}}}{}.
\newblock
{\BBOQ}\APACrefatitle {{Jupiter's Unearthly Jets: A New Turbulent Model
  Exhibiting Statistical Steadiness without Large-Scale Dissipation*}}
  {{Jupiter's Unearthly Jets: A New Turbulent Model Exhibiting Statistical
  Steadiness without Large-Scale Dissipation*}}.{\BBCQ}
\newblock
\APACjournalVolNumPages{Journal of Atmospheric Sciences}{73}{}{1119-1141}.
\newblock
\begin{APACrefDOI} \doi{10.1175/JAS-D-14-0370.1} \end{APACrefDOI}
\PrintBackRefs{\CurrentBib}

\bibitem [\protect \citeauthoryear {%
{Thorpe}%
}{%
{Thorpe}%
}{%
{\protect \APACyear {1986}}%
}]{%
86thorpe}
\APACinsertmetastar {%
86thorpe}%
\begin{APACrefauthors}%
{Thorpe}, A\BPBI J.%
\end{APACrefauthors}%
\unskip\
\newblock
\APACrefYearMonthDay{1986}{{\APACmonth{01}}}{}.
\newblock
{\BBOQ}\APACrefatitle {{Synoptic Scale Disturbances with Circular Symmetry}}
  {{Synoptic Scale Disturbances with Circular Symmetry}}.{\BBCQ}
\newblock
\APACjournalVolNumPages{Monthly Weather Review}{114}{7}{1384}.
\newblock
\begin{APACrefDOI} \doi{10.1175/1520-0493(1986)114<1384:SSDWCS>2.0.CO;2}
  \end{APACrefDOI}
\PrintBackRefs{\CurrentBib}

\bibitem [\protect \citeauthoryear {%
{Vasavada}%
\ \protect \BOthers {.}}{%
{Vasavada}%
\ \protect \BOthers {.}}{%
{\protect \APACyear {2006}}%
}]{%
06vasavada}
\APACinsertmetastar {%
06vasavada}%
\begin{APACrefauthors}%
{Vasavada}, A\BPBI R.%
, {H{\"o}rst}, S\BPBI M.%
, {Kennedy}, M\BPBI R.%
, {Ingersoll}, A\BPBI P.%
, {Porco}, C\BPBI C.%
, {Del Genio}, A\BPBI D.%
\BCBL {}\ \BBA {} {West}, R\BPBI A.%
\end{APACrefauthors}%
\unskip\
\newblock
\APACrefYearMonthDay{2006}{}{}.
\newblock
{\BBOQ}\APACrefatitle {{Cassini imaging of Saturn: Southern hemisphere winds
  and vortices}} {{Cassini imaging of Saturn: Southern hemisphere winds and
  vortices}}.{\BBCQ}
\newblock
\APACjournalVolNumPages{Journal of Geophysical Research
  (Planets)}{111}{E10}{5004}.
\newblock
\begin{APACrefDOI} \doi{10.1029/2005JE002563} \end{APACrefDOI}
\PrintBackRefs{\CurrentBib}

\bibitem [\protect \citeauthoryear {%
{Wong}%
\ \protect \BOthers {.}}{%
{Wong}%
\ \protect \BOthers {.}}{%
{\protect \APACyear {2023}}%
}]{%
23wong}
\APACinsertmetastar {%
23wong}%
\begin{APACrefauthors}%
{Wong}, M\BPBI H.%
, {Bjoraker}, G\BPBI L.%
, {Goullaud}, C.%
, {Stephens}, A\BPBI W.%
, {Luszcz-Cook}, S\BPBI H.%
, {Atreya}, S\BPBI K.%
\BDBL {}{Brown}, S\BPBI T.%
\end{APACrefauthors}%
\unskip\
\newblock
\APACrefYearMonthDay{2023}{{\APACmonth{01}}}{}.
\newblock
{\BBOQ}\APACrefatitle {{Deep Clouds on Jupiter}} {{Deep Clouds on
  Jupiter}}.{\BBCQ}
\newblock
\APACjournalVolNumPages{Remote Sensing}{15}{3}{702}.
\newblock
\begin{APACrefDOI} \doi{10.3390/rs15030702} \end{APACrefDOI}
\PrintBackRefs{\CurrentBib}

\bibitem [\protect \citeauthoryear {%
{Wong}%
\ \protect \BOthers {.}}{%
{Wong}%
\ \protect \BOthers {.}}{%
{\protect \APACyear {2025}}%
}]{%
25wong_epsc}
\APACinsertmetastar {%
25wong_epsc}%
\begin{APACrefauthors}%
{Wong}, M\BPBI H.%
, {Oyafuso}, F\BPBI A.%
, {Imai}, M.%
, {Kolma{\v{s}}ov{\'a}}, I.%
, {Mizumoto}, S.%
, {Levin}, S\BPBI M.%
\BDBL {}{Bolton}, S\BPBI J.%
\end{APACrefauthors}%
\unskip\
\newblock
\APACrefYearMonthDay{2025}{{\APACmonth{09}}}{}.
\newblock
{\BBOQ}\APACrefatitle {{Radio pulse power distribution of lightning in
  Jupiter's 2021-2022 stealth superstorms}} {{Radio pulse power distribution of
  lightning in Jupiter's 2021-2022 stealth superstorms}}.{\BBCQ}
\newblock
\BIn{} \APACrefbtitle {EPSC-DPS Joint Meeting 2025} {Epsc-dps joint meeting
  2025}\ (\BVOL\ 2025, \BPG~EPSC-DPS2025-1054).
\newblock
\begin{APACrefDOI} \doi{10.5194/epsc-dps2025-1054} \end{APACrefDOI}
\PrintBackRefs{\CurrentBib}

\bibitem [\protect \citeauthoryear {%
{Wong}%
\ \protect \BOthers {.}}{%
{Wong}%
\ \protect \BOthers {.}}{%
{\protect \APACyear {2020}}%
}]{%
20wong}
\APACinsertmetastar {%
20wong}%
\begin{APACrefauthors}%
{Wong}, M\BPBI H.%
, {Simon}, A\BPBI A.%
, {Tollefson}, J\BPBI W.%
, {de Pater}, I.%
, {Barnett}, M\BPBI N.%
, {Hsu}, A\BPBI I.%
\BDBL {}{Fletcher}, L\BPBI N.%
\end{APACrefauthors}%
\unskip\
\newblock
\APACrefYearMonthDay{2020}{{\APACmonth{04}}}{}.
\newblock
{\BBOQ}\APACrefatitle {{High-resolution UV/Optical/IR Imaging of Jupiter in
  2016-2019}} {{High-resolution UV/Optical/IR Imaging of Jupiter in
  2016-2019}}.{\BBCQ}
\newblock
\APACjournalVolNumPages{ApJ Supplement}{247}{2}{58}.
\newblock
\begin{APACrefDOI} \doi{10.3847/1538-4365/ab775f} \end{APACrefDOI}
\PrintBackRefs{\CurrentBib}

\bibitem [\protect \citeauthoryear {%
{Zhang}%
\ \BBA {} {Marcus}%
}{%
{Zhang}%
\ \BBA {} {Marcus}%
}{%
{\protect \APACyear {2024}}%
}]{%
24zhang}
\APACinsertmetastar {%
24zhang}%
\begin{APACrefauthors}%
{Zhang}, A.%
\BCBT {}\ \BBA {} {Marcus}, P\BPBI S.%
\end{APACrefauthors}%
\unskip\
\newblock
\APACrefYearMonthDay{2024}{{\APACmonth{04}}}{}.
\newblock
{\BBOQ}\APACrefatitle {{Stable three-dimensional vortex families consistent
  with Jovian observations including the Great Red Spot}} {{Stable
  three-dimensional vortex families consistent with Jovian observations
  including the Great Red Spot}}.{\BBCQ}
\newblock
\APACjournalVolNumPages{Journal of Fluid Mechanics}{984}{}{A61}.
\newblock
\begin{APACrefDOI} \doi{10.1017/jfm.2024.132} \end{APACrefDOI}
\PrintBackRefs{\CurrentBib}

\bibitem [\protect \citeauthoryear {%
{Zhang}%
\ \protect \BOthers {.}}{%
{Zhang}%
\ \protect \BOthers {.}}{%
{\protect \APACyear {2020}}%
}]{%
20zhang}
\APACinsertmetastar {%
20zhang}%
\begin{APACrefauthors}%
{Zhang}, Z.%
, {Adumitroaie}, V.%
, {Allison}, M.%
, {Arballo}, J.%
, {Atreya}, S.%
, {Bjoraker}, G.%
\BDBL {}{Wong}, M\BPBI H.%
\end{APACrefauthors}%
\unskip\
\newblock
\APACrefYearMonthDay{2020}{{\APACmonth{09}}}{}.
\newblock
{\BBOQ}\APACrefatitle {{Residual Study: Testing Jupiter Atmosphere Models
  Against Juno MWR Observations}} {{Residual Study: Testing Jupiter Atmosphere
  Models Against Juno MWR Observations}}.{\BBCQ}
\newblock
\APACjournalVolNumPages{Earth and Space Science}{7}{9}{e01229}.
\newblock
\begin{APACrefDOI} \doi{10.1029/2020EA001229} \end{APACrefDOI}
\PrintBackRefs{\CurrentBib}

\bibitem [\protect \citeauthoryear {%
Zhang%
\ \protect \BOthers {.}}{%
Zhang%
\ \protect \BOthers {.}}{%
{\protect \APACyear {2016}}%
}]{%
16zhang_ocean}
\APACinsertmetastar {%
16zhang_ocean}%
\begin{APACrefauthors}%
Zhang, Z.%
, Tian, J.%
, Qiu, B.%
, Zhao, W.%
, Chang, P.%
, Wu, D.%
\BCBL {}\ \BBA {} Wan, X.%
\end{APACrefauthors}%
\unskip\
\newblock
\APACrefYearMonthDay{2016}{}{}.
\newblock
{\BBOQ}\APACrefatitle {Observed 3D Structure, Generation, and Dissipation of
  Oceanic Mesoscale Eddies in the South China Sea} {Observed 3d structure,
  generation, and dissipation of oceanic mesoscale eddies in the south china
  sea}.{\BBCQ}
\newblock
\APACjournalVolNumPages{Scientific Reports}{6}{1}{24349}.
\newblock
\begin{APACrefURL} \url{https://doi.org/10.1038/srep24349} \end{APACrefURL}
\newblock
\begin{APACrefDOI} \doi{10.1038/srep24349} \end{APACrefDOI}
\PrintBackRefs{\CurrentBib}

\bibitem [\protect \citeauthoryear {%
Zhang%
, Wang%
, Wang%
\BCBL {}\ \BBA {} Liu%
}{%
Zhang%
\ \protect \BOthers {.}}{%
{\protect \APACyear {2024}}%
}]{%
24zhang_ocean}
\APACinsertmetastar {%
24zhang_ocean}%
\begin{APACrefauthors}%
Zhang, Z.%
, Wang, G.%
, Wang, H.%
\BCBL {}\ \BBA {} Liu, H.%
\end{APACrefauthors}%
\unskip\
\newblock
\APACrefYearMonthDay{2024}{}{}.
\newblock
{\BBOQ}\APACrefatitle {Three-Dimensional Structure of Oceanic Mesoscale Eddies}
  {Three-dimensional structure of oceanic mesoscale eddies}.{\BBCQ}
\newblock
\APACjournalVolNumPages{Ocean-Land-Atmosphere Research}{3}{}{0051}.
\newblock
\begin{APACrefURL} \url{https://spj.science.org/doi/abs/10.34133/olar.0051}
  \end{APACrefURL}
\newblock
\begin{APACrefDOI} \doi{10.34133/olar.0051} \end{APACrefDOI}
\PrintBackRefs{\CurrentBib}

\end{thebibliography}

%
%
%
%
%

\end{document}